\documentclass[12pt, oneside, paper=letter, DIV=15, BCOR=0mm, abstract=true, headings=small]{scrartcl}
\usepackage[utf8]{inputenc}

\usepackage[headsepline 
]{scrlayer-scrpage}

\usepackage[toc,page]{appendix}

\newcommand{\versionnumber}{1.4}

\newcommand{\version}{\versionnumber~(20th October 2024)}

\newcommand{\ifootertext}{IG 2.0 Codebook (Version:~\versionnumber)}

\newcommand{\ofootertext}{}

\automark[subsubsection]{subsubsection}
\ifoot{\ifootertext}
\ofoot{\ofootertext}

\usepackage{graphicx}

\usepackage[x11names]{xcolor}

\usepackage{makeidx}         

\usepackage{booktabs}

\usepackage{threeparttable}

\usepackage{multirow}

\usepackage[hidelinks]{hyperref}        

\usepackage[style=apa,natbib,uniquename=false]{biblatex} 
\DeclareSourcemap{ 
    \maps[overwrite, datatype=bibtex]{
        \map{ 
            \step[fieldsource=doi,
                  match=\regexp{\{\\\_\}|\{\_\}|\\\_},
                  replace=\regexp{\_}]
        }
        \map{ 
            \step[fieldsource=doi,
                  match=\regexp{\{\$\\sim\$\}|\{\~\}|\$\\sim\$},
                  replace=\regexp{\~}]
        }
        \map{ 
            \step[fieldsource=doi,
                  match=\regexp{\{\\\x{26}\}},
                  replace=\regexp{\x{26}}]
        }
    }
}

\usepackage{longtable}

\usepackage{mathtools}
\usepackage{bm}

\usepackage{newtxmath}

\usepackage[capitalize, noabbrev]{cleveref}

\newcommand*\subtxtNormal[1]{_{\textup{#1}}} 

\newcommand{\corecolor}{blue}
\newcommand{\corecolorname}{blue}
\newcommand{\corecolorconst}{purple}
\newcommand{\corecolorconstname}{purple}
\newcommand{\extendedcolor}{Green4}
\newcommand{\extendedcolorname}{green}
\newcommand{\logicocolor}{OrangeRed2}
\newcommand{\logicocolorname}{orange}
\newcommand{\neutralcolor}{black}

\newcommand{\format}[1]{\textbf{\textcolor{\corecolor}{#1}}} 
\newcommand{\formatConst}[1]{\textbf{\textcolor{\corecolorconst}{#1}}} 
\newcommand{\formatExt}[1]{\textbf{\textcolor{\extendedcolor}{#1}}} 
\newcommand{\formatLog}[1]{\textbf{\textcolor{\logicocolor}{#1}}} 
\newcommand{\formatNeutral}[1]{\textbf{\textcolor{\neutralcolor}{#1}}} 
\newcommand{\formatNeutralIt}[1]{\textit{\textcolor{\neutralcolor}{#1}}} 

\newcommand{\A}[1]{\textcolor{\corecolor}{$\textbf{A#1}$}} 

\newcommand{\Bext}[1]{\textcolor{\extendedcolor}{$\textbf{B#1}$}} 

\newcommand{\Bdirraw}{Bdir} 
\newcommand{\Bdir}[1]{\textcolor{\corecolor}{$\textbf{Bdir#1}$}} 

\newcommand{\Bindraw}{Bind} 
\newcommand{\Bind}[1]{\textcolor{\corecolor}{$\textbf{Bind#1}$}} 

\newcommand{\Cacraw}{C$\subtxtNormal{ac}$} 
\newcommand{\Cac}{\textcolor{\corecolor}{$\textbf{Cac}$}} 
\newcommand{\Cacconst}{\textcolor{\corecolorconst}{$\textbf{Cac}$}} 

\newcommand{\Cexraw}{C$\subtxtNormal{ex}$} 
\newcommand{\Cex}{\textcolor{\corecolor}{$\textbf{Cex}$}} 
\newcommand{\Cexconst}{\textcolor{\corecolorconst}{$\textbf{Cex}$}} 

\newcommand{\I}{\format{I}} 

\newcommand{\D}{\format{D}} 

\newcommand{\M}{\formatConst{M}} 

\newcommand{\Pp}{\formatConst{P}} 

\newcommand{\PpExt}[1]{\textcolor{\extendedcolor}{$\textbf{P#1}$}} 

\newcommand{\E}{\formatConst{E}} 

\newcommand{\F}{\formatConst{F}} 

\newcommand{\Oe}{\format{O}} 

\newcommand{\Oeconst}{\formatConst{O}} 

\newcommand{\prop}[1]{\format{p#1}} 
\newcommand{\propconst}[1]{\formatConst{p#1}} 
\newcommand{\propExt}[1]{\formatExt{p#1}} 

\newcommand{\lp}{\formatNeutral{(}} 
\newcommand{\rp}{\formatNeutral{)}} 
\newcommand{\lpconst}{\formatNeutral{(}} 
\newcommand{\rpconst}{\formatNeutral{)}} 
\newcommand{\lb}{\formatExt{\{}} 
\newcommand{\rb}{\formatExt{\}}} 
\newcommand{\ls}{\formatNeutral{[}} 
\newcommand{\rs}{\formatNeutral{]}} 
\newcommand{\lsi}{\formatNeutralIt{[}} 
\newcommand{\rsi}{\formatNeutralIt{]}} 

\newcommand{\customExtRaw}[1]{\formatExt{#1}} 
\newcommand{\customExt}[1]{\formatExt{\ls{}#1\rs{}}} 
\newcommand{\customLog}[1]{\formatLog{\ls{}#1\rs{}}} 

\newcommand{\AND}{\formatNeutral{[AND]}}
\newcommand{\ANDconst}{\formatNeutral{[AND]}}
\newcommand{\OR}{\formatNeutral{[OR]}}
\newcommand{\ORconst}{\formatNeutral{[OR]}}
\newcommand{\XOR}{\formatNeutral{[XOR]}}
\newcommand{\XORconst}{\formatNeutral{[XOR]}}
\newcommand{\NOT}{\formatNeutral{[NOT]}}
\newcommand{\NOTconst}{\formatNeutral{[NOT]}}
\newcommand{\ORELSE}{\formatNeutral{OR ELSE}}

\newcommand{\Ureg}{U$\subtxtNormal{reg}$}
\newcommand{\Ucon}{U$\subtxtNormal{con}$}

\newcommand{\EntryDecompositionOfComponentLevelCombinations}{\begin{emp}Decomposition of component-level combinations\end{emp}}

\newcommand{\EntryLogicalRelationshipsAmongComponents}{\begin{emp}Logical relationships among statement components\end{emp}}

\newcommand{\EntryCrossComponentSemanticAnnotations}{\begin{emp}Cross-Component Semantic Annotations\end{emp}}

\newcommand{\ctx}{ctx}
\newcommand{\anim}{anim}
\newcommand{\metatype}{metatype}
\newcommand{\role}{role}
\newcommand{\polref}{ref}
\newcommand{\confunc}{confunc}
\newcommand{\regfunc}{regfunc}

\renewcommand{\sp}{\vspace{0.5cm}}

\newenvironment{hil} 
{\bfseries}{}

\newenvironment{emp} 
{\itshape}
{}

\newcommand{\fullref}[1]{\begin{emp}\cref{#1} \nameref{#1}\end{emp}} 

\newenvironment{ptit} 
{
\vspace{0.2cm}
\bfseries
\noindent}
{\vspace{0.2cm}}

\newenvironment{exa} 
{   
    \vspace{0.1cm}
    \bgroup
    \par
    \itshape
}
{%
    \par\egroup
    \vspace{0.1cm}
}

\newenvironment{exi} 
{\itshape}
{}

\usepackage{soul} 

\newcommand{\udl}[1]{\ul{#1}}

\usepackage[most]{tcolorbox}

\newcommand{\framedfigure}[1]{
\tcbox[colframe=black,
               colback=Snow1!30]{#1}
}

\newcommand{\boxbgcolor}{Snow1}

\usepackage{pdflscape}

\usepackage[vmargin=1.4cm,hmargin=2cm,
headsep = \dimexpr2\baselineskip-5mm\relax,
footskip = \dimexpr1\baselineskip+5mm\relax]{geometry}

\newcommand{\tabledims}{
p{3cm} p{7cm} p{7cm} p{6cm} 
}

\newcommand{\tablespan}{
p{25cm}
}

\newcommand{\tablespanNarrow}{p{17cm}}

\title{Institutional Grammar 2.0 Codebook}
\author{Christopher K.~Frantz\footnote{Affiliation: Norwegian University of Science and Technology, E-mail: christopher.frantz@ntnu.no}\qquad Saba N.~Siddiki\setcounter{footnote}{6}\footnote{Affiliation: Syracuse University, E-mail: ssiddiki@maxwell.syr.edu}}

\date{Version: \version}

\begin{document}

\maketitle

\section*{Abstract}
The Grammar of Institutions, or Institutional Grammar, is an established approach to encode policy information in terms of institutional statements based on a set of pre-defined syntactic components. This codebook provides coding guidelines for a revised version of the Institutional Grammar, the Institutional Grammar 2.0 (IG 2.0)~\citep{Frantz2021,Frantz2022InstitutionalGrammar}. IG 2.0 is a specification that aims at facilitating the encoding of policy to meet varying analytical objectives. To this end, it revises the grammar with respect to comprehensiveness, flexibility, and specificity by offering multiple levels of expressiveness (IG Core, IG Extended, IG Logico). In addition to the encoding of regulative statements, it further introduces the encoding of constitutive institutional statements, as well as statements that exhibit both constitutive and regulative characteristics. As a supplementary resource to the IG 2.0 specification, the codebook initially covers fundamental concepts of IG 2.0, before providing an overview of pre-coding steps relevant for document preparation. Detailed coding guidelines are provided for both regulative and constitutive statements across all levels of expressiveness, along with the encoding guidelines for statements of mixed form -- hybrid and polymorphic institutional statements. The document further provides an overview of taxonomies used in the encoding process and referred to throughout the codebook. The codebook concludes with a summary and discussion of relevant considerations to facilitate the coding process. An initial Reader's Guide helps the reader tailor the content to her interest.

\tableofcontents

\newpage

\clearpairofpagestyles
\chead{\headmark}
\automark[section]{section}
\ifoot{\ifootertext}
\ofoot{\ofootertext}
\cfoot{\thepage}

\section*{Acknowledgements}

The authors wish to explicitly acknowledge the valuable input and feedback provided by Ute Brady, Edella Schlager, Matia Vannoni, and the IGRI interns as part of ongoing evaluation activities. We are further grateful for the constructive feedback on earlier codebook drafts received from Seth Frey, Bartosz Pieli\'{n}ski, and Marcello Ceci.

\section*{Document Revisions}

All publicly released versions of this document can be retrieved under \url{https://arxiv.org/abs/2008.08937}.

\begin{table}[h]
\begin{tabular}{p{2.2cm} p{14.3cm}}
\toprule
\textbf{Version} & \textbf{Description} \\
\toprule

\version & Refined selected IG Script examples to reflect current feature set, extended study checklist in Appendix A to offer clearer guidance for the development of project-specific coding instructions \\

\midrule

1.3~(3rd March 2022) & Recoded examples in Section 4 using the IG Script notation, aligned taxonomies with book information, revised selected figures, conceptual and methodological overview \\

\midrule

1.2~(20th June 2021) & 

Clarified definitions, restructured Context Taxonomy, introduced explicit guidelines for Attribute/Attribute Property differentiation, revised selected examples, added statement-level annotation characterizations, refined selected figures \\

\midrule

1.1~(6th December 2020) &

Refinement of component definitions and conceptual background; constitutive `Modal'; refinement of institutional function characterization as regulative and constitutive functions respectively; addition of differentiation heuristics for constitutive and regulative statements; addition of `Domain', `Cause/Reason' and `Event' categories in Context Taxonomy, Metatype Taxonomy; clarification of component-level nesting; extended example coding for constituted entity; introduction of Reader's Guide
\\

\midrule

1.0~(15th August 2020) &

Initial codebook version (underlying concepts, pre-coding steps, coding guidelines for regulative and constitutive statements, constitutive-regulative statements, polymorphic statements, annotation taxonomies) \\

\toprule
\end{tabular}
\end{table}

\section{Introduction}
\label{sec:Introduction}

\subsection{The Codebook at a Glance}

The following instructions are intended to provide guidance for the coding of institutional statements, the focal unit of analysis in the Institutional Grammar~\citep{Crawford1995,Crawford2005}, with specific focus on the Institutional Grammar 2.0\index{Institutional Grammar 2.0} (IG 2.0)~\citep{Frantz2021,Frantz2022InstitutionalGrammar}, to which this text is supplementary.\footnote{In addition to \citet{Crawford1995,Crawford2005}, the specification draws on the original IG codebook \citep{Brady2018InstitutionalGuidelines}, and conceptual refinements \citep{Siddiki2011,Frantz2013} comprehensively integrated into the IG 2.0 \citep{Frantz2022InstitutionalGrammar}.} \begin{emp}An institutional statement\index{Institutional Statement} describes expected actions for actors within the presence or absence of particular constraints, or parameterizes features of an institutional system.\end{emp}~\citep{Frantz2021} Institutional statements convey information that contextualizes their applicability. They vary in prescriptiveness and force, as reflected by the presence of information that more or less strongly compels behavior and by the presence of information that specifies payoffs for compliance, or non-compliance, with statements instructions. Varying in the inclusion of these various kinds  of information, institutional statements typically take two functional forms: constitutive and regulative. \begin{emp}Constitutive statements\end{emp} parameterize an institutional setting, and in this process, introduce, modify or otherwise constitute features of an institutional system and action situations embedded therein, including aspects such as actor positions and roles, associated affordances, processes, environmental characteristics, objects and artifacts inasfar as they are institutionally relevant. \begin{emp}Regulative statements\end{emp} describe actors' duties and discretion linked to specific actions within certain contextual parameters. 

According to the IG 2.0, institutional statements are commonly comprised of a set of syntactic components, with individual components associating with unique information, and which combine to convey a statement's institutional meaning. Broadly characterized,\footnote{A precise definition of each component is provided in \cref{subsec:Syntax}.} \begin{emp}regulative statements\end{emp} are composed of some or all of the following components with the corresponding syntactic labels: 
 (i) a responsible actor, referred to as \begin{emp}Attributes\end{emp}; 
 (ii) action regulated by the statement, referred to as an \begin{emp}Aim\end{emp}; 
 (iii) statement context, referred to as \begin{emp}Context\end{emp}; 
 (iv) a receiver of action, referred to as an \begin{emp}Object\end{emp};
 (v) a prescriptive operator that describes how strongly an action is compelled or restrained, referred to as a \begin{emp}Deontic\end{emp}; and 
 (vi) a consequence of violating the regulated action, referred to as an \begin{emp}Or else\end{emp}.

\begin{emp}Constitutive statements\end{emp} are composed of some or all of the following components with the corresponding syntactic labels: 
%
 (i) the entity that is being constituted or directly modified within a statement, referred to as a \begin{emp}Constituted Entity\end{emp}; 
 (ii) a parameterizing activity or function that introduces or otherwise characterizes the Constituted Entity in relation to the institutional setting and potential Constituting Properties, called the \begin{emp}Constitutive Function\end{emp}; 
 (iii) the statement context, referred to as \begin{emp}Context\end{emp}; 
 (iv) properties that serve as input to the constitutive function, called \begin{emp}Constituting Properties\end{emp}; 
 (v) a modal operator that defines to which extent the constitutive function of an institutional statement is required (necessary) or merely possible (optional), referred to as a \begin{emp}Modal\end{emp}; and 
 (vi) a consequence linked to the non-fulfillment of the function referenced in the constitutive function, referred to as an \begin{emp}Or else\end{emp}. 

The operational definition of an institutional statement is tied to the presence of certain syntactic components, or \begin{emp}necessary components\end{emp}. 
To qualify as a complete regulative institutional statement, the statement must at least contain Attributes, Aim, and Context components (\begin{emp}necessary components\end{emp}; however, they can be contextually implied or inferred\footnote{This is discussed in greater detail as part of the component definitions provided in \cref{sec:Definitions}.}). The \begin{emp}Object\end{emp}, \begin{emp}Deontic\end{emp}, and \begin{emp}Or else\end{emp} components are deemed \begin{emp}sufficient components\end{emp}. Similarly, the presence of \begin{emp}Constituted Entity\end{emp}, \begin{emp}Constitutive Function\end{emp} and \begin{emp}Context\end{emp} components is necessary to qualify as a constitutive institutional statement; \begin{emp}Constituting Properties\end{emp}, \begin{emp}Modal\end{emp} and \begin{emp}Or else\end{emp} are sufficient components. 

The elementary form of an institutional statement is the \begin{emp}atomic institutional statement\end{emp}. An atomic institutional statement is a statement of regulative or constitutive kind that contains the corresponding necessary components, alongside potential optional components other than the \begin{emp}Or else\end{emp}, and in which none of the components contains \begin{emp}component-level combinations\end{emp} (e.g., multiple attributes, combinations of aims, etc.). Building on this concept, IG 2.0 introduces the concept of nested institutional statements in order to capture complex institutional configurations expressed in terms of institutional statements of various kinds interlinked in various forms, alongside additional features to facilitate structural and semantic decomposition of institutional arrangements at various levels of detail.

In this codebook, guidance is offered for the encoding of regulative and constitutive institutional statements along the aforementioned syntactic components at three levels of expressiveness: (1) \begin{emp}IG Core\end{emp}; (2) \begin{emp}IG Extended\end{emp}; and (3) \begin{emp}IG Logico\end{emp}. The core definitions of syntactic components remain the same across levels of expressiveness. However, the level of granularity with which components are parsed differs across levels.  \cref{sec:Definitions} provides elaborated definitions of syntactic components that generalize across levels of expressiveness. This is followed by a terminological overview of concepts used in IG 2.0. \cref{sec:PrecodingSteps} provides an overview of the pre-coding, or pre-processing steps relevant for document preparation prior to coding. \cref{sec:CodingGuidelines} specifies the syntactic conventions used in this document, followed by the coding guidelines by syntactic component and level of expressiveness. This further includes the discussion of hybrid forms of institutional statements. \cref{sec:Taxonomies} provides a comprehensive overview of taxonomies used for semantic annotations and referenced throughout the different sections. \cref{sec:Conclusion} offers a concluding discussion of considerations of relevance in the coding process, alongside references to additional resources.

To better guide the reader through the specification and guidelines provided in this document, the following section sketches potential `pathways' through the codebook that emphasize aspects of interest tailored to specific audiences.

\subsection{Reader's Guide}
\label{subsec:ReadersGuide}

The codebook is intended to cover features of IG 2.0 comprehensively. While underlying conceptual principles will be discussed elsewhere,\footnote{A more comprehensive treatment of motivation and conceptual underpinnings will be provided in forthcoming companion literature.} a specific emphasis lies on providing concrete instructions to ensure a practical use, while aiming at being comprehensive at the same time. This codebook thus highlights features that are of primary concern for different audiences, such as the researcher aiming at gaining a comprehensive picture of the opportunities provided by IG 2.0, the study designer who is primarily concerned with methodological aspects, as well as operational coders of different backgrounds and experience. Finally, researchers from diverse disciplines may find interest in identifying opportunities that IG 2.0 provides more generally, but also to develop insights how  the principles of the institutional grammar can be linked to concepts established in the researcher's domain. With those audiences in mind, in the following, we highlight selected \begin{emp}pathways through the codebook\end{emp} to increase accessibility and streamline the contents for a given readership. Note that all references to sections, tables and other items throughout this document are clickable to support efficient navigation.

\paragraph{Pathway `General Overview'}

Where readers intend to gather a high-level understanding of IG 2.0, the basic concepts, but limited concern for specifics of coding, we suggest consulting the following sections:

\begin{itemize}
\item \fullref{sec:Definitions} provides a high-level conceptual backdrop to provide a basic understanding of all the involved concepts. 
\item To develop an intuitive understanding syntax and features across different levels, the review of \fullref{subsec:RegulativeStatementCoding} and \fullref{subsec:ConstitutiveStatementCoding} is recommended. 
\item Where specifics related to operational coding are of interest, readers may narrow down specific features of interest using \cref{tab:FeatureSectionCombinations}.\footnote{The pathway `Coder' highlights the exemplary use of \cref{tab:FeatureSectionCombinations}.}
\item \fullref{sec:Conclusion} offers a summative overview, alongside important considerations for the use of IG 2.0, as well as pointers to further resources.
\end{itemize}

\paragraph{Pathway `Research Design'}

Where readers seek to consider important aspects for study design for IG 2.0, such as a background for considering coding levels and statement types, as well as pre-processing guidelines for documents, we suggest consulting the following sections:

\begin{itemize}
\item \fullref{sec:Definitions} offers an overview of the essential concepts and high-level operationalization. 
\item \fullref{sec:PrecodingSteps} highlight all preprocessing steps the designer needs to consider for document preparation.
\item \fullref{sec:CodingGuidelines} should be reviewed at high level in order to 
	\begin{itemize}
	\item determine the suitable level of encoding (IG Core, IG Extended, IG Logico, or combinations thereof -- see \fullref{subsec:IgCodingLevels}), and 
	\item identify the types of institutional statements to be considered in the study (regulative (\fullref{subsec:RegulativeStatementCoding}), constitutive (\fullref{subsec:ConstitutiveStatementCoding}. Where combinations of both are relevant, the reader should further consider \fullref{subsec:ConstitutiveRegulativeHybrids}. 
	\end{itemize}
	Note that \cref{tab:FeatureSectionCombinations}\footnote{The pathway `Coder' highlights the exemplary use of \cref{tab:FeatureSectionCombinations}.} supports the selection of relevant subsections corresponding to the desired feature set. 

Especially where both constitutive and regulative statements are considered, a specific emphasis should lie on \fullref{subsec:StatementTypesHeuristics}, to inform the development of project-specific coding guidelines for both statement types.
\item Review \fullref{subsec:IgProfiles} to identify the specific feature set to be used in the study.
\item Where the handling of specific data structures is of concern (e.g., extraction of numerical information, collections), \fullref{subsec:DataStructurePatterns} offers specific directions.
\item \fullref{sec:Conclusion} highlights some considerations for the use of IG 2.0.
\end{itemize}

\paragraph{Pathway `Coder'}

Where the reader is primarily concerned with the operational coding of statements in a given study, the selection of consulted sections varies based on background and study design. The `Coder' pathway thus offers the most variable configuration of all suggested paths.

To tailor the reading accordingly, the coder should reflect on the following questions:

\begin{itemize}
\item Is the coder fundamentally acquainted with concepts of IG 2.0? If not, consult relevant sections in \fullref{sec:Definitions} as identified in conjunction with the next question.

\item Does the coding involve regulative and/or constitutive statements? 

\item Which level of expressiveness is the coding performed at (IG Core, IG Extended, IG Logico)?

Depending on the combination of statements and level of expressiveness, choose the corresponding sections within this document by consulting \cref{tab:FeatureSectionCombinations} (Usage instructions are provided alongside the table). 

Example: Readers who wish to encode regulative statements on IG Extended level, should first consult relevant foundation sections, followed by \fullref{subsec:CodingSyntax} for an overview of the syntax employed in all coding examples. To engage in the actual coding, the reader should review \fullref{subsec:RegulativeStatementCoding}, and therein specifically \fullref{subsubsec:IGCoreRegulative} and \fullref{subsubsec:IGExtendedRegulative}. These sections provide instructions for IG Core as a coding basis, as well as IG Extended, which builds on the IG Core coding. 


Where, for example, both regulative and constitutive statements are of concern, the reader is encouraged to consult \fullref{subsec:ConstitutiveRegulativeHybrids}, in addition to the specific guidelines corresponding to statement type (\fullref{subsec:RegulativeStatementCoding} and \fullref{subsec:ConstitutiveStatementCoding}) and sections relevant for specific levels of expressiveness. 

\item Note: If both regulative and constitutive statements are coded, \fullref{subsec:StatementTypesHeuristics} provides essential heuristics to ensure unambiguous characterization of statements, in addition to guidelines related to the coding of hybrid forms of institutional statements (\fullref{subsec:ConstitutiveRegulativeHybrids}). 

\item Additional information related to the coding of specific data structures is provided in \fullref{subsec:DataStructurePatterns}. 

\end{itemize}

\begin{table}
\begin{threeparttable}
\begin{tabular}{p{3.5cm} | c c c | c c c}
\toprule
{\textbf{Relevant Section}} &
{\textbf{Regulative}} &
{\textbf{Constitutive}} &
{\textbf{Hybrids}} &
{\textbf{IG Core}} &
{\textbf{IG Extended}} &
{\textbf{IG Logico}} \\
\toprule

Sections \ref{subsec:Syntax}, \ref{subsec:InstitutionalStatementAssumptions}, \ref{subsec:ActionSituation} \& \ref{subsec:IgCodingLevels} & \textbullet & \textbullet & \textbullet & \textbullet & \textbullet & \textbullet \\
\cref{subsec:ComponentLevelNesting} & & & \textbullet & & \textbullet & $\circ$ \\
\cref{subsec:AttributeObjectHierarchy} & & & & & \textbullet & $\circ$ \\
\cref{subsec:StatementTypesHeuristics} & \textbullet & \textbullet & \textbullet & & & \\

\midrule
\cref{subsec:CodingSyntax} & \textbullet & \textbullet & \textbullet & \textbullet & \textbullet & \textbullet \\

\midrule
\cref{subsec:RegulativeStatementCoding} & \textbullet & & \textbullet & \textbullet & \textbullet & \textbullet \\
\cref{subsubsec:IGCoreRegulative} & \textbullet & & \textbullet & \textbullet & $\circ$ & $\circ$ \\
\cref{subsubsec:IGExtendedRegulative} & \textbullet & & \textbullet & & \textbullet & $\circ$ \\
\cref{subsubsec:IGLogicoRegulative} & \textbullet & & $\circ$ & & & \textbullet \\

\midrule
\cref{subsec:ConstitutiveStatementCoding} & & \textbullet & \textbullet & \textbullet & \textbullet & \textbullet \\
\cref{subsubsec:IGCoreConstitutive} & & \textbullet & \textbullet & \textbullet & $\circ$ & $\circ$ \\
\cref{subsubsec:IGExtendedConstitutive} & & \textbullet & \textbullet & & \textbullet & $\circ$ \\
\cref{subsubsec:IGLogicoConstitutive} & & \textbullet & $\circ$ & & & \textbullet \\

\midrule
\cref{subsec:ConstitutiveRegulativeHybrids} & & & \textbullet & & & \\

\midrule
\cref{sec:Taxonomies} & & & & & $\circ$\tnote{1} & \textbullet \\

\toprule
\end{tabular}
\begin{tablenotes}
\item[1] The Context Taxonomy in \cref{subsec:ContextTaxonomy} is relevant for IG Extended coding.

\vspace{0.3cm}
Usage Instructions: This table indicates the sections relevant to the coder based on feature sets of interest. Select the feature(s) of interest (types of statements; level of expression) in the columns to identify the relevant sections. Filled circles signal strong relevance for the subject of concern; hollow circle indicate indirect or partial relevance. 
\end{tablenotes}

\end{threeparttable}

\caption{Relevant Coding Instructions by Feature(s) of Interest}
\label{tab:FeatureSectionCombinations}
\end{table}

\paragraph{Pathway `Coder experienced in the Original IG'}

Where coders are experienced in the coding of the original institutional grammar, and intend to adapt their coding to IG 2.0 without consideration of extended feature sets, the following sections are of relevance:

\begin{itemize}
\item \fullref{sec:Definitions} develops a basic understanding of the core concepts; specifically \fullref{subsec:Syntax}, and \fullref{subsec:InstitutionalStatementAssumptions} are important.
\item To understand the basic coding of regulative syntax, review \fullref{subsec:CodingSyntax}, as well as the overview of statement structure in \fullref{subsec:RegulativeStatementCoding}, and the coding described in \fullref{subsubsec:IGCoreRegulative} are relevant.
\item To capture `Context' (in loose correspondence to the Conditions characterization in the original IG (see \cite{Brady2018InstitutionalGuidelines}), the Item `Context' in \cref{tab:CodingIgExtendedRegulative} (in \cref{subsubsec:IGExtendedRegulative}) offers important clarifications, alongside the Context Taxonomy in \cref{subsec:ContextTaxonomy}.
\end{itemize}

\paragraph{Pathway `Analytical Linkage'}

Where readers are primarily concerned with analytical opportunities associated with semantic annotations introduced in IG 2.0, e.g., in the form of domain-specific annotation sets, or attempt to draw relationships to concepts within their domain, the reader is encouraged to consult the following sections:

\begin{itemize}
\item \fullref{sec:Definitions} develops a basic understanding of the core concepts.
\item The forthcoming companion literature (linked here once available) will provide the analyst with a detailed overview of levels of expressiveness and associated analytical opportunities. 
\item The analyst should be clear about the level that corresponds to analytical needs and read sections accordingly, guided by the trade-off of shallow (IG Core) vs.~deep structural coding (IG Extended), the need for semantic annotations (IG Logico), as well as the involved statement types, or any mix thereof (see \fullref{subsec:IgCodingLevels}). To identify relevant sections, the reader is encouraged to draw on \cref{tab:FeatureSectionCombinations}. 
\item Specifically relevant for the conceptual linkage to domain-specific frameworks or concepts is the consideration of semantic annotations more generally: IG Logico coding is shown in \fullref{subsubsec:IGLogicoRegulative} for regulative statements, and in \fullref{subsubsec:IGLogicoConstitutive} for constitutive statements; associated taxonomies can be found in \fullref{sec:Taxonomies}.
\end{itemize}

Where the reader cannot identify with a specific audience, a general guidance is to follow the `General Overview' pathway, with focus on conceptual foundations initially (\fullref{sec:Definitions}), followed by a selective overview of the coding principles for specific features in \cref{sec:CodingGuidelines}, and finally, the concluding section (\cref{sec:Conclusion}). 

\newpage

\section{Conceptual Foundations of the Institutional Grammar 2.0}
\label{sec:Definitions}

\clearpairofpagestyles
\chead{Conceptual Foundations of the Institutional Grammar 2.0}
\automark[subsection]{subsection}
\ifoot{\ifootertext}
\ofoot{\ofootertext}
\cfoot{\thepage}

IG 2.0 is premised on a set of syntactic definitions, conceptualizations of institutional statements, and assumptions regarding institutional statements, as well as various forms of nesting (statement-level and component-level nesting). While these definitions, conceptualizations, and assumptions generalize across levels of IG encoding, how they are captured depends on at which level the encoder is working. In this section, we will thus lay out these foundational syntactic definitions, concepts, and assumptions. We start by offering complete definitions of IG components more generally, and then move to defining key institutional statement concepts, including the various notions of nested institutional statements and associated assumptions. We further provide principles and operational guidelines for the identification of statements in as far as they relate to the fundamental concept \begin{emp}Action Situation\end{emp}, namely context characterizations embedded in institutional statements and identification of institutional statement types in the first place. Concluding this section, we organize the coding levels based on involved features, along with providing principal guidelines for the coding process. Note that most concepts in this section are introduced with a pragmatic focus on providing essential foundational understanding to perform the coding of institutional statements. For an extended treatment of the conceptual foundations, refer to \citet{Frantz2022InstitutionalGrammar}.

\subsection{Syntactic Definitions of Institutional Statement Components}
\label{subsec:Syntax}

The IG structure as referred to in this document relies on the elementary syntactic components of regulative and constitutive statements highlighted in Tables \ref{tab:SyntacticElementsRegulative} and \ref{tab:SyntacticElementsConstitutive}. Definitions of these components that hold across IG 2.0 encoding levels are provided alongside each syntactic component.

\begin{table}[h!]
\centering
\begin{tabular}{p{2.3cm} p{14cm}}
\toprule
\textbf{Syntactic Element} & 
\textbf{Definition} \\
\toprule
Attributes & An actor (individual or corporate) that carries out, or is expected to/to not carry out, the action (i.e., Aim) of the statement. The Attributes component may also contain descriptors of the actor. 

The presence of this component is \begin{emp}necessary\end{emp} for any regulative statement.\\
\midrule
Deontic & A prescriptive or permissive operator that defines to what extent the action of an institutional statement is compelled, restrained, or discretionary. 

In institutional statements varying levels of prescriptiveness or strength thereof are represented with different terms or language that situate along a continuum.

The presence of this component is \begin{emp}optional\end{emp} for regulative statements.\\
\midrule
Aim & The goal or action of the statement assigned to the statement Attribute. 

The presence of this component is \begin{emp}necessary\end{emp} for any regulative statement.\\
\midrule
Object & The inanimate or animate part of an institutional statement that is the receiver of the action captured in the Aim. Objects can be of \begin{emp}direct\end{emp} or \begin{emp}indirect\end{emp} nature. Indirect objects are objects that are affected or targeted by the application of the Aim to direct objects. Objects can both be real-world entities, or abstract ones (e.g., beliefs, concepts, facts, procedures). Abstract objects here further include complex constructs, such as conditions/processes that the responsible actor is or will be acting upon. Such complex objects  can potentially take the structural form of institutional statements themselves. 

The presence of this component is \begin{emp}optional\end{emp} for regulative statements.\\
\midrule
Context & The context component instantiates settings in which the focal action of a statement applies, or qualifies the action indicated in an institutional statement. The former type of Context is referred to as an \begin{emp}``Activation Condition''\end{emp}. The latter type of Context is referred to as an \begin{emp}``Execution Constraint''\end{emp}. Both can occur in a given institutional statement, including multiples of either type. 

The presence of this component is \begin{emp}necessary\end{emp} for any regulative statement, but can be implied (as with the original institutional grammar). Where no explicit Activation Condition is specified, the context clause is by default ``under all conditions''. Where no explicit Execution Constraints are specified, the context clause is by default ``no constraints''. 

It is important to note that \begin{emp}Context\end{emp} in institutional statements reflects the context specific to the coded statement (Statement Context), as opposed to capturing context in the wider sense, making reference to the context of the policy or the domain more generally.\\
\midrule
Or else & A sanctioning provision associated with the action indicated in a particular institutional statement represented in a nested institutional statement (as defined in the following discussion). Where combinations of regulative and constitutive statements are applied (hybrid institutional statements), the consequence can be existential in kind (see \cref{tab:SyntacticElementsConstitutive} and \cref{subsec:ConstitutiveRegulativeHybrids}).

The presence of this component is \begin{emp}optional\end{emp} for regulative statements.\\
\toprule
\end{tabular}
\caption{Definitions of Syntactic Elements for Regulative Statements}
\label{tab:SyntacticElementsRegulative}
\end{table}

\begin{table}[h!]
    \centering
\begin{tabular}{p{2.3cm} p{14cm}}
\toprule
\textbf{Syntactic Element} & 
\textbf{Definition} \\
\toprule
Constituted Entity & The entity being constituted, reconstituted, modified or otherwise directly affected within an institutional statement as characterized by the constitutive function. This can be an entity as relevant in the institutional setting, or the policy itself. 

The presence of this component is \begin{emp}necessary\end{emp} for any constitutive statement.\\
\midrule
Modal & An operator that signals necessity or (im-)possibility of constitution or modification of system features captured in the constitutive function. 

In institutional statements varying levels of necessity or strength thereof are represented with different terms or language that situate along a continuum. 

The presence of this component is \begin{emp}optional\end{emp} for constitutive statements.\\
\midrule
Constitutive Function & An expression that relates the constituted entity and the institutional setting, e.g., by defining, establishing, or modifying the former, including the conferral of status. Where constituting
properties exist, constitutive functions functionally link Constituted Entity and Constituting Properties. 

The presence of this component is \begin{emp}necessary\end{emp} for any constitutive statement.\\
\midrule
Constituting Properties & Constituting Properties parameterize the Constituted Entity via the Constitutive Function, and can both be physical or abstract in kind. Constituting Properties may or may not be present in constitutive statements. 

The presence of this component is \begin{emp}optional\end{emp} for constitutive statements.\\
\midrule
Context & The Context instantiates settings in which the focal Constitutive Function of a statement applies, or qualifies the function indicated in an institutional statement. The former type of Context is referred to as an \begin{emp}``Activation Condition.''\end{emp} The latter type of Context is referred to as an \begin{emp}``Execution Constraint.''\end{emp} Both can occur in a given institutional statement, including multiples of either type. 

The presence of this component is \begin{emp}necessary\end{emp} for any constitutive statement, but can be implied (as with the original institutional grammar). Where no explicit Activation Condition is specified, the context clause is by default ``under all conditions''. Where no explicit Execution Constraints are specified, the context clause is by default ``no constraints''. 

It is important to note that \begin{emp}Context\end{emp} in institutional statements reflects the context specific to the coded statement (Statement Context), as opposed to capturing context in the wider sense, making reference to the context of the policy or the domain more generally.
\\
\midrule
Or else & A consequence characterization associated with the non-fulfilment or -enactment of the Constitutive Function indicated in a separate  institutional statement nesting on the leading institutional statement (as defined in the following discussion). Consequences, as understood in the context of constitutive statements, can be existential in kind (e.g., invalidating policy), as opposed to the non-existential sanctioning provisions on the regulative side. 

The presence of this component is \begin{emp}optional\end{emp} for constitutive statements.\\
\toprule
\end{tabular}
\caption{Definitions of Syntactic Elements for Constitutive Statements}
\label{tab:SyntacticElementsConstitutive}
\end{table}

\newpage

\begin{minipage}{5cm} 

\end{minipage}

\subsection{Institutional Statements}
\label{subsec:InstitutionalStatementAssumptions}

Capturing both the regulative and constitutive nature of statements, in IG 2.0 \begin{emp}an institutional statement\index{Institutional Statement} describes expected actions for actors within the presence or absence of particular constraints, or parameterizes features of an institutional system.\end{emp}

To accommodate a more comprehensive representation of structure and content, the construction of institutional statements in IG 2.0 requires a differentiated understanding of institutional statements, both in terms of \begin{emp}atomic\end{emp} and \begin{emp}nested institutional statements\end{emp} of various forms. 

\subsubsection{Atomic Institutional Statements}
\label{subsubsec:AtomicInstitutionalStatements}

Before highlighting notions of nested institutional statements, we reiterate the notion of an atomic institutional statement as the elementary form of an institutional statement: an \begin{emp}atomic institutional statement\end{emp}\index{Atomic Institutional Statement} is a statement of regulative or constitutive kind that contains the corresponding necessary components, alongside potential optional components other than the \begin{emp}Or else\end{emp}, and in which none of the components contains multiple values (\begin{emp}component-level combinations\end{emp}) or is otherwise nested (\begin{emp}component-level nesting\end{emp} -- see \cref{subsec:ComponentLevelNesting}). Examples for component-level combinations are multiple attributes or combinations of aims in a given institutional statement. Examples for component-level nesting include context characterizations (with details to be introduced at a later stage) expressed as institutional statements themselves.

Illustrating this concept, the statement \begin{exi}``Organic farmers must commit to organic farming standards''\end{exi} is \begin{emp}atomic\end{emp} in kind. It identifies the following elements:

\begin{itemize}
\item Attributes: Organic farmers
\item Deontic: must
\item Aim: commit to
\item Direct object: organic farming standards
\item (Implied) Context: under all conditions.
\end{itemize}

It does not include an \begin{emp}Or else\end{emp}, and neither of the components contains value combinations.\footnote{Note that a component can contain phrases; however, the terms need to represent a semantic unit in the context of the concerned institutional statement.} 

In contrast, the statement \begin{exi}``Organic farmers must commit to their organic farming standards and accommodate regular reviews of their practices''\end{exi} combines multiple distinctive activities and associated objects, namely \begin{emp}``commit to organic farming standards''\end{emp} and \begin{emp}``accommodate regular reviews of their practices''\end{emp}, as illustrated below by aggregating all phrases by component type (multiple phrases are separated by semicolon). 

\begin{itemize}
\item Attributes: Organic farmers
\item Deontic: must
\item Aim: commit to; accommodate
\item Direct Object:\footnote{The distinction between direct and indirect object is discussed in greater detail at a later stage.} organic farming standards; regular reviews of their practices
\item (Implied) Context: under all conditions.
\end{itemize}

We refer to any such combination of component values as a \begin{emp}component-level combination\end{emp}.

In consequence, this statement is \begin{emp}not atomic\end{emp} in kind.

\subsubsection{Nested Institutional Statements}
\label{subsubsec:NestedInstitutionalStatements}

Motivating the conception of nested institutional statements\index{Nested Institutional Statements} is the pragmatic observation that statements can reflect a linkage or embedding of multiple actors or regulated activities, reflecting complex configurational settings; rules in form are generally not expressed in the atomic form specified above. Lack of specificity of these linkages undermines the coder's ability to comprehensively, and accurately, capture institutional content for versatile downstream analysis.

IG 2.0 accommodates two forms of institutional statement nesting: \begin{emp}horizontal nesting\end{emp} and \begin{emp}vertical nesting\end{emp}. Generally, horizontal nesting\index{Horizontal Nesting} allows for the representation of multiple institutional statements that convey co-occurring or alternative actions, actor or object involvement, i.e., any case in which multiple of the same syntax element occur in an institutional statement. 

Generally, vertical nesting\index{Vertical Nesting} allows for the representation of multiple institutional statements that convey coupled actions that follow from one another in the form of a consequential relationship. It is particularly suited to representing the case of consequentially linked statements in which statement A delineates permitted, required, or forbidden activity (for regulative statements), or is constitutive in kind, and statement B delineates sanctions for non-conformance with statement A. In the IG 2.0 parlance, statement A is considered a ``\begin{emp}monitored statement\index{monitored statement}\end{emp},'' and statement B a ``\begin{emp}consequential statement\index{consequential statement}\end{emp}.'' Horizontal nesting and vertical nesting are described in more detail below.

\sp

\textbf{Horizontal Nesting\index{Horizontal Nesting}}: Horizontal nesting describes a logical combination of two or more statements to capture institutional content comprehensively. Exemplified in narrative form, a horizontally nested statement can combine two or more statements. Borrowing the example introduced earlier (\begin{exi}``Organic farmers must commit to their organic farming standards and accommodate regular reviews of their practices''\end{exi}), horizontal nesting reflects the decomposition of such statement into two logically-combined atomic institutional statements, such as \begin{exa}``Organic farmers must commit to organic farming standards'' \AND ``Organic farmers must accommodate regular reviews of their practices''\end{exa}

Horizontal nesting allows for more complex constructs linking an, in principle, arbitrary number of atomic institutional statements and combinations thereof, such as \begin{exa}(``Organic farmers must commit to organic farming standards'' \AND 

``Organic farmers must accommodate regular reviews of their practices'') \XOR 

(``Organic farmers must \NOT~sell their produce under the organic farming label'').\end{exa}

In this example, we can observe the linkage of three statements, in which two AND-combined statements apply as an alternative (XOR) to a third statement. 

Note the use of parentheses to signal the precedence of individual statements where multiple logical operators are used. 

Possible logical operators for horizontal nesting, or \begin{emp}statement combinations\end{emp}, are \AND~(conjunction), \OR~(inclusive disjunction; colloquially: \begin{hil}AND/OR\end{hil}), or \XOR~(exclusive disjunction; colloquially: \begin{hil}EITHER/OR\end{hil}). Where negation is involved, those can be combined with the operator \NOT~(as highlighted in the previous example). An alternative equivalent representation is \begin{exa}(``Organic farmers must commit to organic farming standards'' \AND 

``Organic farmers must accommodate regular reviews of their practices'') \XOR

\NOT~(``Organic farmers must sell their produce under the organic farming label'').\end{exa}

A common use case in practice is the decomposition of action combinations (e.g., \begin{exi}``\dots establish and maintain \dots''\end{exi}) or multiple actors (e.g., \begin{exi}
``\dots producers or inspectors of the facility \dots''\end{exi}).\footnote{The decomposition of such component-level combinations is discussed as part of the coding guidelines in \cref{sec:CodingGuidelines}, specifically in  \hyperlink{tablink:DecompositionOfComponentLevelCombinationsCoreRegulative}{\cref{tab:CodingIgCoreRegulative}, Item \EntryDecompositionOfComponentLevelCombinations}, as well as \hyperlink{tablink:DecompositionOfComponentLevelCombinationsExtendedRegulative}{the same item in \cref{tab:CodingIgExtendedRegulative}} for regulative statements. \hyperlink{tablink:DecompositionOfComponentLevelCombinationsCoreConstitutive}{\cref{tab:CodingIgCoreConstitutive}, Item \EntryDecompositionOfComponentLevelCombinations} provides guidelines for constitutive statements. General operational guidelines for fine-grained decomposition can be found in \hyperlink{tablink:LogicalRelationshipsAmongStatementComponents}{\cref{tab:CodingIgLogicoRegulative}, Item \EntryLogicalRelationshipsAmongComponents}. Note: All preceding table item references are clickable links.}

\sp

\textbf{Vertical Nesting\index{Vertical Nesting}}: Vertical nesting describes a relationship of two or more statements, in which the leading statement (\begin{emp}monitored statement\end{emp}) describes an action that is regulated by a second statement nested in the Or else component (\begin{emp}consequential statement\end{emp}), rendering the Or else an abstract component backed by a separate institutional statement. More accurately, the Or else does not reflect an IG component, but rather a logical connective that links two institutional statements. In this linkage, the second statement reflects a consequence of violating the instructions captured in the monitored statement. Consequences generally involve some pay-off for non-compliance or compliance.\footnote{Where constitutive statements are involved, consequences can further be existential in kind.} Exemplifying vertical nesting in narrative form, we can write \begin{exa}``Organic farmers must comply with organic farming regulations'', \begin{hil}OR ELSE\end{hil} ``Certifiers must revoke the organic farming certification''.\end{exa} Note that both forms of nesting can be combined, i.e., monitored and consequential statements can embed horizontal nesting. Extending the previous example, we can state (visually supported by indentation to signal the corresponding forms of nesting) 

\begin{exa}(``Organic farmers must comply with organic farming regulations'' \begin{hil}AND\end{hil}\\
\-\hspace{0.43cm}\ ``Organic farmers must accommodate regular review of their practices''), \\
\-\hspace{1.5cm}\ \begin{hil}OR ELSE\end{hil} (``Certifiers must suspend the organic farming certification'' \begin{hil}XOR\end{hil} \\ 
\-\hspace{3.6cm}\ ``Certifiers must revoke the organic farming certification'').\end{exa} Note the use of parentheses to signal precedence of the respective statements. 
Vertical nesting can occur across an arbitrary number of levels (i.e., a consequential statement may be a monitored statement in deeper levels of nesting). Exemplifying multi-level nesting, we can state 
\begin{exa}(``Organic farmers must comply with organic farming regulations'' \begin{hil}AND\end{hil}\\
\-\hspace{0.43cm}\ ``Organic farmers must accommodate regular review of their practices''), \\ 
\-\hspace{1.5cm}\ \begin{hil}OR ELSE\end{hil} (``Certifiers must suspend the organic farming certification'' \begin{hil}XOR\end{hil} \\
\-\hspace{3.8cm}``Certifiers must revoke the organic farming certification''),\\ 
\-\hspace{3cm}\ \begin{hil}OR ELSE\end{hil} ``USDA may revoke certifier’s accreditation''.\end{exa}

The combination of both forms of nesting affords the representation of complex institutional arrangements, both in terms of institutional content (horizontal nesting) and enforcement (vertical nesting). The principles are schematically highlighted in \cref{fig:NestingPrinciples}.\footnote{An formal discussion of nesting can be found in \citet{Frantz2013} and an extended conceptual discussion is provided in \citep{Frantz2022InstitutionalGrammar}.} While exemplified here for regulative statements, the principles equally apply to constitutive, and in consequence, hybrid institutional statements (see \cref{subsec:ConstitutiveRegulativeHybrids}).

\begin{figure}[h!]
    \centering
    \framedfigure{\includegraphics[width=0.8\textwidth]{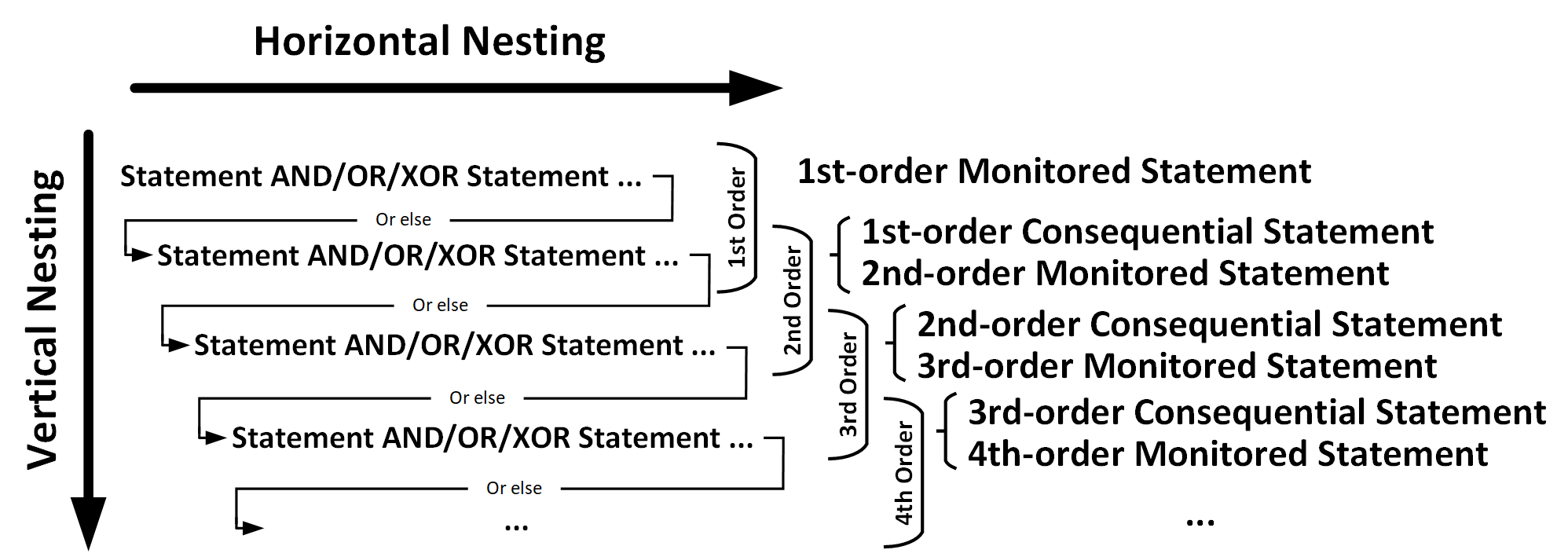}}
    \caption{Institutional Grammar Nesting Principles}
    \label{fig:NestingPrinciples}
\end{figure}

\subsubsection{Component-Level Nesting}
\label{subsec:ComponentLevelNesting}

In addition to the nesting of statements, IG 2.0 further recognizes the possibility of nesting on individual components. Specific components may be substituted with entire institutional statements, or generalizable instances of activity (e.g., statements of fact, observed behavior) expressed in the syntactic form of institutional statements. For regulative statements, those may characterize individuals with respect to their actor properties, embed statements as abstract objects (e.g., the belief that citizens must comply with the law), or express preconditions, i.e., statements whose fulfillment is a precondition for the application of a given statement (as exemplified below).

To motivate this approach explicitly, we will use the following example: \begin{exi}``Organic farmers may sell their produce under the organic label \lb under the condition that organic farmers apply for certification\rb''\end{exi}, with the precondition (the \begin{emp}Activation Condition\end{emp} as a more specific variant of the Context component, as to be discussed in \cref{subsec:ActionSituation}) identified in braces. Rewriting this statement for illustration, we find the following component characterization: 

\begin{itemize}
\itemsep0em
\item Attributes: Organic farmers
\item Deontic: may
\item Aim: sell
\item Direct Object: their produce
\item Execution Constraint (Context): under the organic farming label
\item Activation Condition (Context):\footnote{The activation condition and execution constraint components are a specialization of the Context component, which will be introduced at greater detail in \cref{subsec:ActionSituation}.} under the condition that organic farmers apply for certification
\end{itemize}

Here the instance of behavior embedded in the activation condition reflects the precondition, but, upon closer inspection, may in itself be expressed in terms of syntactic components of institutional statements, either reflecting a specific action configuration (as statement of fact, behavioral instance, or observation) or an institutional statement in its own right (including combinations of statements). 

We refer to cases where individual components of an institutional statement can be replaced by institutional statements (or statements that follow the syntactic structure of institutional statements) as \begin{emp}component-level nesting\end{emp}\index{Component-Level Nesting}. An institutional statement that contains any such nested structure (as initially indicated in \cref{subsubsec:AtomicInstitutionalStatements}) is not atomic in kind. 

Expanding the statement above correspondingly, we arrive at the following structure exemplifying the nesting on the activation condition:

\begin{itemize}
\itemsep0em
\item Attributes: Organic farmers
\item Deontic: may
\item Aim: sell
\item Direct Object: their produce
\item Execution Constraint (Context): under the organic farming label
\item Activation Condition (Context): under the condition that 
\begin{itemize}
\item Attributes: organic farmers 
\item Aim: apply
\item Direct Object: for certification
\end{itemize}
\end{itemize}

The use case for such component-level nesting further includes the articulation of abstract concepts, including non-physical concepts (e.g., beliefs, suspicions, etc.) about individuals' behaviors inasfar as those are relevant to capture the institutional setting accurately (e.g., an official may sanction an individual if the official believes that the individual has performed a violation).

Exemplifying nesting on the object component, we can use the statement \begin{emp}``Inspectors must ensure that organic farmers comply with organic farming regulations''\end{emp}. In this instance, the nesting occurs on the object component, i.e., the receiver of the action \begin{emp}ensure\end{emp}, as visualized below.

\begin{itemize}
\itemsep0em
\item Attributes: Inspector
\item Deontic: must
\item Aim: ensure
\item Direct Object: 
\begin{itemize}
\itemsep0em
\item Attributes: organic farmers 
\item Aim: comply
\item Direct Object: organic farming regulations
\end{itemize}
\item (Implied) Context: under all conditions
\end{itemize}

While not motivated further at this stage, component-level nesting (as statement-level nesting) equally applies to regulative and constitutive statements. Component-level nesting can apply to Attributes, Objects, Context components (Activation conditions and Execution constraints) in regulative statements, as well as to Constituted Entity, Constituting Properties and Context components in constitutive statements. It can further be applied in combination with all forms of statement-level nesting, including the further nesting within component-level nested statements. (For example, imagine a chain of preconditions expressed in nested statements). 

The coding for all forms of nesting and combinations thereof, as conceptually described here, is specified in \cref{sec:CodingGuidelines}.

\subsection{Object-Property Hierarchy}
\label{subsec:AttributeObjectHierarchy}

In addition to the nesting concepts, advanced coding relies on decomposing actors and objects into core descriptors and associated properties. For this purpose, we rely on the conceptual representation of an \begin{emp}Object-Property Hierarchy\end{emp}\index{Object-Property Hierarchy} as exemplified in \cref{fig:ObjectPropertyHierarchyExample}. In this visualization, statements such as \begin{exi}``\dots~a written notification of proposed suspension or revocation of certification \dots''\end{exi} reflect an involved Object hierarchy centering on the \begin{exi}``notification''\end{exi}, that has a property \begin{exi}``written''\end{exi}. Looking at the context of the notification we recognize the concept of \begin{exi}``certification''\end{exi} that has the property of being \begin{exi}``suspended''\end{exi} or \begin{exi}``revoked''\end{exi}, expressed as dependent Objects (\begin{exi}``suspension''\end{exi}, \begin{exi}``revocation''\end{exi}), whereas the latter concepts themselves have a shared property of being \begin{exi}``proposed''\end{exi} in the first place. However, while the property \begin{exi}``written''\end{exi} functionally depends on the \begin{exi}``notification''\end{exi}, that is, writtenness alone does not make sense with an Object it refers to, the existence of a certification does not rely on the notification (i.e., it is functionally independent), and has a self-contained property hierarchy (suspended, revoked, proposed) as described above. 

Interpreting complex Object specifications with this decomposition hierarchy in mind affords a uniform coding approach. The structure of this hierarchy for the specific statement is shown in \cref{fig:ObjectPropertyHierarchyExample} (the dashed line signals relationships between functionally independent Objects). Note specifically the potential use of logical operators (\begin{exi}``XOR''\end{exi}), as well as the ability to reflect shared properties (\begin{exi}``proposed''\end{exi}), to disambiguate the logical relationship between the identified properties, an aspect that is implicit in the textual representation. 

\begin{figure}[h!]
    \centering
    \framedfigure{\includegraphics[width=0.65\textwidth]{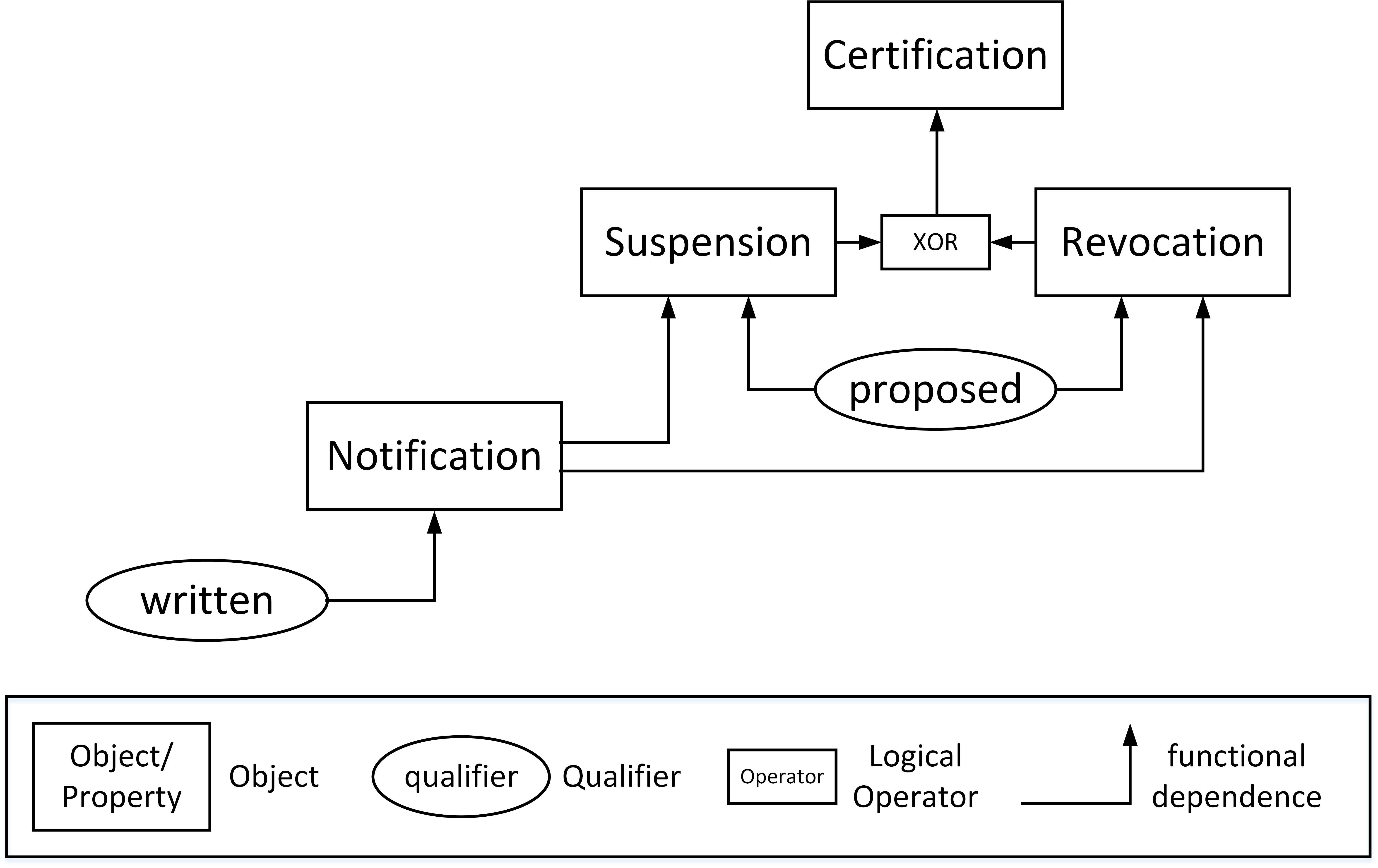}}
    \caption{Object-Property Hierarchy for the given Example}
    \label{fig:ObjectPropertyHierarchyExample}
\end{figure}

Capturing all potential decomposition approaches, we can apply this scheme to functionally dependent properties (\begin{exi}``written''\end{exi} in the previous example), functionally independent Objects or properties (\begin{exi}``certification''\end{exi} in the previous example), and furthermore afford the substitution of Objects by complete institutional statements. Since the latter aspect relies on richer contextualization, it will be exemplified in the context of the coding instructions. The stylized general form of the \begin{emp}Object-Property Hierarchy\end{emp} is shown in \cref{fig:ObjectPropertyHierarchy}. Note that logical operators apply to both functionally dependent and independent properties and on any level of decomposition.

\begin{figure}[h!]
    \centering
    \framedfigure{\includegraphics[width=0.55\textwidth]{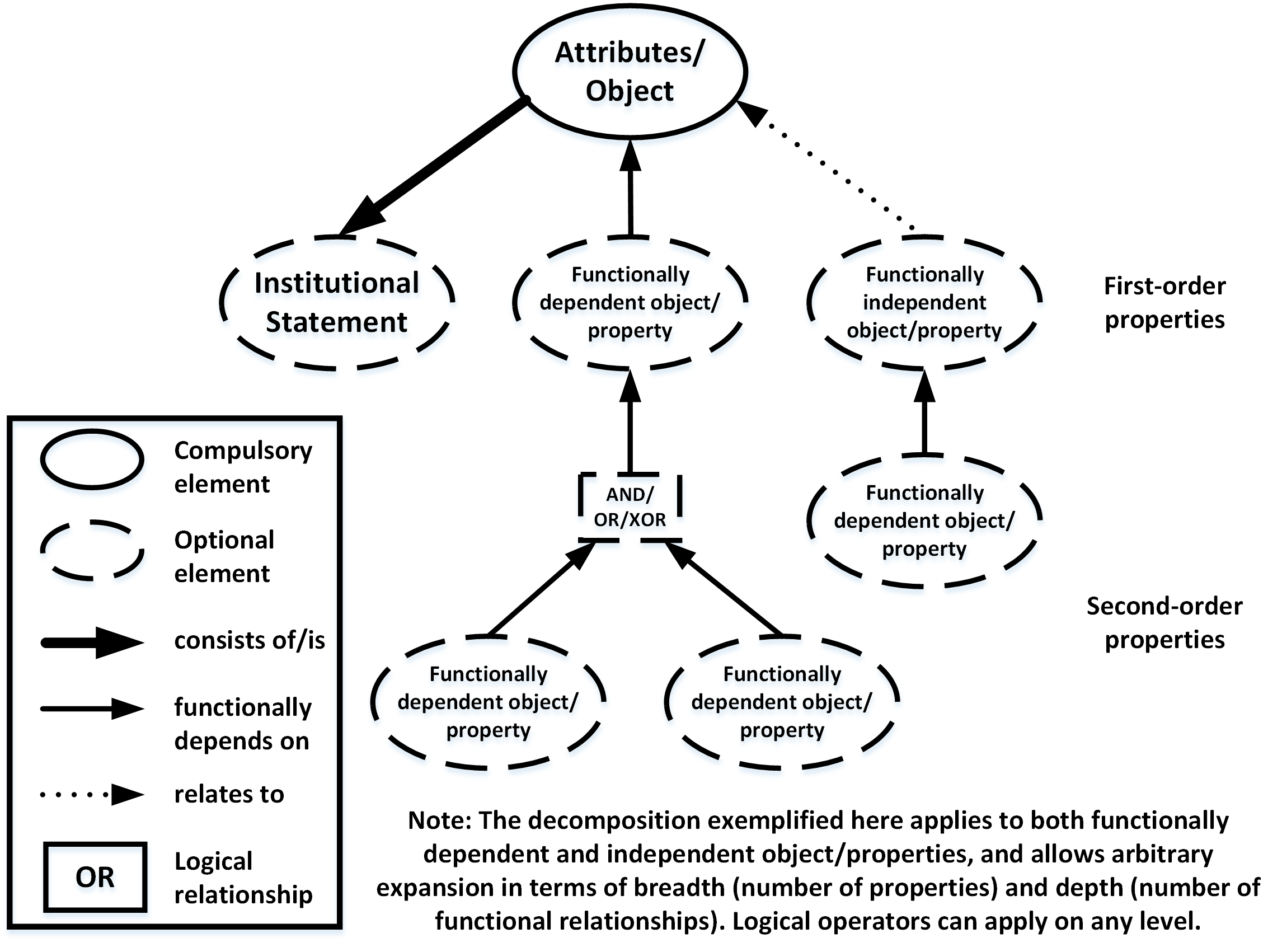}}
    \caption{Object-Property Hierarchy}
    \label{fig:ObjectPropertyHierarchy}
\end{figure}

\newpage

\subsection{Relating Institutional Statements to Action Situations}
\label{subsec:ActionSituation}

A concept central to the coding with IG 2.0 is the action situation~\citep{Crawford2005}. Basically, an action situation is defined as an institutionally governed setting in which two or more actors interact, in relation to which specific outcomes emerge. Action situations are governed by a configuration of different types of institutional statements, which can be regulative or constitutive in kind, with distinctive functional properties. The seven types of institutional statements\index{Rule types}, labeled in parentheses in terms of different types of ``rules'', convey: positions that actors can occupy within an action situation (position rules), eligibility criteria for occupying those positions (boundary rules), operational actions linked to actors occupying certain positions (choice rules), situational outcomes (scope rules), channels of information flow (information rules), guidance on collective decision making (aggregation rules), and incentives tied to particular actions (pay-off rules). Each action situation can be governed by multiple statements of a particular type.  Action situations, and key action situation components, are schematically visualized in \cref{fig:ActionSituation}. In the widest sense, action situations describe the context in which institutional statements operate, and in the context of regulative statements, specifically the mapping between actors, actions, outcomes and the associated payoffs.

\begin{figure}[h!]
    \centering
    \framedfigure{\includegraphics[width=0.8\textwidth]{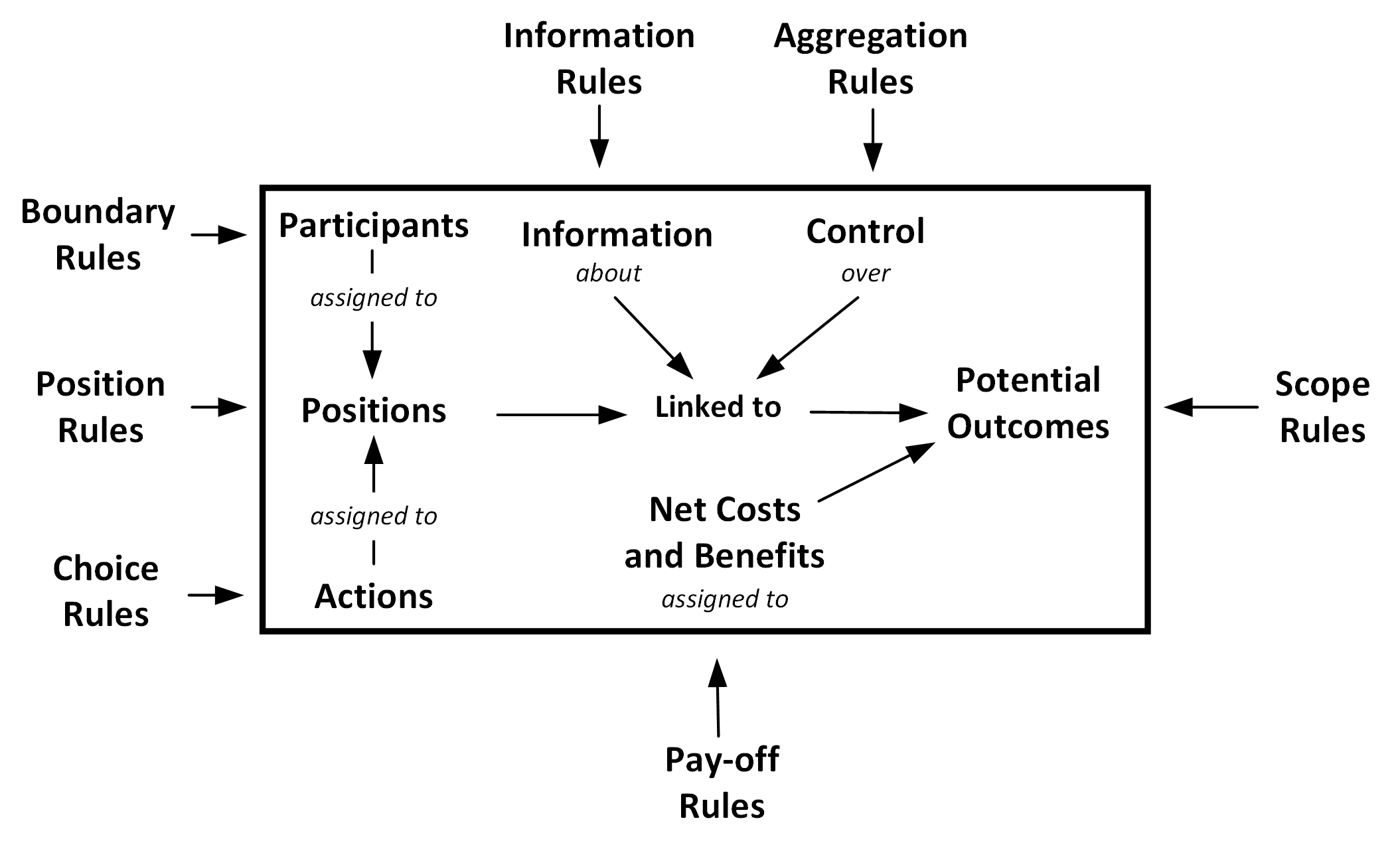}}
    \caption{Schematic Visualization of the Action Situation (as per \citet{Crawford2005})}
    \label{fig:ActionSituation}
\end{figure}

Again, the rule type taxonomy associated with the action situation concept links to the IG insofar as whole institutional statements can be classified accordingly, depending on their functional properties. 

The IG 2.0 specification leverages the action situation concept in recognizing generally that institutional statements characterize activity occurring in action situations, and in accounting through syntactic classification for ways that institutional statement information corresponding to the Context component contextualizes intra- and inter-statement activities. 

Operationalized in the context of regulative statements, Context clauses can instantiate an action situation in which an Attribute acts on Objects in a particular manner, and which are governed by some configuration of institutional statements. By way of contrast, the Context clauses of other statements may simply constrain an Attribute’s behavior in some way within a given action situation. As noted in \cref{subsec:Syntax}, context clauses which serve an instantiation function, as well as Attribute or Object changes are referred to as Activation Conditions\index{Activation conditions}. Context clauses which qualify action are referred to as Execution Constraints\index{Execution constraints}. \cref{fig:ActionSituationConditionsConstraints} schematically represents how Activation Conditions and Execution Constraints situate relative to action situations. 

\begin{figure}[h!]
    \centering
    \framedfigure{\includegraphics[width=0.5\textwidth]{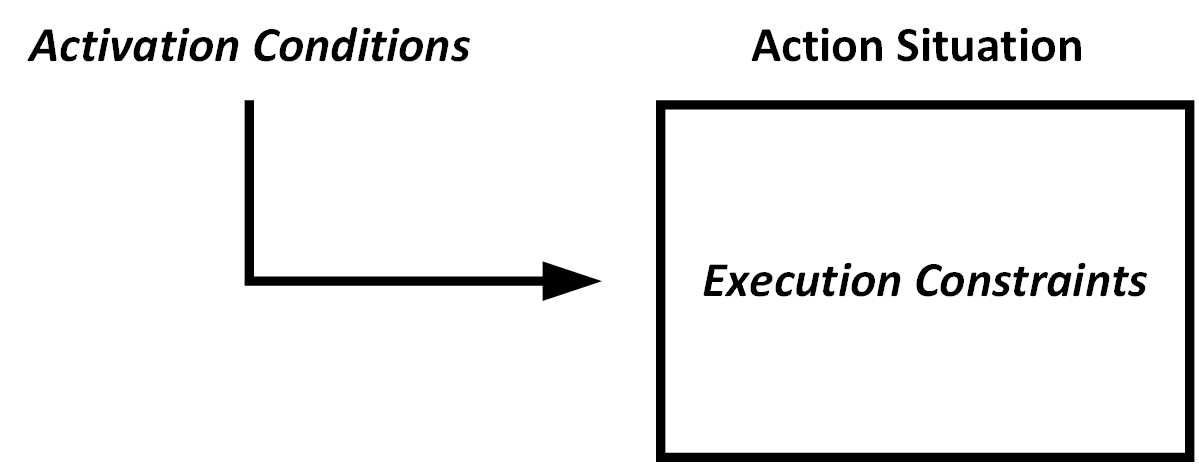}}
    \caption{Activation Conditions and Execution Constraints Principles}
    \label{fig:ActionSituationConditionsConstraints}
\end{figure}

Naturally, explicit characterization of how institutional statements relate to action situations necessitates understanding of the institutional domain. Without this, the coder may encounter difficulty in determining as to whether a specific Context descriptor refers to the action situation more generally, or the action specifically in the form of an action property. While we offer further elaboration as part of the coding guidelines in \cref{sec:CodingGuidelines}, we provide a brief example to motivate the distinction at this stage. 

Inherent to the Activation Condition is reference to a set of exogenous variables; exogenous in the sense that it references states or actions that are beyond the actions that can be qualified within certain Execution Constraints in an instantiated environment (e.g., a new action situation, an environment in which Attributes change or take on new roles, or an environment in which Attributes act in an altered way upon Objects). In other words, Activation Conditions precede the regulated action and activate a given institutional statement in the first place. Conversely, Execution Constraints describe constraints on actions once enacted (and implicitly on actors and associated pre-/proscription as visualized in \cref{fig:ActionSituationContextRelationships}). 

\begin{figure}[h!]
    \centering
    \framedfigure{\includegraphics[width=0.9\textwidth]{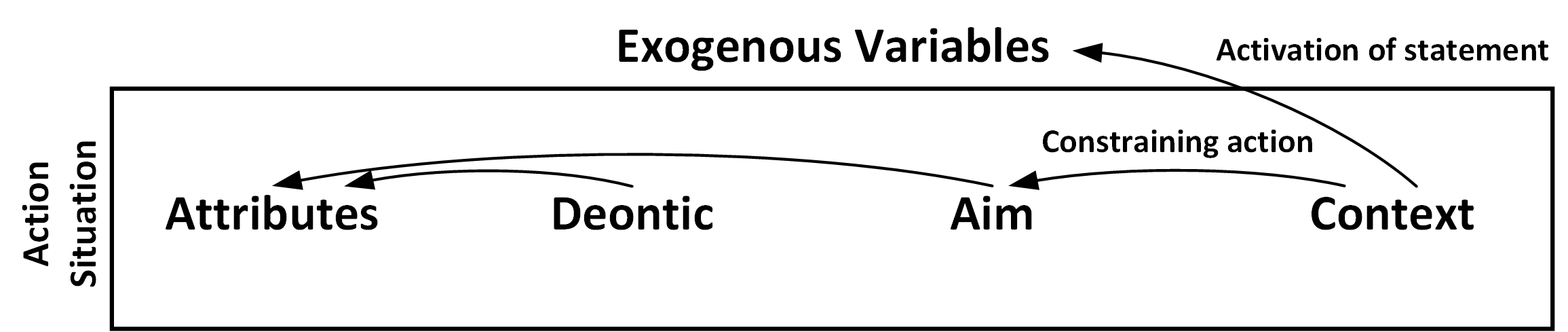}}
    \caption{Context Relationships in the Action Situation}
    \label{fig:ActionSituationContextRelationships}
\end{figure}

Procedurally, this implies different semantics for Activation Conditions and Execution Constraints. Activation Conditions represent Context external to the action situation an institutional statement is embedded in that activates the non-conditional part of an institutional statement, and possibly leading to the activation or modification of an action situation. Execution Constraints, in contrast, are directly attached to the institutional statement and thus reflect context embedded within the action situation itself. The discussed distinction is summarized in \cref{fig:ActionSituationContextRelationships}. 

Offered here is more operational guidance on the differentiation highlighted above, which starts with  the identification of Context clauses in an institutional statement. To systematize the differentiation, we firstly provide a terminological basis. Linguistically, context clauses are generally modifiers, specifically qualifiers (\begin{emp}``usually''\end{emp}, \begin{emp}``some''\end{emp}, \begin{emp}``annually''\end{emp}), adverbial clauses (\begin{emp}``When the traffic light turns from red to green, \dots''\end{emp}) and prepositional clauses (\begin{emp}``after midnight''\end{emp}). Whereas qualifiers reliably signal Execution Constraints (action properties in the narrow sense), and adverbial clauses generally indicate Activation Conditions, depending on contextual interpretation, prepositional clauses can fall in either category and selectively signal Activation Conditions or reflect Execution Constraints (action properties in the wider sense). 

The differentiated treatment for prepositional clauses is best described with an example: \begin{emp}``At 8am, farmers may begin selling their goods in accordance with market rules,''\end{emp} contains two context clauses (\begin{emp}at 8am, in accordance with market rules\end{emp}), one of which is a conditions clause (\begin{emp}at 8am\end{emp}) and one of which is a constraints clause (\begin{emp}in accordance with market rules\end{emp}). Context clauses may be implicit, and institutional statements are not constrained in the number of clauses for condition and constraint type. The remainder of the statement is the non-context clause of an institutional statement. \cref{fig:ContextClausesInInstitutionalStatement} highlights this decomposition of regulative institutional statements with respect to the Context component.

\begin{figure}[h!]
    \centering
    \framedfigure{\includegraphics[width=0.95\textwidth]{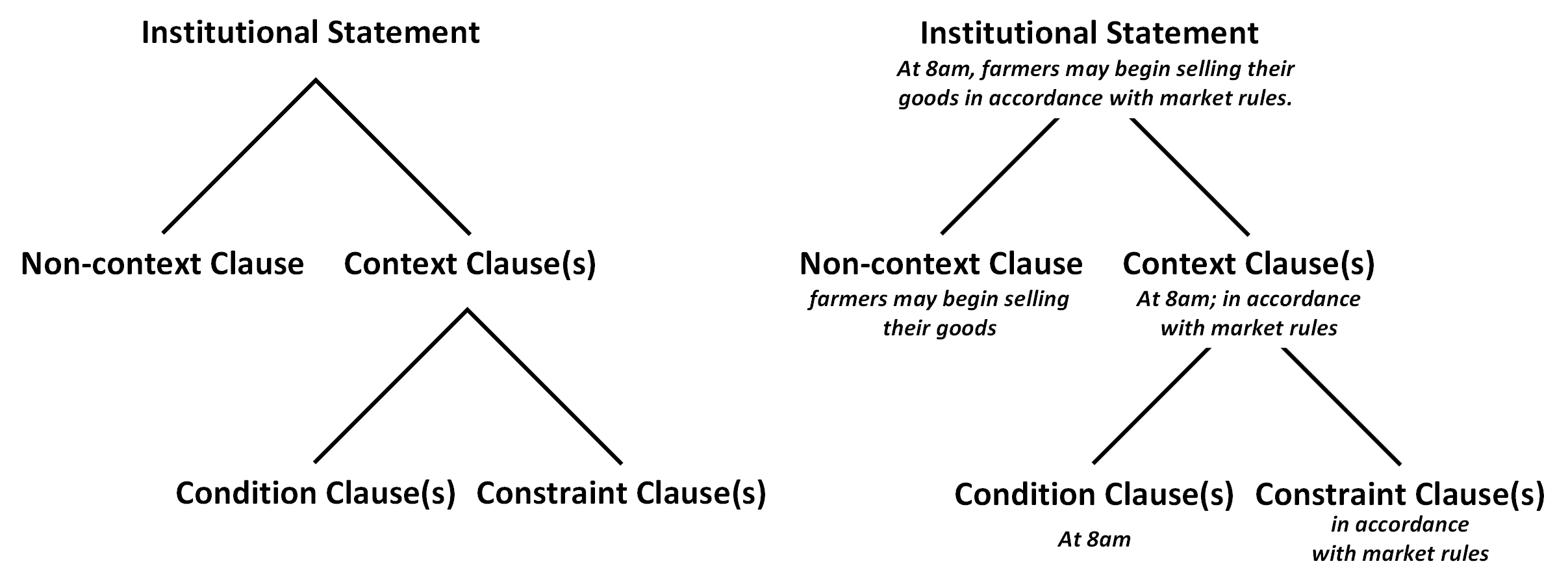}}
    \caption{Context Clauses in Institutional Statements}
    \label{fig:ContextClausesInInstitutionalStatement}
\end{figure}

Decision heuristics can be employed to aid in the identification of activation conditions and execution constraints. The following heuristics are particularly designed to help the analyst determine if a context clause in question is an activation condition, leaving the resultant classification of the clause as an execution constraint, if it is determined that it is not. Offered first is a heuristic that generalizes across regulative and constitutive statements. This is followed by heuristics specific to two different types of statements. 

\begin{ptit}General Heuristic for Identifying Activation Conditions:\end{ptit} 

The clause instantiates a discrete setting (constrained temporally, spatially, or otherwise as shown below) and/or event that activates the non-condition clauses of the institutional statement (i.e., non-context clauses along with potential constraint clauses) as a whole.
The following example statements contain activation conditions (underlined) for illustration.:
\begin{itemize}
    \item \begin{emp}``Upon receiving final notice of non-compliance, farmers shall cease sale of any product bearing the USDA organic farming label.''\end{emp} This statement signals the instantiation of a novel attribute-object link described by the activity, and is positioned within a discrete temporal setting.
    \item \begin{emp}``Starting January 1, the Department of Agriculture is the certifying authority.''\end{emp} Here the statement makes explicit reference to an event that leads to the activation of an associated role change in a constitutive statement.
    \item \begin{emp}``Upon entry into the house, visitors must remove shoes.''\end{emp} (Event) vs. \begin{emp}``At home individuals must not wear shoes.''\end{emp} (Discretized setting). Whereas the first statement references a specific event (entry), the second statement describes a general discretized setting in which the statement holds at all times, i.e., the statement is activated at any time.
\end{itemize}

\begin{ptit}Heuristics for Identifying Activation Conditions in Regulative Statements:\end{ptit}

\begin{emp}Attributes\end{emp}: The clause instantiates a) a change in attributes linked to a statement's activity or b) a change in attribute role.
\begin{itemize}
    \item \begin{emp}``Between the hours of 6pm and 6am on Mondays, members of neighborhood watch residing in blocks 7-10 will assume night patrol activities.''\end{emp} This example signals a change in attribute role within a specified time frame.
\end{itemize}

\begin{emp}Objects\end{emp}: The clause instantiates a change of the object(s) linked to the statement's activity.
\begin{itemize}
    \item \begin{emp}``Starting Dec.~15th, inspectors must exclusively use the revised inspection form.''\end{emp} (novel object use)
\end{itemize}

To support the classification more generally, we can further offer practical considerations to aid in the decision making: 

\begin{itemize}
    \item Regulative statements: More generally, in a regulative context with a given specification of pre/proscriptions, activation conditions constitute a discretized setting in which institutional content can in principle be adhered to or violated.
    \item Context-clause interdependencies: If the  application of the clause of concern is contingent on the prior activation of another context clause, the former is an execution constraint, whereas the latter describes an activation condition in the context of the analyzed institutional statement. If only the satisfaction of both clauses leads to the activation of the non-context clause, both are sensibly identified as activation conditions.
    \begin{itemize}
        \item Example: \begin{emp}``When live fish or viable gametes are sold, traded, taken or otherwise disposed of from an aquaculture facility, the permittee or operator shall, at the time of transfer of possession, give an invoice to the person receiving such fish or viable gametes.''\end{emp} Here \begin{emp}``when live fish \dots facility''\end{emp} represents the activation condition; the subsequent provision of an invoice \begin{emp}``at the time of transfer of possession''\end{emp} is an execution constraint.
    \end{itemize}
\end{itemize}

Returning to the initial example \begin{emp}``At 8am, farmers may begin selling their goods in the farmer's market.''\end{emp}, the condition clause \begin{emp}``At 8am''\end{emp} signals the instantiation of a discrete temporal setting in which the remaining statement is activated as a whole (i.e., permitting the sales of goods on the market). The clause \begin{emp}``in the farmer's market''\end{emp} complements the description of the regulated content and neither affects attribute/role, object nor does it define a setting that activates the remaining statement. The resulting statement would thus be coded as follows:

\begin{exa}
Coded statement: ``At 8am (Activation statement), farmers (Attribute) may (Deontic) begin selling (Aim) their goods (Object) at the farmer’s market (Execution constraint).''
\end{exa}

To highlight the distinction between activation conditions and execution constraints more clearly, we can review the following statement: \begin{emp}``Farmers may sell non-organic goods in the organic farmer's market only between the hours of 3 and 5pm.''\end{emp}

Here the time frame \begin{emp}``between the hours of 3 and 5pm''\end{emp} signals a distinctively different relationship between attributes (farmers), aim and object, whereas the \begin{emp}``in the farmer's market''\end{emp} complements the characterization of the institutional setting in which the permission holds. 

This is in contrast to the following statement:
\begin{exa}``\underline{Farmers must perform inventory of goods sold} at farmers market daily.'' (execution constraint)\end{exa} 
This statement signals a general obligation to provide inventory information (i.e., is activated at all times), but does not establish a specific discretized setting or event that triggers the obligation.

Contrasting this, the following example highlights such event, leading to the characterization of the conditional clause as activation condition:
\begin{exa}``At the close of market each day (activation condition), \underline{farmers must perform inventory of goods sold}.''\end{exa}

\begin{ptit}Heuristics for Identifying Activation Conditions in Constitutive Statements:\end{ptit}

\begin{itemize}
    \item \begin{emp}Entity\end{emp}: The clause instantiates a change in the Entity that is being constituted. 
    \begin{itemize}
        \item Example: \begin{exa}``In the event that the Board Chair position becomes vacant, the Vice-Chair is the chief executive of the Council.''\end{exa} (change in entity specification under event)
    \end{itemize}
    \begin{emp}Properties\end{emp}: The clause instantiates a change in the constituting properties of the entity that is constituted, reconstituted or otherwise affected in the institutional statement.
    \begin{itemize}
        \item Example: \begin{exa}``Starting Dec.~15th, organic farming is agricultural production that does not involve the use of synthetic chemicals or genetically modified organisms.''\end{exa} (change in constituting properties of constituted entity)
    \end{itemize}
\end{itemize}

\begin{ptit}Considerations for challenges cases:\end{ptit}

In written institutional statements, it is not uncommon to find constructions that elevate a statement's pre- or proscriptiveness embedded in activation condition or execution constraint. A stylized example is the following statement: \begin{exa}``Drivers must comply with traffic rules, especially when encountering congested traffic.''\end{exa}

A specific characteristic of such statement is the elevated prescriptiveness based on the context clause, with specific focus on the contextual characterization \begin{exi}``\dots~especially when encountering congested traffic''\end{exi}. As such, the context characterization signals a situational increase of prescriptiveness, beyond the generally expected compliance. Facing such statements, the analyst will need to evaluate, whether this statement exclusively highlights such contextual elevation, which is interpreted as an execution constraint, or characterizes a specific condition under which the non-context part of a statement applies as a whole. Typical terms associated with elevated prescriptiveness mediated via execution constraints include 

\begin{itemize}
    \item especially
    \item particularly
    \item specifically
\end{itemize}

Activation conditions are conventionally signaled using terms such as

\begin{itemize}
    \item where possible
    \item under the condition
    \item when \dots
\end{itemize}

\newpage

\subsection{Institutional Statement Type Heuristics}
\label{subsec:StatementTypesHeuristics}

IG 2.0 has been developed in response to observed challenges in the encoding of policy, such as the inability to sensibly capture institutional statements of constitutive kind. In addition to affording the syntactic parsing of constitutive institutional statements, their unambiguous characterization is of central importance so as to ensure coding reliability.

While a characterization based on syntactic alignment has an intuitive appeal, not in all instances does this lead to correct statement classification, e.g., based on contextually implied actor associations or stylistic artifacts. To afford a reliable characterization of statements, we rely on a set of first principles expressed in a set of general guidelines that seek to identify both constitutive and regulative statements, alongside potential combinations of those, the expression of which IG 2.0 explicitly supports. This is then followed by a more specific set of questions, supporting the identification of individual features relevant for a distinctive statement characterization.

To lend best possible guidance, the guidelines, or heuristics, are expressed as closed questions, alongside clarifying explanations.

\begin{ptit}General Heuristics for Identifying Institutional Statement Types:\end{ptit}

\begin{itemize}
\item Does the statement introduce or parameterize fundamental aspects of the action situation (boundaries, definition or modification of actors, actions, objects, artifacts and associated affordances; endowment of rights or authority/power), and in doing so, define positioning and constellation of entities in an institutional setting in which potentially regulated behavior is enacted? $\rightarrow$ If so, the statement is \begin{emp}constitutive\end{emp} in kind.
\item Does the statement signal unambiguously implied or explicit agency, while  specifying duty and discretion, or sanctions for transgression? In doing so, does the statement draw on (i.e., makes implicit or explicit reference to) actors, objects or artifacts, including a potential modification of the institutional setting as an outcome of actors exercising activities regulated as part of the institutional statement? $\rightarrow$ If so, the statement is \begin{emp}regulative\end{emp} in kind.
\item If both criteria are satisfied (e.g., introduction of novel entity as byproduct of enacted agency), test for combination of regulative and constitutive statements under \begin{emp}consideration of the primary and secondary objective/purpose of the statement in the institutional setting as commonly signaled by aim or constitutive function respectively\end{emp} (e.g., essential focus on regulation of behavior vs.~parameterization of institutional setting/action situation -- with further operational criteria provided in \cref{tab:SemanticStatementTypeHeuristics}). 

Statements are then characterized as regulative-constitutive (where regulation is of primary concern), or constitutive-regulative hybrid (where parameterization is of primary concern), respectively (see \cref{subsec:ConstitutiveRegulativeHybrids} for more details on hybrid institutional statements).

A special case are statements that can coded according to both syntactic forms. Where dual coding of statements is possible and desirable (i.e., coding as \begin{emp}both\end{emp} constitutive and regulative statements), statements can be encoded as polymorphic institutional statements (see \cref{subsec:SyntacticPolymorphs}). 

Where a specific preference for regulative or constitutive forms exists based on analytical preferences, this poses challenges to coding reliability, an aspect that can be alleviated by defining a specific \begin{emp}interpretational scope\end{emp} prior to coding (as described in the following).

\end{itemize}

\fcolorbox{black}{\boxbgcolor}{\begin{minipage}{16.5cm} 
\hypertarget{tablink:InterpretationalScopeOfInstitutionalStatements}{\begin{ptit}Interpretational Scope of Institutional Statement Coding\end{ptit}}

To establish a reliable and consistent encoding of institutional statements, an important concern is the appropriate characterization of the intended \begin{emp}scope of interpretation\end{emp} coders should put forth when establishing the structure and type of institutional statements. More specifically, while the encoding of text centers around individual institutional statements as units of analysis, tacit preferences in interpretation can become overt when distinguishing between constitutive and regulative statements. 

Using the following set of (stylized) example statements for illustration, we can observe two leading statements of regulative kind, which, in alignment with the general heuristics expressed above, is signaled by the primary focus on behavior regulation. The last statement, in contrast, offers further contextualization for the previous statements.

\vspace{0.2cm}
\begin{exa}
(1) Organic farming operations must not utilize genetically-modified seeds.

(2) Organic farming operations may not process crops other than the ones specified in Appendix A.

\dots

(10) Paragraphs in this section do not apply to traces of genetically modified material.
\end{exa}
\vspace{0.2cm}

Exploring the statements structurally, the last statement does not directly regulate behavior, but instead imposes operational constraints on the preceding statements. Given its broader overarching function and associated capacity for re-configuration, it may be interpreted as \begin{emp}parameterizing in nature\end{emp}, since it affects the wider institutional setting regulated in the referenced statements that offer specific operational guidance. Such characterization would lead to a qualification of the statement as \begin{emp}constitutive\end{emp} in kind. 

Conversely, however, the statement can be reconstructed in a form that captures the operational character. Absent specific coding instructions (which are discussed in \cref{sec:CodingGuidelines} onwards), the last statement could be reformulated as

\vspace{0.2cm}
\begin{exa}
Organic handling operations may not apply paragraphs in this section to traces of genetically modified material.
\end{exa}
\vspace{0.2cm}

Following the general heuristics for statement classification, the reformulated statement primarily focuses on behavior regulation, suggesting its characterization as \begin{emp}regulative\end{emp}. 

Without further specification, a potential inconsistent interpretation of statements of such character can lead to reliability challenges, an aspect IG 2.0 seeks to alleviate.

To afford a consistent differentiation that a) establishes reliability, and b) ensures methodological consistency in response to analytical objectives, we differentiate forms of coding by \begin{emp}scope of interpretation\end{emp}, where interpretation of \begin{emp}narrow scope\end{emp} emphasizes an immediate interpretational focus on the statement of concern, including its semantics as well as the function it holds with respect to other statements expressed in the policy, or the policy at large; however, semantic linkages of the statement are not resolved beyond the statement itself. 

Contrasting the narrow scope, applying a \begin{emp}wide scope\end{emp} of interpretation expands the focus of analysis by resolving implied or explicit semantic links to other statements, and thereby introduces potential inferences that can affect the interpretation and representation of the analyzed statement both structurally (as showcased above for the reconstruction in regulative form) and semantically (e.g., by drawing deeper inferences draw from and even extend beyond the linked statement). While rooted in the original institutional statement of concern, coding based on wide scope implicitly invokes the interpretation from a systemic perspective, and thereby introduces a variable unit of analysis; this interpretation is necessarily anchored in, but invariably extends beyond, the originating institutional statement. 

\end{minipage}}

\fcolorbox{black}{\boxbgcolor}{\begin{minipage}{16.5cm}

The choice of interpretational scope is an important consideration and determined as part of the design of the coding exercise, alongside aspects such as pre-coding steps (see \cref{sec:PrecodingSteps}), prior to operational coding. Justifications for either choice of interpretational scope can be based on the analytical objectives, such as an actor-centric perspective that favors the uniform representation in behavioral terms, as opposed to systemic analyses that emphasize conceptual and relational aspects of the institutional system. Justification can further occur on methodological grounds, such as reliability concerns as well as coder experience and contextual knowledge.  

Where coders seek encoding that is agnostic of specific analytical biases, such statement can be systematically coded both with narrow and wide interpretational scope, leading to the characterization as \begin{emp}polymorphic institutional statements\end{emp}, which are discussed at greater detail in \cref{subsec:SyntacticPolymorphs}.

\end{minipage}}

\sp
\sp

\begin{ptit}Specific Heuristics for Identifying Institutional Statement Types:\end{ptit}

With focus on cases for which the general characteristics are insufficiently precise, \cref{tab:SemanticStatementTypeHeuristics} lists various operational and in part interrelated heuristics that reflect on the distinctive purpose and semantics of individual institutional statements. The heuristics are ordered by specificity. Earlier heuristics are more broadly applicable, whereas later ones offer guidance relevant for more specific cases. Where useful and applicable, the table further includes explanatory support, while drawing links to related heuristics. 


\begin{longtable}{p{2.2cm} p{9.1cm} p{2.2cm} p{2.3cm}}
\toprule
\textbf{Heuristic} & 
\textbf{Description} & 
\textbf{Statement Type} & 
\textbf{Related Heuristic(s)} \\

\toprule

Function & 
Is the statement's focal purpose to  explicitly define, introduce, or modify an entity (e.g., actor, action, role, object), or otherwise parameterize the analyzed institutional system?

$\rightarrow$ If so, the statement is \begin{emp}constitutive\end{emp} in kind.

Is the statement's focal purpose to compel, restrain, permit, or otherwise assign expectations about, an individual or collective (e.g., corporate) actor's performance of a particular action or set of actions (i.e., regulate behavior)? 

$\rightarrow$ If so, the statement is \begin{emp}regulative\end{emp} in kind.
&
constitutive or regulative &
-- \\

\midrule

%
%
%
%
%
%
%
%
%

Consequence of Violation &

Does the institutional statement refer to an activity the violation of which leads to a consequence that is, if specified, localized or systemic in nature, i.e., does the violation modify, or otherwise \begin{emp}re-parameterize\end{emp}, the system? 

$\rightarrow$ If the consequence is systemic, i.e., signals a modification or re-parameterization of the system (e.g., an entity does not come about), then the statement is \begin{emp}constitutive\end{emp} in kind. 

$\rightarrow$ Otherwise, the statement is \begin{emp}regulative\end{emp} in kind.
%
%
%
%
%
%
%
%
%
&
constitutive or regulative
&
Function \\



%

\midrule
Deontic vs.~Non-deontic Modal &
Does the statement contain a modal that explicitly characterizes discretionary actions or signals duty as imposed on the responsible actor? 

$\rightarrow$ If so, the statement is \begin{emp}regulative\end{emp} in kind.

Does the statement contain a modal that describes the general possibility or necessity of a constitution or modification of an entity described by the action?

$\rightarrow$ If so, the statement is \begin{emp}constitutive\end{emp} in kind.

\udl{Example:} Oftentimes, statements of such type attach obligations to objects (e.g., \begin{exi}``Access to mediation shall be maintained \dots ''\end{exi}). While these may contextually allow for the inference of the responsible actor, interpreted on statement level the modal `shall' here signals the epistemic necessity of access to mediation (not an individual's duty to provide it). &
constitutive or regulative &
Consequence of Violation \\

\midrule
Actor/Entity as Target of Conferral &
Is the actor or constituted entity referred to in the statement an explicit recipient of a right, role and associated authority (power), or other form of status? 

$\rightarrow$ If so, the statement is \begin{emp}constitutive\end{emp} in kind.

\udl{Note:} There may exist separate corresponding regulative statements expressing specific duties associated with a role or authority, or other actors' complementary duties ensuring the satisfaction of a right. However, the statement of concern is exclusively focused on conferral in terms of rights, authority or other forms of status, and is thus constitutive in kind. &
constitutive &
Function, Deontic vs.~Non-deontic Modal \\

%

\toprule
\caption{Institutional Statement Type Heuristics}
\label{tab:SemanticStatementTypeHeuristics}
\end{longtable}

%
%
%
%
%
%

\newpage

\subsection{IG Coding Levels}
\label{subsec:IgCodingLevels}

The IG 2.0 identifies three levels of encoding to provide flexible accommodation of coding necessities based on the complexity of encoded data, as well as the analytical objectives of the coder: \begin{hil}IG Core\index{IG Core}\end{hil}, \begin{hil}IG Extended\index{IG Extended}\end{hil}, and \begin{hil}IG Logico\index{IG Logico}\end{hil}.

\begin{hil}IG Core\end{hil}: IG Core facilitates coding following a fundamental syntactic structure. This level best accommodates comparatively simple institutional statements that largely follow the basic regulative or constitutive structure, along with analytical objectives that involve the statistical assessment of references to the individual components (e.g., distribution of actor, action, object or deontic references).

\begin{hil}IG Extended\end{hil}: The next higher level, IG Extended, focuses on capturing the syntactic structure of institutional statements in greater detail (\begin{emp}deep structure\end{emp}). For regulative statements, this involves the fine-granular encoding of actors and objects, along with complex property relationships. Furthermore, it enables for both regulative and constitutive statements, a detailed encoding of context, such as the characterization of statement dependencies, and categorization based on circumstantial aspects of conditions and constraints (e.g., temporal, spatial, procedural aspects). Choosing to encode on this level may be motivated by the complexity of the encoded institution regulation (e.g., complex statements involving Object-Property Hierarchies (\cref{subsec:AttributeObjectHierarchy}), or extensive statement interdependencies), but also by the analytical objectives, such as the operationalization of the extracted structure in advanced computational models that require the explicit representation of actor properties and context characterization.

\begin{hil}IG Logico\end{hil}: The highest level of expressiveness, IG Logico, aims at enabling the analyst to derive more sophisticated understanding of semantic relationships embedded in and among institutional statements based on institutional statement classification across syntactic categories; for example, improved understanding of actor roles, explicit references between statements, as well as inference of actor obligations tacitly expressed in the coded document. As a point of contrast, whereas at the IG Core and IG Extended levels, syntactic classification of institutional statements is a final goal of the encoding exercise, at the IG Logico level the goal is to build on syntactic classification by leveraging this coding toward identification of institutional semantics that relay functional and/or relational information of interest to the institutional analyst.

\begin{ptit}Shared Assumptions\end{ptit}:

All coding levels are backward-compatible, i.e., statements coded at higher levels of expressiveness can be reduced to any lower level of expressiveness. In other words, statement information corresponding to different syntactic components is simply more finely decomposed as one moves from lower to higher levels of expressiveness (e.g., IG Core to Extended), with the effect that moving in the other direction, the coder can simply collapse decomposed information. Methodologically, this accommodates a multi-pass approach towards coding: coding can commence at the lowest level, before being incrementally refined to accommodate syntactic parsing associated with the respective next higher level(s) of expressiveness, and conclude at the desired level of expressiveness set out during study design (which is informed by the nature of the coded document and analytical objectives as discussed above).

\begin{ptit}Mapping Prerequisites:\end{ptit}

The different coding levels make varying use of the concepts highlighted in \cref{sec:Definitions} as outlined in \cref{tab:ConceptsAndCodingLevels}. Concepts specified at the IG Core level, apply to IG Extended and IG Logico, and concepts that apply to IG Extended apply to IG Logico. Statement level nesting applies at all levels. Given the multi-pass coding approach, concepts specified for respective lower levels apply to all higher levels (e.g, statement-level nesting applies to all levels). 
 
\begin{table}[h!]
\centering
\begin{tabular}{p{3cm} p{12cm}}
\toprule
\textbf{Coding Level} & 
\textbf{Relevant Concepts} \\
\toprule
\multirow{6}{*}{IG Core} & \begin{itemize}
    \item Horizontal and vertical nesting (Statement-level nesting) 
    \item Activation conditions, execution constraints 
    \item Conceptual understanding of Hybrid Institutional Statements (\cref{subsec:ConstitutiveRegulativeHybrids})
\end{itemize} \\
\midrule
\multirow{10}{*}{IG Extended} & \begin{itemize}
    \item Component-level nesting 
    \item Object-Property Hierarchy in regulative statements; equally applies to Constituted Entities/Properties in constitutive statements (both of which may have properties on their own)
    \item Hybrid \& Polymorphic Institutional Statements (\cref{subsec:ConstitutiveRegulativeHybrids})
\end{itemize} \\
\toprule
\end{tabular}
\caption{Relevant concepts on different coding levels}
\label{tab:ConceptsAndCodingLevels}
\end{table}

\vspace{3cm}

A high-level overview of the individual levels, along with the discussed objectives is captured in \cref{fig:IgBibCoding}, highlighting both the coder perspective as well as analytical perspective. The principles and objectives of the individual codings are discussed in  detail in \cref{sec:CodingGuidelines} in this Codebook; the analytical perspectives are covered elsewhere.\footnote{A detailed introduction of the levels of expressiveness, their underlying motivation and appropriate analytical techniques are covered in \citet{Frantz2022InstitutionalGrammar}.}

\begin{figure}[h!]
    \centering
	\framedfigure{\includegraphics[width=0.5\textwidth]{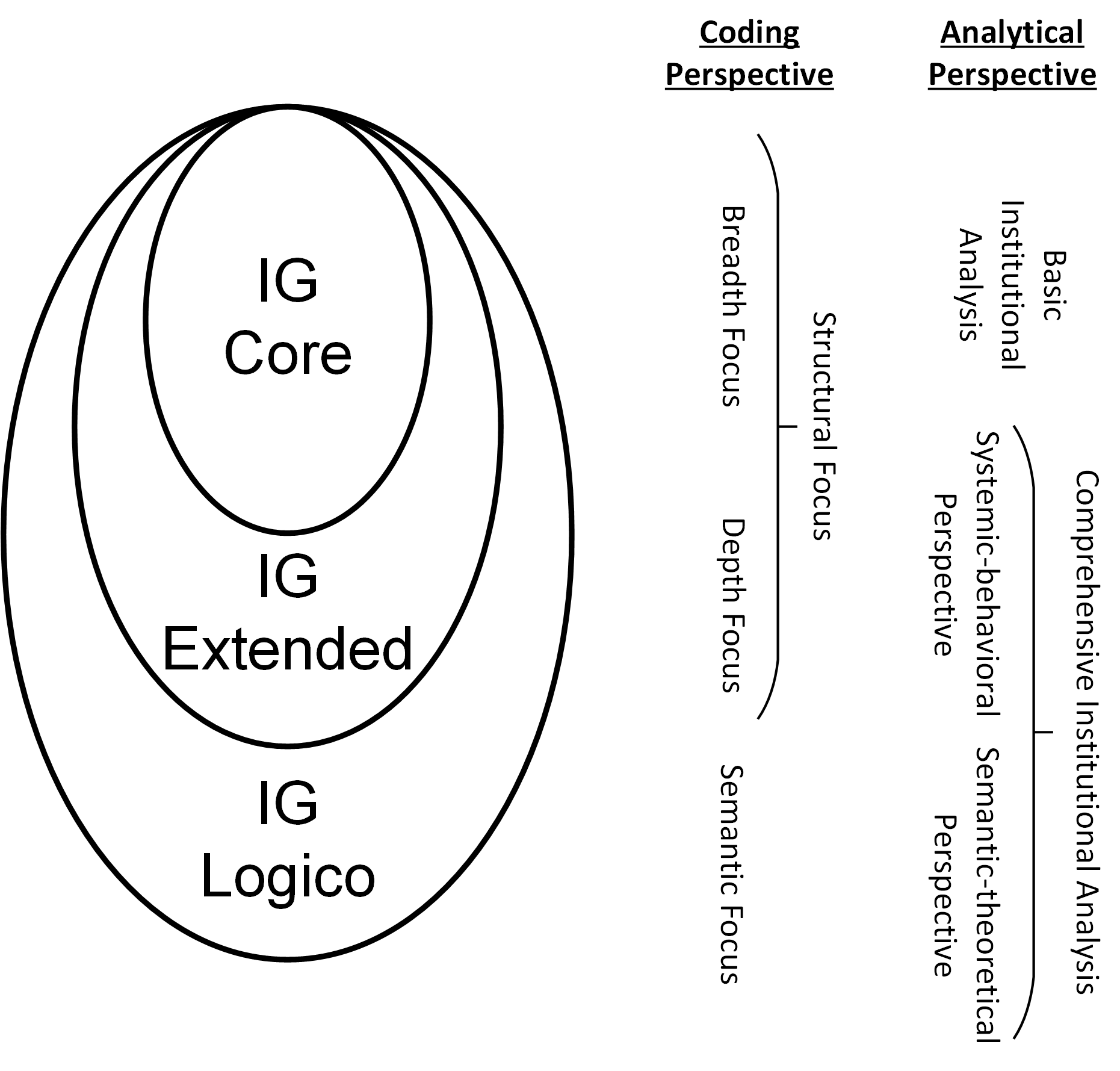}}
    \caption{Overview of IG Coding Levels and Associated Objectives}
    \label{fig:IgBibCoding}
\end{figure}

\clearpairofpagestyles
\chead{\headmark}
\automark[subsection]{subsection}
\ifoot{\ifootertext}
\ofoot{\ofootertext}
\cfoot{\thepage}

\section{Pre-coding Steps}
\label{sec:PrecodingSteps}

Before discussing the coding of institutional statements in detail, in this section we lay out ``pre-coding'' steps that relate to familiarization with the institutional setting and document preparation, commencing with general pre-processing, followed by considerations specific to distinctive levels of expressiveness.

\subsection{General Steps}

\begin{enumerate}
    \item \underline{Familiarization with institutional setting:} Prior to embarking on any coding, the institutional analyst should carefully review the institution to be coded. A thorough pre-coding review (e.g., reading) of the institution to be coded is necessary for gaining a high-level understanding of institutional actors, actions, and institutional statement relationships that can be leveraged in the encoding process.
    
    \item \underline{Selection of coding platform:} One of the first steps the institutional analyst should engage in as she gains familiarity with the institutional setting is identifying the coding platform in which institutional data will be stored. The selection of a coding platform will be informed by the analyst’s expectations regarding at which level of expressiveness institutional statements will be encoded, and related assessment of institutional complexity, as certain platforms are better equipped to capture institutional statement complexity. The selection of a coding platform will also be informed by the analyst’s anticipated usage of institutional data; for example, whether stored data will later be engaged in computational applications. Platforms store data in forms that are more or less computer readable. 
    
    \item \underline{Initial organization of institutional information:} Once the institutional analyst reviews the institution to be coded as part of step 1, she can start to organize its contents. Though variable across jurisdictional setting, policy content is typically organized according to (i) a preamble, that describes the motivation for the policy; (ii) key definitions, that provide descriptions for actors (e.g., \begin{exi}```Secretary' means the Secretary of Agriculture''\end{exi}) and other terms (e.g., \begin{exi}```Prohibited substances' are substances that have been banned by the Dept.~of Agriculture for use in organic food production''\end{exi}) and abbreviations (e.g., \begin{exi}```NOSB' means National Organic Standards Board''\end{exi}) that aid in the effective interpretation of policy content; and (ii) Policy instructions organized by topic according to section and subsection headers. 

    All three types of content (preambles, definitions, and policy instructions) should be coded. Preambles are likely to be comprised of self-referential statements that can convey the purpose of, or contextualize, the institution under examination more generally as well as regulative and/or constitutive statements.  In most cases, definitions are constitutive, and can be useful for encoding institutional statements encountered in a policy document; e.g., when some statement clause references something that is defined in the definitions section of the policy document. This is computationally useful [in the coding process] because it allows the computer to reference particular definitions when certain terms inline. It also allows the computer to link statements that share common definitional information in the analysis process.
    
    The critical aspect of this coding step is to ensure that the institutional analyst identifies all relevant, codeable information -- i.e., all information that is comprised of codeable institutional statements. 
    
    \item \underline{Verification and pre-processing of institutional statements:} Following the identification of candidate statements in step 3, the analyst should engage in verification and pre-processing of institutional statements to enable their syntactic decomposition in step 5. Verification in this case means ascertaining that candidate statements accord with defining syntactic and semantic features of regulative and constitutive statements. Principally, this means verifying that statements presumed to be regulative in kind at least contain an Attribute, Aim, and Context, and that statements presumed constitutive in kind at least contain a Constituted Entity, Constitutive Function, and Context component. Institutional statements often do not align with sentences encountered in formal institutions, as a result of writing style (e.g., compound sentences) and punctuation (e.g., bulleted lists). Examples of excerpts of formal institutions that do and do not accord with the definition of regulative and constitutive statements are provided below. Importantly, text that does not classify as institutional statements should be retained and annotated as domain specific background. This information can be useful for institutional interpretation and implication, but has oftentimes informational character, without parameterizing or regulating function per se.

    \underline{Institutional Statements:}
    
    \begin{exa}Organic farming is hereby established as a practice regulated under the Department of Agriculture.\end{exa}
    
    \begin{exa}The Department of Agriculture shall promulgate regulations governing the practice of organic farming.\end{exa}

Pre-processing in this step also means organizing the content of institutional statements to both remove extraneous content from statements (i.e., punctuation that accompanies statements often reflecting institutional style or organization; for example, roman numerals, bullet points) as well as to begin to arrange statement content to offer additional clarity about how statements are to be coded in step 5. Provided below is text pre-processed to remove extraneous punctuation. 

\fcolorbox{black}{\boxbgcolor}{\begin{minipage}{15.5cm}

\underline{Unprocessed Excerpt}

\begin{exa}``(a) The producer of an organic livestock operation must establish and maintain year-round livestock living conditions which accommodate the health and natural behavior of animals, including:

(1) Year-round access for all animals to the outdoors, shade, shelter, exercise areas, fresh air, clean water for drinking, and direct sunlight, suitable to the species, its stage of life, the climate, and the environment: Except, that, animals may be temporarily denied access to the outdoors in accordance with §§ 205.239(b) and (c). Yards, feeding pads, and feedlots may be used to provide ruminants with access to the outdoors during the non-grazing season and supplemental feeding during the grazing season \dots''\end{exa}

\vspace{0.3cm}

\underline{Pre-processed Text for Institutional Statement Delineation and Punctuation Removal}

\begin{exa}The producer of an organic livestock operation must establish and maintain year-round livestock living conditions which accommodate the health and natural behavior of animals, including: Year-round access for all animals to the outdoors, shade, shelter, exercise areas, fresh air, clean water for drinking, and direct sunlight, suitable to the species, its stage of life, the climate, and the environment. 
Except, that, animals may be temporarily denied access to the outdoors in accordance with §§ 205.239(b) and (c). 
Yards, feeding pads, and feedlots may be used to provide ruminants with access to the outdoors during the non-grazing season and supplemental feeding during the grazing season.\end{exa}

\end{minipage}}

\item \underline{Decomposition of institutional statements:} Following the verification and preprocessing of institutional statements, the analyst should commence the syntax-based encoding process according to a selected level of expressiveness.
\end{enumerate}

Pre-processing is an optional step in the encoding process but can significantly reduce the time and cognitive load associated with subsequent decomposition. Further, the analyst can choose degrees of pre-processing. More extensive pre-processing is particularly useful for encoding at higher levels of expressiveness. Below are pre-processing guidelines that are useful for encoding at any level of expressiveness, as well as guidelines that are level specific. The level specific guidelines build upon each other, rather than being exclusive, meaning that guidelines applicable for IG Core are relevant for IG Extended and IG Logico, and IG Extended guidelines are applicable for IG Logico. The specific steps outlined here pertain to a manual process. Where tool support, e.g., in the form of an automated parser, is employed, the steps may deviate (specifically for the aspects that are automated).

Generally, pre-processing, particularly of a more extensive kind, will be easier for analysts with greater familiarity with the IG, as they will likely be able to detect statement structure, components, and relations without engaging in even a preliminary decomposition of statements. Some degree of formal decomposition might be required of analysts less familiar with the IG to be able to discern these.

\begin{ptit}General pre-processing guidelines:\end{ptit}

\begin{itemize}
    \item Data cleaning: dealing with extraneous punctuation, fixing typos
    \item Delineation of text into institutional statements
    \item Delineation of nested statements (e.g., Or else statements) 
    \item Preliminary classification of institutional statements as regulative, constitutive, regulative or-else, constitutive or-else, or combinations thereof (see \cref{subsec:StatementTypesHeuristics} and  \cref{subsec:ConstitutiveRegulativeHybrids})
    \item Preliminary organization of statements by identifiers capturing the institutional structure/ordering of institutional statements in the document. These identifiers can uniquely identify institutional statements, and statement linkages (i.e., nested statements), as well as policy sections or parts that can facilitate understanding of statement context and cross-statement or policy references.
\end{itemize}

\subsection{Pre-processing Guidelines for IG Core} 

\begin{itemize}
    \item To accommodate encoding of institutional statements at the core level, preliminary decomposition of institutional statements to account for multiple values within individual syntactic fields should be entertained during the preprocessing of institutional documents. Institutional statements often contain multiple Attributes, Aims, and/or Objects. For example: \begin{exa}``The producer of an organic livestock operation must establish and maintain year-round livestock living conditions which accommodate the health and natural behavior of animals.''\end{exa} This statement contains multiple Aims, \begin{emp}``establish''\end{emp} and \begin{emp}``maintain.''\end{emp} Neither of these Aims is associated with unique values in other syntactic fields, and therefore, they can both be captured within a single institutional statement. However, downstream coding is facilitated by capturing multiple values individually within separate statements. With the recommended decomposition, the statement above is reflected as two:\footnote{The decomposition of \begin{emp}component-level combinations\end{emp}, as identified here, is further discussed in \cref{subsec:RegulativeStatementCoding}.} \begin{exa}``The producer of an organic livestock operation must establish year-round livestock living conditions which accommodate the health and natural behavior of animals.''\end{exa} and \begin{exa}``The producer of an organic livestock operation must maintain year-round livestock living conditions which accommodate the health and natural behavior of animals.''\end{exa}

    \item In preprocessing institutional statements for coding at the IG Core level, the analyst may consider reformulating statements into active form (where statements are originally captured in passive form while being careful to retain statement meaning from an institutional perspective). Note that conversion of statements from passive to active form typically requires some implication of values according with different syntactic fields. For example, the passive statement \begin{emp}``Notifications of compliance must be sent to farmers within 30 days of facility inspections''\end{emp} converts to \begin{emp}``[Certifier] must send farmers notifications of compliance within 30 days of facility inspection,''\end{emp} prompting the implication of \begin{emp}``Certifier''\end{emp} as the relevant Attribute, or actor in charge of performing action. This implication requires understanding of institutional context obtained through step 1, as well potentially of the identification of a convention for notating implied information, such as for example, the use of brackets ([ ]) in the example included here.

    \item Where actions or actors are implied, those are inferred from the context and additionally specified as part of the coding in terms of the institutional statement structure. While found across a wide range of statements, this is commonly necessary in the context of statement combinations (combination of two actions performed by the same actor). The same applies to implied logical relationships (e.g., AND, OR, XOR).

    \item When facing complex sentence structures, statements should be thought of in terms of sequentially applied actions. For example, if statements report outcomes of actions without making reference to such actions, the coder should reconstruct the action sequence leading to such outcome in terms of institutional statements (see \cref{subsec:Syntax}).

    \item At this stage, the analyst should flag additional semantic information that she wishes to capture in the syntactic decomposition of statements and associated label.
    
\end{itemize}

\subsection{Pre-processing Guidelines for IG Extended} 

Additional pre-processing of institutional statements to accommodate their downstream coding at the IG Extended level involves some preliminary characterization of institutional statement linkages, particularly to capture action sequences. The coding in IG Extended affords richer decomposition of institutional statements into action sequences. Composite actions are often represented as exemplified in the following: \begin{emp}``When an inspection of an accredited certifying agent by the Program Manager reveals any noncompliance with the Act or regulations in this part, a written notification of noncompliance shall be sent to the certifying agent.''\end{emp}, where \begin{emp}``When an inspection of an accredited certifying agent by the Program Manager reveals any noncompliance with the Act or regulations in this part''\end{emp} represents a conditional clause that does not overtly reflect an institutional statement due to the expression of actions in terms of nouns (conceptual reification). From a semantic perspective, this clause captures two linked action statements, namely the fact that a \begin{emp}``Program Manager inspects accredited certifying agents''\end{emp} and that the \begin{emp}``Program Manager reveals non-compliance in this process''\end{emp}. Retaining the essential institutional semantics, the original statement can thus be rewritten as (with inference of implied components) \begin{emp}``When Program Manager inspects accredited certifying agents and [the Program Manager] reveals non-compliance in this process, the Program Manager shall send a written notification of noncompliance to the certifying agent.''\end{emp} While possible to identify as part of the coding process, a specific consideration of IG Extended is to identify such action sequences, and potentially offload their reconstruction to the pre-processing process, the decision on which is subject to the analytical objective, nature of the coded policy, as well as coder background, used coding platform, and opportunities for automation of the reconstruction.

\subsection{Pre-processing Guidelines for IG Logico} 

Additional pre-processing of institutional statements to accommodate their downstream coding at the IG Logico level involves some more extensive, albeit still preliminary, capturing of inter-statement relationships and embedded actions within institutional statements that the analyst may want to fully reconstruct in terms of institutional statements during the encoding process. Inter-statement relationships are often indicated with referential clauses that embed within institutional statements. In policy documents, these references are often to statement collections, the coded document at large, or third-party documents. The example statements below include types of referential clauses that embed within statements that the institutional analyst may wish to capture during the pre-processing phase. Embedded actions can generally be thought of as actions ancillary to that represented in the focal action of an institutional statement (reflected in the Aim or Constitutive Function for regulative and constitutive statements, respectively). Embedded actions, while referenced, are incompletely described. However, reconstruction of these embedded actions into complete institutional statements can afford a more complete depiction and understanding of the institutional domain being evaluated. The particular reconstruction the analyst pursues will depend on her analytical objectives, but also the specific types of institutional functions [constitutive (see \cref{subsec:confunc}) or regulative (see \cref{subsec:regfunc})] she wants to capture within explicit and reconstructed statements. In the pre-processing phase, the analyst might consider constructing a dictionary of terms they observe through preliminary review of institutional statements that signal different institutional functions. This prompts consideration of how different types of observed actions might link to different institutional functions of interest to the analyst. The example statements below include embedded actions that can be fully reconstructed during the encoding process. 

\sp

\fcolorbox{black}{\boxbgcolor}{\begin{minipage}{15.5cm}

\underline{Example Statements with Referential Clauses}

\begin{exa}Any operation that: (1) Knowingly sells or labels a product as organic, except in accordance with the Act, shall be subject to a civil penalty of not more than 3.91(b)(1)(xxxvii) of this title per violation.\end{exa}

\begin{exa}A production or handling operation that sells agricultural products as ``organic'' but whose gross agricultural income from organic sales totals \$5,000 or less annually is exempt from certification under subpart E of this part.\end{exa}

\begin{exa}Any agricultural product that is sold, labeled, or represented as ``100 percent organic,'' ``organic,'' or ``made with organic (specified ingredients or food group(s))'' must be: (a) Produced in accordance with the requirements specified in §205.101 or §§205.202 through 205.207 or §§205.236 through 205.240 and all other applicable requirements of part 205.\end{exa}

\vspace{0.3cm}

\underline{Example Statements with Embedded Actions}

\begin{exa}A handler of organic products may use information provided by the certified operation to determine percentage of organic ingredients.\end{exa}

$\rightarrow$ Embedded action: information provided by the certified operation (i.e., provision of information by certified operation)

\begin{exa}A certifying agent must provide an applicant with a copy of the on-site inspection report, as approved by the certifying agent, for any on-site inspection performed.\end{exa}

$\rightarrow$ Embedded action: approved by the certifying agent (i.e., approval of report by certifying agent)

\begin{exa}A certifying agent whose accreditation is suspended by the Secretary under this section may at any time submit a request for reinstatement of its accreditation.\end{exa}

$\rightarrow$ Embedded action: accreditation is suspended by the Secretary (i.e., accreditation suspension by Secretary)

\begin{exa}The Program Manager may initiate suspension proceedings against a certified operation, when a certifying agent fails to take appropriate action to enforce the Act.\end{exa}

$\rightarrow$ Embedded action: certifying agent fails to take appropriate action to enforce the Act (i.e., failure to act by certifying agent). In this case, the failure to act is also signalling a violation, or non-compliance of some kind, which could be marked as an institutional function of interest and even used in downstream coding toward the reconstruction of both direct statements and their logical inverses. 

\end{minipage}}

\sp

Concluding this overview on activities relevant for the design and planning of the coding exercise, the following section turns to the coding of institutional statements. 

\newpage

\clearpairofpagestyles
\chead{\headmark}
\automark[subsection]{subsection}
\ifoot{\ifootertext}
\ofoot{\ofootertext}
\cfoot{\thepage}

\section{Coding Guidelines}
\label{sec:CodingGuidelines}

\clearpairofpagestyles
\chead{\headmark}
\automark[subsection]{subsection}
\ifoot{\ifootertext}
\ofoot{\ofootertext}
\cfoot{\thepage}

In this section, we provide guidelines for coding institutional statements at the IG Core, IG Extended, and IG Logico Levels of Expressiveness. Following the specification of utilized syntax, we specify encoding principles for regulative and constitutive statements.

\subsection{Coding Symbols \& Syntax}
\label{subsec:CodingSyntax}

\subsubsection{Coding Symbols}
\label{subsubsec:CodingSymbols}

The syntactic coding for examples in the remainder of this document relies on specific symbols, whose function depends on the applied context, i.e., grammar component vs. institutional statement, and respective coding level (IG Core, IG Extended, IG Logico). An overview of all symbols along with application context, minimum level of applicable encoding, description and examples is provided in \cref{tab:SymbolReference}. The specific syntactic notation referenced in this codebook (introduced in \citep{Frantz2022InstitutionalGrammar}) is referred to as \begin{emp}IG Script\end{emp}\footnote{A more detailed discussion of all syntactic features of IG Script are provided under \url{https://github.com/chrfrantz/IG-Parser?tab=readme-ov-file\#ig-script}.}, a notation that manages the trade-off between computational representation on the one hand, and retaining general human readability of the encoded text on the other. Before exploring the encoding in practice more generally, initially all commonly referenced symbols are introduced, alongside specific features that are applicable in particular levels of expressiveness.

Throughout the remainder of this section we use color coding to signal the association/introduction of specific symbols for syntactic components or features with specific levels of expressiveness (as introduced in  \cref{subsec:IgCodingLevels}). Symbols associated with IG Core features are held in \textcolor{\corecolor}{\corecolorname} for regulative statements, and in \textcolor{\corecolorconst}{\corecolorconstname} for constitutive statements (specifically relevant from \cref{tab:CodingIgCoreConstitutive} onward). Symbols associated with IG Extended are held in \textcolor{\extendedcolor}{\extendedcolorname}, and features associated with IG Logico are called out in \textcolor{\logicocolor}{\logicocolorname}. Symbols of general relevance across levels and regulative and constitutive statements (e.g., parentheses to signal precedence or nesting) are held in bold \textbf{black}. Naturally, the examples draw on features not introduced to this stage, but offer an illustration of the representations used throughout the subsequent guidelines. 

\begin{longtable}{p{1.5cm} p{2cm} p{2.2cm} p{4cm} p{4.5cm}}
\toprule
\textbf{Symbol/ Symbol Pairs} & 
\textbf{Coding Context} & 
\textbf{Lowest applicable Level of Expressiveness} \& \textbf{Statement Type} & 
\textbf{Description} & 
\textbf{Example} \\
\midrule
\lp~~\rp & 
Component & 
IG Core, Regulative \& Constitutive Statements & 
Component classification: The characterization of an expression as a component type is signaled through parentheses that contain the component type. & 
\begin{exi}\A{}(Certifier) \dots\end{exi} , where \A{}~identifies the \begin{exi}certifier\end{exi} as an attribute in a given institutional statement. \\
 & 
 & 
 &
\\
 &
 &
 &
 Where used to combine individual components of the same type (in addition to annotation or the indication of statement combinations), parentheses signal component-level combinations (with logical operators discussed at the end of this table). Inner parentheses are only needed to indicate precedence for three or more elements linked by different logical operators. &
 \begin{exi}Attendees must not \I{}\lp eat \AND{}~drink\rp~on the train.\end{exi}

 Note that this is equivalent to `Attendees must not \I{}(\lp eat \AND{}~drink\rp)~on the train.', but the additional parentheses can be omitted, unless relevant to indicate precedence (e.g., combinations of three or more actions with differing logical operators), such as \I{}\lp firstAction \AND{}~\lp secondAction \OR{}~thirdAction\rp \rp.
 \\
\midrule
 \textit{[}~~\textit{]} & 
 Component & 
 IG Core, Regulative \& Constitutive Statements & 
 Tacit components: The explicit specification of implied components (e.g., actor(s)) is signaled with brackets. & 
 \begin{exi}They [\A{}(farmers)] must comply with the certification regulation \dots, where [\A{}(farmers)] characterizes the inferred actor.\end{exi} \\
 & 
 & 
 &
 \\
 {[~~]} &
 &
 &
 Where component annotations (e.g., animate, inanimate) are used, those are embedded in square brackets following the component symbol, but preceding the component content. & 
 \begin{exi}\A{}\customLog{type=animate}(Certifier) \dots\end{exi}, where \A{}~identifies the \begin{exi}certifier\end{exi} as an attribute in a given institutional statement, and \begin{exi}animate\end{exi} as a semantic annotation, a feature of specific relevance in IG Extended (applied in \cref{subsubsec:IGExtendedRegulative}) and IG Logico (from \cref{subsubsec:IGLogicoRegulative} onward).\\
\midrule
\{~~\} & 
Statement & 
IG Core, Regulative \& Constitutive Statements & 
Horizontally nested statements are represented using surrounding parentheses to emphasise the precedence of combined individual statements. & 
\begin{exi}\{ stmt \AND~stmt \}; \{ stmt \AND~\{ stmt \OR~stmt \}\}\end{exi}, where stmt represents an institutional statement combined with other institutional statements using logical operators (\AND, \OR, \XOR, and potentially \NOT) -- more details on logical operators below. \\
& 
& 
&
&
\\
&
&
&
Vertically nested statements are represented using braces that embed the respective consequential statement & 
\begin{exi}stmt1\{stmt2\}\end{exi}, where \begin{emp}stmt1\end{emp} represents a monitored statement, and \begin{emp}stmt2\end{emp} the corresponding consequential statement. \\
\midrule
\lb~~\rb & 
Component & 
IG Extended, Regulative \& Constitutive Statements & 
Component-level nesting is represented by embedding the component-substituting nested institutional statement in braces. In the case of component-level nesting, the component type specification follows the embedded nested statement. & 
\begin{exi}\A{}(Certifier) \I{}(believes) \Bdir{}\lb \A{}(farmer) \I{}(violates) \Bdir{}(code of conduct)\rb\end{exi} 

\vspace{0.5cm}
 In this example, the direct object (\Bdir{}) of a given institutional statement is substituted with another institutional statement. \\
\midrule
 \A{} & 
 Component & 
 IG Core, Regulative Statements & 
 Identifies the preceding expression as Attributes component.[2] & \begin{exi}\A{}(Certifier)\end{exi} \\
\midrule
 \I & 
 Component & 
 IG Core, Regulative Statements & 
 Identifies the preceding expression as aim component. & 
 \begin{exi}\A{}(Certifier) \I{}(monitors) \Bdir{}(farmers).\end{exi} \\
\midrule
 \Bdir{} & 
 Component & 
 IG Core, Regulative Statements & 
 Identifies the preceding expression as direct object component. &
 \begin{exi}\A{}(Certifier) \I{}(administers) \Bdir{}(certifications).\end{exi} \\
\midrule
 \Bind{} & 
 Component & 
 IG Core, Regulative Statements & 
 Identifies the preceding expression as indirect object component. &
 \begin{exi}\A{}(Certifier) \I{}(registers) \Bdir{}(certification) for \Bind{}(organic farmer).\end{exi} \\
%
\midrule
 \D{} & 
 Component & 
 IG Core, Regulative Statements &
 Identifies the preceding expression as a deontic modal representing duty. &  
 \begin{exi}\A{}(Certifier) \D{}(must) \I{}(monitor) \Bdir{}(farmers).\end{exi}\\

\midrule

\Cac{}/\Cacconst{} & 
Component & 
IG Core, Regulative \& Constitutive Statements & Identifies the preceding expression as an activation condition component. & 
 \udl{Regulative:} \newline
 \begin{exi}\Cac{}(Upon accreditation) \A{}(certifier) \D{}(must) \I{}(monitor) \Bdir{}(farmers).\end{exi}
 \newline \newline \udl{Constitutive:} \newline
 \begin{exi}\Cacconst{}(From 1st January onwards), \E{}(Council) \M{}(shall) \F{}(include) \Pp{}(organic farming representatives) \Cexconst{}(to review chemical allowances within organic food production standards).\end{exi} \\
\midrule

\Cex{}/\Cexconst{} & 
Component & 
IG Core, Regulative \& Constitutive Statements & Identifies the preceding expression as an execution constraint component. & 
\udl{Regulative:} \newline 
\begin{exi}\A{}(Certifier) \D{}(must) \I{}(monitor) \Bdir{}(farmers) \Cex{}(at any time).\end{exi} 
\newline\newline \udl{Constitutive:} \newline 
\begin{exi}\Cacconst{}(From 1st January onwards), \E{}(Council) \M{}(shall) \F{}(include) \Pp{}(organic farming representatives) \Cexconst{}(to review adherence with food production standards).\end{exi} \\

\midrule

\E{} & 
Component & 
IG Core, Constitutive Statements & 
Identifies the preceding expression as constituted entity & 
\begin{exi}\Cacconst{}(From 1st January onwards), \E{}(Council) \M{}(shall) \F{}(include) \Pp{}(organic farming representatives) \Cexconst{}(to review chemical allowances within organic food production standards).\end{exi}\\

\midrule

\Pp{} & 
Component & 
IG Core, Constitutive Statements & 
Identifies the preceding expression as constituting property & 
\begin{exi}\Cacconst{}(From 1st January onwards), \E{}(Council) \M{}(shall) \F{}(include) \Pp{}(organic farming representatives) \Cexconst{}(to review chemical allowances within organic food production standards).\end{exi} \\

\midrule

\F{} & 
Component & 
IG Core, Constitutive Statements & 
Identifies the preceding expression as constitutive function & 
\begin{exi}\Cacconst{}(From 1st January onwards), \E{}(Council) \M{}(shall) \F{}(include) \Pp{}(organic farming representatives) \Cexconst{}(to review allowances within organic food production standards).\end{exi} \\

\midrule

\M{} & 
 Component & 
 IG Core, Constitutive Statements &
 Identifies the preceding expression as a modal representing existential necessity (in contrast to duty or permission in regulative statements). & 
 \begin{exi}\Cacconst{}(From 1st January onwards), \E{}(Council) \M{}(shall) \F{}(be responsible) \Cexconst{}(for adherence with food production standards).\end{exi}
 \newline \newline \udl{Alternative example:} \newline
 \begin{exi}\Cacconst{}(From January 1st onward), there \M{}(shall) \F{}(be) a \E{}(National Organic Standards Advisory Council) \Pp{}(within the Department of Agriculture).\end{exi} \\
 
\midrule

\prop{}/ \propconst{} & 
Attributes, Object, Entity and Property components &
IG Core, Regulative \& Constitutive Statements &
Identifies properties of attributes and objects respective. The \begin{emp}prop\end{emp} symbol is used in conjunction with the respective component identifier. If coding on IG Extended level, where multiple properties for a given component exist, they receive a numeric index suffix. 

\vspace{1.05cm}

IG Extended further supports the explicit encoding of object and property hierarchies.

Where multiple levels of object/properties exist in the property hierarchy, those are contextualized with the objects/properties they refer to (i.e., they are appended to the component specification). Further details on property coding are provided in \cref{tab:CodingIgExtendedRegulative} and illustrated below.

& 

\udl{IG Core:}\newline\newline
\udl{Regulative:}\newline
\begin{exi}\A{},\prop{}(Certified organic) \A{}(farmers) \D{}(must) \I{}(respond) to \Bdir{},\prop{}(formal) \Bdir{}(certification requirements).\end{exi}
\newline\newline
\udl{Constitutive:}\newline
\begin{exi}The \E{}(Council) \F{}(consists of) \Pp,\propconst{}(elected)  \Pp{}(officials) \Pp,\propconst{}(resident in the electorate).\end{exi}
\newline\newline
\udl{IG Extended:}\newline\newline
\udl{Regulative:}\newline
\begin{exi}\A{},\propExt{1}(Certified) \A{},\propExt{2}(organic) \A{}(farmers) \D{}(must) \I{}(respond) to \Bdir,\propExt{1}(formal) \Bdir{}(certification requirements).\end{exi}
\newline\newline
\udl{Constitutive:}\newline
\begin{exi}The \E{}(Council) \F(consists of) \Pp,\propExt{1}(elected) \Pp(officials) \Pp,\propExt{2}(resident in the electorate).\end{exi}

\\
 & 
 & 
 &
 & 
\\

 & 
 &
 &
 &
\\
 &
 &
 &
 &

\\

 \prop{}/ \propconst{} \formatNeutral{(ctd.)}

 & Component
 & IG Extended, Regulative \& Constitutive Statements
 &
 The example on the right highlights a complex object hierarchy structure previously discussed in the context of the Object-Property Hierarchy (\cref{subsec:AttributeObjectHierarchy}). As mentioned above, where multiple objects on a given hierarchy level exist, they are uniquely identified with a numeric index (e.g., \begin{exi}\Bext{1}\end{exi}, \begin{exi}\Bext{2}\end{exi}, etc.). Where multiple properties on a given hierarchy level exist (where properties can be objects by themselves), they are uniquely identified with a numeric index (e.g., \begin{exi}\propExt{1}\end{exi}, \begin{exi}\propExt{2}\end{exi}, etc.). & 
 
 \begin{exi}\dots~\Bext{1},\propExt{1},\propExt{};\Bext{1},\propExt{2},\propExt{}(proposed) \Bext{1},\propExt{1}(suspension) or \Bext{1},\propExt{2}(revocation) of \Bext{1}(certification) \dots\end{exi}

 \\

 & 
 &
 &
 Where a single property applies to multiple properties, references to both objects/properties are maintained on this property (separated by semicolon). & 
 \\

\midrule

\AND{}, \OR{}, \XOR{}, \NOT{} & 
Statement, Component & 
IG Core, Regulative \& Constitutive Statements &
The logical operators identify the relationship between statement and/or components as either conjunction (\AND{}), inclusive disjunction (\OR{}), or exclusive disjunction (\XOR{}), as well as any combinations thereof. Where negation is involved, the \NOT{}~ operator is used (e.g., in deontics: must not; combination of exceptions: \NOT{}~\lp option 1 \AND{}~option 2\rp). & 
\begin{exi}\A{}(Certifiers) \D{}(must) \I{}\lp review \AND{}~ assess\rp~\Bdir{}(applications).\end{exi} 

\vspace{0.5cm}

\begin{exi}\A{}(Certifiers) \D{}(must) \I{}\lp review \AND{}~ \lp approve \XOR{} reject \rp \rp~\Bdir{}(applications).\end{exi} 

\vspace{0.5cm}

\begin{exi}\A{}(Certifiers) \D{}(must) \NOT{}~\I{}(review) \Bdir{}(applications) by \Bdir,\prop{}(offenders).\end{exi}

\\
\toprule

\caption{Symbol Reference for IG Coding as applied in this document}
\label{tab:SymbolReference}
\end{longtable}

\subsubsection{Coding Syntax}
\label{subsubsec:IGScript}

Following the intuitive introduction of the individual symbols annotating particular features of an institutional statement (e.g., actors, actions), at this stage, we will highlight selected syntactic principles of IG Script more explicitly, since these will be referenced specifically in the more advanced coding examples. The base syntax of IG Script is as follows:

\vspace{0.2cm}
\begin{exa}componentSymbol\lp~\dots~encoded content~\dots~\rp\end{exa}
\vspace{0.2cm}

\dots~where \begin{emp}componentSymbol\end{emp} is a symbolic reference to the component of choice as introduced in the previous \cref{subsubsec:CodingSymbols} (e.g., \begin{exi}\A{}(actor)\end{exi}). 

Where component values are combined as part of horizontal nesting, the linkage and the scope of the linkage is explicitly captured by parentheses (e.g., to indicate elements that apply to both linked values): 

\vspace{0.2cm}
\begin{exa}componentSymbol(shared value \lp left value \ls logicalOperator\rs~right value\rp~shared value)\end{exa}
\vspace{0.2cm}

Applied operationally, this can be exemplified as \begin{exi}\I{}(to \lp revise \AND{}~resubmit\rp )\end{exi}

All relational logical operators can be applied in this fashion, including a nested representation, in which additional parentheses indicate precedence in the case of different logical operators, such as exemplified below:

\vspace{0.2cm}
\begin{exa}\I{}\lp \lp revise \AND{}~resubmit\rp~\OR{}~revoke\rp\end{exa}
\vspace{0.2cm}

For logical operators of the same kind the indication of precedence can be omitted.

\vspace{0.2cm}
\begin{exa}\I{}\lp revise \OR{}~resubmit \OR{}~revoke\rp\end{exa}
\vspace{0.2cm}

The logical operator \NOT{}~plays a particular value, and can annotate individual components as well as statements as a whole.

Taking the following example, 

\begin{exa}
\A{}(Author) \D{}(must \NOT{}) \I{}(review) \Bdir{},\prop{}(own) \Bdir{}(paper).
\end{exa}

the operator can invert a single component value. 

\vspace{0.2cm}

Statements can be scoped using braces (\lb, \rb), e.g., to delimit multiple statements, or to reflect nesting structures. An example for a scope statement is 

\vspace{0.2cm}
\begin{exa}
\lb\A{}(Actor) \D{}(must) \I{}(conform) \Bdir{}(with policy).\rb
\end{exa}
\vspace{0.2cm}

Augmenting this statement with the inversion operator (where of analytical value) leads to the negated interpretation of the statement entirely:

\vspace{0.2cm}
\begin{exa}
\lb\A{}(Actor) \D{}(must) \I{}(conform) \Bdir{}(with policy)  \NOT{}\rb
\end{exa}
\vspace{0.2cm}

A central feature linked to the use of braces is the concept of vertical nesting and component-level nesting, i.e., the substitution of a given component of an institutional statement with the syntactic form of an institutional statement (e.g., preconditions expressed in terms of institutional state characterizations). Where components are augmented with braces instead of parentheses, this signals component-level nesting. 

In the case of vertical nesting, the \ORELSE~ (syntactically represented as \Oe{}) is essentially substituted with an entire statement (identifying the consequences of violating the monitored statement), as shown in the following example:

\vspace{0.2cm}
\begin{exa}
\A{}(citizen) \D{}(must) \I{}(conform) with \Bdir{}(policy) \Oe\lb \A{}(official) \D{}(may) \I{}(sanction) \Bdir{}(citizen)
\Cex{}(immediately)\rb.
\end{exa}
\vspace{0.2cm}

Where nesting occurs on other components, this is annotated in equivalent form. Component-level nesting is commonly found in the context of the activation condition by expressing preconditions for a particular activity in the same structure as the activity itself. This is exemplified in the following: 

\vspace{0.2cm}
\begin{exa}
\A{}(Official) \D{}(must) \I{}(impose) \Bdir{}(fine) \Cac{}\lb \A{}(citizen) \I{}(violates) \Bdir{}(rule)\rb.
\end{exa}
\vspace{0.2cm}

Implied component values are indicated with brackets surrounding the inferred value. For example, the observed statement ``they must comply'' may be contextually interpreted as ``citizens must comply''. To make this inference explicit, it can be encoded as follows:

\vspace{0.2cm}
\begin{exa}
\A{}([citizens]) \D{}(must) \I{}(comply).
\end{exa}
\vspace{0.2cm}

A common observation in the context of complex statements is the unique linkage between particular components and associated properties, an aspect related to the Object-Property Hierarchy (\cref{subsec:AttributeObjectHierarchy}). To indicate the direct linkage, the encoding relies on suffices associated with particular components in order to identify the linkage. 

\vspace{0.2cm}
\begin{exa}
\A{}(Official) \D{}(must) \I{}(impose) \Bdir{1},\prop{}{}(monetary) \Bdir{1}(fine) 

\Cac{}\lb \A{}(citizen) \I{}(violates) \Bdir{}(rule)\rb.
\end{exa}
\vspace{0.2cm}

A feature referenced in the context of IG Logico is the augmentation of the encoded institutional statements with semantic annotations that facilitate epistemological linkages to concepts drawn from theory of interest, or general categorizations introduced as part of the taxonomies discussed in \cref{sec:Taxonomies}. Semantic annotations are captured in squared brackets following the component symbol, but preceding the parentheses scoping the component content specification. The following example showcases the use of semantic annotations:

\vspace{0.2cm}
\begin{exa}
\A{}\customLog{type=actor}(Official) \D{}\customLog{stringency=prescription}(must) \I{}\customLog{act=administer}(impose) 

\Bdir1,\prop{}{}\customLog{prop=quals}(monetary) \Bdir{1}\customLog{object=sanction}(fine).
\end{exa}
\vspace{0.2cm}

Where annotations pertain to the statement entirely, the specification precedes the statement scope as shown below (here exemplified based on the qualification of the statement as regulative):

\vspace{0.2cm}
\begin{exa}
\customLog{regulativeStatement}\lb\A{}\customLog{type=actor}(Official) \D{}\customLog{stringency=prescription}(must) 

\I{}\customLog{act=administer}(impose) \Bdir{1},\prop{}{}\customLog{prop=quals}(monetary) \Bdir{1}\customLog{object=sanction}(fine).\rb
\end{exa}
\vspace{0.2cm}

\subsection{Regulative Statement Coding}
\label{subsec:RegulativeStatementCoding}

The base syntax of regulative statements (as shown in \cref{fig:RegulativeStatementLevels}) consist of necessary (in solid boxes) and sufficient components (in dashed boxes) as defined in \cref{tab:SyntacticElementsRegulative} using the symbols introduced in \cref{tab:SymbolReference}. The figure organizes feature refinements across different levels of expressiveness.

\begin{figure}[h!]
    \centering
    \framedfigure{\includegraphics[width=0.95\textwidth]{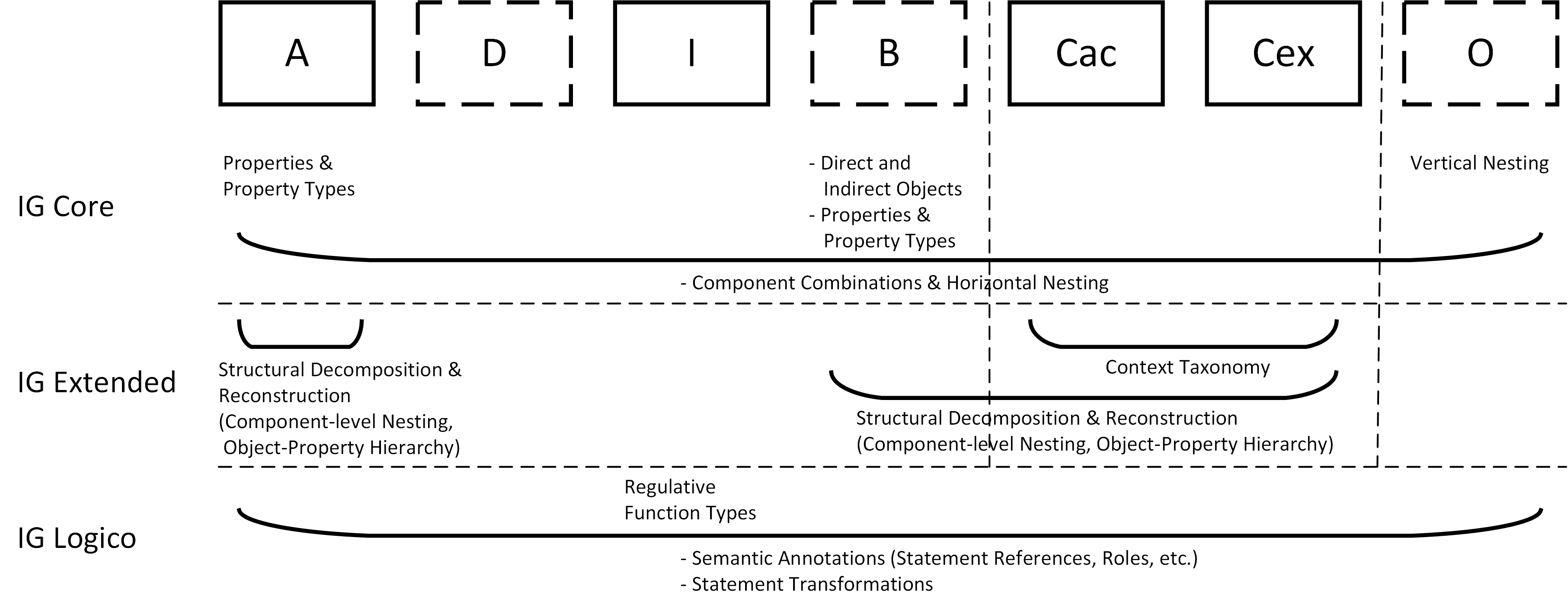}}
    \caption{Syntax and Features of Regulative Statements by Level of Expressiveness}
    \label{fig:RegulativeStatementLevels}
\end{figure}

Motivating the principal coding of regulative statements (without further elaboration at this stage), a stylized atomic regulative statement, coded in shorthand form, thus reads: 

\begin{exa}
\A{},\prop{}(Certified) \A{}(farmers) \D{}(must) \Cex{}(strictly) \I{}(adhere) to \Bdir{}(organic farming practices) 

\Cac{}(following their certification).
\end{exa}

\sp

The following Tables \ref{tab:CodingIgCoreRegulative} to \ref{tab:CodingIgLogicoRegulative} provide detailed coding guidelines for regulative statements for all individual components, organized by level of expressiveness, starting with IG Core through IG Logico, leveraging the notation convention captured in \cref{subsec:CodingSyntax}. As the institutional analyst commences coding at any level of expressiveness, the following general coding principles should be entertained.

\begin{itemize}
    \item Where possible, the targeted level of encoding should be clarified at the beginning (see \cref{subsec:IgCodingLevels}).
    \item The coder should acquaint oneself with the concepts relevant for the target level of encoding (see \cref{subsec:IgCodingLevels}), as well as coding conventions, such as applied notation. A specific concern discussed as part of the planning process is the handling of absent syntactic components that are either implied or can be inferred from policy texts.
    \item Recall that coding can (but does not have to) occur iteratively, starting at one level that prompts less granular syntactic expressiveness (e.g., IG Core) moving with a subsequent coding pass to another level that prompts more granular coding (e.g., IG Extended). However, the specific approach is subject to coder experience, and of course available tool support, all of which should be considered in the planning phase (see \cref{sec:PrecodingSteps} for further considerations). 
\end{itemize}

\newpage

\begin{landscape}

\subsubsection{IG Core Coding of Regulative Statements}
\label{subsubsec:IGCoreRegulative}

\begin{longtable}{\tabledims}

\toprule
\multicolumn{4}{l}{\textbf{Level of Expressiveness: IG Core}} \\
\multicolumn{4}{\tablespan}{ 

IG Core enables basic, structural analysis of institutional statements. Encoding at this level is designed to be human readable and moderately comprehensive in the detail with which syntactic properties of institutional statements are captured.} \\

\toprule
{\textbf{Syntactic\newline Component}} & 
{\textbf{Treatment of Syntactic Components by Level of Encoding}} &
{\textbf{Relevant Examples}} &
{\textbf{Complete Syntactic Classification of Examples}} \\
\toprule

\begin{emp}Attributes\end{emp} & {The encoding of Attributes, which can include an animate actor (individual or organizational) only or an animate actor and a property of this actor, differentiates between actor (Attributes) and actor property (Attribute Properties).

\sp

An important decision to establish reliability is a shared understanding of the level of decomposition of Attributes and their corresponding properties performed as part of the coding process (or even pre-coding process), an aspect that relies on the domain, context, and nature of the coded policy. 

The example on the right illustrates the principles of decomposition assuming the most fine-grained possible decomposition. 

Guidelines to determine the appropriate level of decomposition are provided below.

} &

{\udl{Example statement:} 

\begin{exi}Certified farmer must submit an organic system plan annually. \end{exi}

\sp

\udl{Attributes encoding:}

\sp

Attributes = \begin{exi}certified farmer\end{exi}
 
\sp
 
Actor = \begin{exi}farmer\end{exi}
 
\sp
 
Actor property = \begin{exi}certified\end{exi}
 
} & 

\vspace{0.03cm}

\begin{exi}{\A{},\prop{}(Certified) \A{}(farmer) \D{}(must) \I{}(submit) an \Bdir{}(organic systems plan) \Cex{}(annually). }\end{exi} \\

\begin{emp}\hypertarget{tablink:AttributePropertyHeuristics}{Attributes (ctd.)}\end{emp} & \multicolumn{3}{p{17cm}}{

\udl{Deciding on the level of decomposition into Attributes and corresponding Attribute Properties:}

\sp

\begin{emp}The guidelines provided below equally apply to Object and Object Properties; Constituted Entity and Constituted Entity Properties; Constituting Properties and Constituting Properties Properties.\end{emp}

\sp

A general challenge across multiple components of the IG is the identification of appropriate heuristics for the decomposition of actor descriptors into Attribute and corresponding properties, and likewise Objects and Object Properties, etc. To support the study design, we propose a set of guidelines to inform the choice of decomposition levels as appropriate for a given study.

\sp

General note: Named entities (e.g., United States Department of Agriculture) are \begin{emp}not\end{emp} decomposed into component properties, i.e., such terms are always annotated as Attributes, Object, Constituted Entity or Constituting Properties, but not their corresponding properties (i.e., Attributes Properties, Object Properties, Constituted Entity Properties, Constituting Properties Properties). 

\sp

Guidelines for determining level of decomposition (in order of consideration):

\begin{itemize}
\item Policy documents often contain a `Definitions' article or section, in which concepts of relevance to the policy are specified. Where this is the case, these definitions represent the desirable level of decomposition, i.e., if `Organic farmer' is defined as part of the definitions, no further decomposition is required, unless analytical objectives motivate further decomposition (see next item).

\item Where absent, the level of decomposition is contingent on the policy focus and context. If, for example, a regulation differentiates between organic and non-organic farmers, the decomposition into descriptor and associated properties is useful, in which both `types' of farmers are distinguished by the properties. If, however, the regulation is exclusively concerned with organic farmers, the decomposition is of little analytical value at this level of encoding.

\item Absent any other guidelines, decomposition should be performed on the most fine-granular level.
\end{itemize}} \\

\newpage

\midrule
\begin{emp}Object\end{emp} & {The encoding of Objects identifies Direct Objects specified within institutional statements, along with their respective properties. Where statements are comprised of both direct and indirect object, it also entails the explicit identification of indirect objects, i.e., objects that are affected by the application of the aim to direct objects, e.g., which the action is targeted to. 

Objects can be concrete or abstract entities. Concrete entities can be animate (e.g., farmers) or inanimate (e.g., notification) in kind. Abstract entities (e.g., beliefs, goals, observations) are generally expressed in terms of statements of facts or institutional statements. The coding of such statements (\begin{emp}component-level nesting\end{emp}) is discussed under IG Extended (see \cref{tab:CodingIgExtendedRegulative}).  

\vspace{0.2cm}

Rules:
\begin{itemize}
    \item Identify object identifier
    \item Identify object properties
    \item Repeat for the indirect object
\end{itemize}

\udl{Note:} The separation of Objects and Object properties operates identical to the decomposition of Attributes and Attributes properties. See corresponding guidelines provided under the Attributes component specification (see \hyperlink{tablink:AttributePropertyHeuristics}{here}).} & 

{\udl{Example statement:}

\begin{exi}Organic certifier must send farmer notification of compliance within 30 days of inspection.\end{exi}

\vspace{0.5cm}

{Direct Object = \begin{exi}notification of compliance\end{exi}}

\vspace{0.5cm}

{Indirect Object = \begin{exi}farmer\end{exi}}

\vspace{0.5cm}

{Object property = \begin{exi}organic\end{exi}}

\vspace{0.5cm}

Note: The interpretation of property depends on the encoded policy (and, of course, the coder’s analytical objectives). If a policy on organic farming exclusively refers to organic farmers as a proper noun, organic is not an attribute property, but part of the attribute; if the policy differentiates between organic and other types of farmers, capturing the specific characterisation as property is suggested. For the example here we highlight the second pathway.

} &

{
\vspace{0.03cm}

\begin{exi}\A{},\prop(Organic) \A{}(certifier) \D{}(must) \I{}(send) \Bind{}(farmer) \Bdir{}(notification of compliance) \Cex{}(within thirty days of inspection).\end{exi}} \\

\begin{emp}Object (ctd.)\end{emp} & \multicolumn{3}{p{13cm}}{
\udl{Potential pitfalls:}
\begin{itemize}
     \item \begin{emp}Objects vs.~Constraints\end{emp}: The introduction of the indirect object offers the benefit of capturing the functional interdependence of objects and implied directionality. This directionality is sometimes explicitly reflected, as in the following statement. When encountering such statements, the coder should be careful to not mischaracterize the indirect object preceded by a preposition as context. Example: ``Parents must take children to school.'', ``school'' is sensibly resolved as indirect object, but based on its prepositional embedding, could be mislabeled as constraint. For the characterization, we thus require an initial consideration of clauses containing objects (other than the direct object) as indirect objects, before characterizing those as a contextual descriptor (which primarily make reference to the contextual embedding of actions).

    \item \begin{emp}Object Properties vs.~Constraints\end{emp}: Differentiation between the object property and execution constraint can also sometimes be challenging in the encoding. When such confusion arises, the coder should ask herself to reflect on whether the statement words or clauses in question are qualifying the object or qualifying the action of the statement. Take the following statement for example that illustrates the referenced potential confusion: ``Certifiers shall perform audits on product stock two times per year.'' The clause that may potentially give rise to confusion is ``on product stock.'' This could be confused as an execution constraint relating to purpose, but in fact it describes the type of audit to be performed. \end{itemize}
} \\

\midrule
\begin{emp}Aim\end{emp} & {The encoding identifies the focal action of the statement.} & \udl{Example statement:}

\begin{exi}Organic certifier must send farmer notification of compliance.\end{exi}
\newline

Aim = send
 & 
 
 \vspace{0.03cm}
 
 \begin{exi}{\A{},\prop{}(Organic) \A{}(certifier) \D{}(must)  \I{}(send) \Bind{}(farmer) \Bdir{}(notification of compliance).}\end{exi} \\

\newpage

\midrule

\begin{emp}Deontic\end{emp} & 

{The encoding identifies the prescriptive operator that indicates whether the Aim (i.e., action) of the statement is required, allowed, or forbidden. Common Deontics indicating varying levels of prescriptive force include \begin{emp}must\end{emp}, \begin{emp}may\end{emp}, and \begin{emp}must not\end{emp}.} & 

{\udl{Example statement:}

\begin{exi}Organic certifier must send farmer notification of compliance.\end{exi} 
\newline

Deontic = must
} & 

\vspace{0.03cm}

\begin{exi}{\A{},\prop{}(Organic) \A{}(certifier) \D{}(must) \I{}(send) \Bind{}(farmer) \Bdir{}(notification of compliance).}\end{exi} \\

\midrule
\begin{emp}Context\end{emp} & 
{The encoding identifies the Context of the institutional statement. The encoding differentiates between \begin{emp}``Activation Conditions,''\end{emp} which are contextual clauses that specify preconditions under which the Aim is expected to occur or not occur, and \begin{emp}``Execution Constraints,''\end{emp} which are contextual descriptors that qualify the Aim by assigning in relation to it temporal, spatial, procedural, and/or other constraining parameters.} & 
\udl{Example statement:} 

\begin{exi}Upon entrance into agreement with organic farmer to serve as his/her certifying agent, organic certifier must inspect farmer’s operation within 60 days.\end{exi}

\sp

Context clauses: \begin{exi}Upon entrance into agreement with organic farmer to serve as his/her certifying agent; within 60 days.\end{exi}

Context encoding: 

\sp

Activation Condition: \begin{exi}Upon entrance into agreement with organic farmer to serve as his/her certifying agent\end{exi}

\sp

Execution Constraint: \begin{exi}within 60 days\end{exi}

\sp

Note: While coding the essential aspects of the statement, IG Core is limited with respect to capturing the nested complexity in the activation condition. We will revisit this statement in the context of IG Extended coding.
 & 
{

\vspace{0.03cm}

\begin{exi}\Cac{}(Upon entrance into agreement with organic farmer to serve as his/her certifying agent), \A{}(organic certifier) \D{}(must) \I{}(inspect) \Bdir{}(farmer’s operation) \Cex{}(within 60 days).\end{exi}} \\

\newpage

\midrule

\begin{emp}Or else\end{emp} 

&

The encoding of Or else statements identifies consequences (e.g., payoffs) of compliance/non-compliance with institutional statements, or the conduct of Aim (i.e., activities) assigned to specific Attributes (i.e., actors) in institutional statements. The encoding captures these consequences, which generally  take the form of regulative institutional statements that nest from the ‘non-Or else’ (monitored)  statements.

\sp

Sometimes, the statements on which Or else, or consequential statements, nest convey a combination of  monitoring activity and the associated  payoff attached to outcomes of monitoring actions. 

\sp

The encoding of Or else statements accommodates both vertical and horizontal nesting. Vertical nesting is applicable when there is one payoff activity that is specified within a distinct institutional statement as a consequence of an action indicated in another institutional statement. Horizontal nesting is applicable when there are two or more payoff activities that can be pursued as consequences of an action indicated in another institutional statement. 

&

\udl{Example:} 

\begin{exi}Certified organic farmers must not apply synthetic chemicals to crops at any time once organic certification is conferred, or else certifier will revoke certification from farmer.\end{exi}

\sp

Or else clause comprising statement: \begin{exi}or else certifier will revoke certification from farmer\end{exi}

\sp

Or else statement nests on: \begin{exi}certified organic farmers must not apply synthetic chemicals to crops at any any time once organic certification is conferred\end{exi}

\sp

The above encoding exemplifies vertical nesting. The following example, which is an extension of the above, exemplifies horizontal nesting within a vertically nested statement. 

\sp

\udl{Example statement:} 

\begin{exi}Certified organic farmers must not apply synthetic chemicals to crops at any time once organic certification is conferred, or else certifier will revoke certification from farmer or fine farmer.\end{exi}

& 

In the following example, horizontal nesting is signaled using parentheses (\lp~and \rp) around statements (as opposed to individual components), and vertical nesting is expressed using braces on the \ORELSE~component (\lb~and \rb). 

\sp

\udl{Vertical nesting:}

\begin{exi}\A{},\prop{1}(Certified) \A{},\prop{2}(organic)  \A{}(farmers) \D{}(must not) \I{}(apply) \Bdir{}(synthetic chemicals) to \Bind{}(crops) \Cex{}(at any time) \Cac{}(once organic certification is conferred), \Oe{}\lb  \A{}(certifier) \D{}(will) \I{}(revoke) \Bdir{}(certification) from \Bind{}(farmer)\rb.\end{exi}

\\

\begin{emp}Or else (ctd.)\end{emp}

&

These statement combinations can signal 

\begin{itemize}
    \item alternative exclusive action options -- XORs -- (e.g., either suspending XOR revoking the certification),
    \item inclusive action options -- ORs -- (e.g., sanctions apply if a driver is caught speeding AND/OR on the phone) or 
    \item co occurring action options -- ANDs -- (e.g., fining a transgression AND reporting to authorities).
\end{itemize}

& 

\udl{Or else clauses comprising statements:} 

\begin{exi}or else certifier will fine farmer or revoke certification from farmer.\end{exi}

\sp

\udl{Vertical and horizontal Or else nesting encoding:}

\sp

\begin{exi}Certified organic farmers must not apply synthetic chemicals to crops at any time once organic certification is conferred\end{exi}

\sp

\udl{[Vertical nesting]}

\sp

\begin{exi}or else certifier will revoke certification from farmer\end{exi}

\sp

\udl{[Horizontal nesting]}

\sp

\udl{\XOR~(signaling exclusive or)}

\sp

\begin{exi}Or else certifier will fine farmer\end{exi}

&

\udl{Horizontal nesting within vertically-nested statement:}

\begin{exa}\A{},\prop{1}(Certified) \A{},\prop{2}(organic) \A{}(farmers) \D{}(must not) \I{}(apply) \Bdir{}(synthetic chemicals) to \Bind{}(crops) \Cex{}(at any time) \Cac{}(once organic certification is conferred), 

\Oe{}~\lb\lb \A{}(certifier) \D{}(will) \I{}(revoke) \Bdir{}(certification) from \Bind{}(farmer)\rb~
\XOR~ 
\lb \A{}(certifier) \D{}(will) \I{}(fine) \Bdir{}(farmer)\rb \rb.\end{exa}

\\

&
\multicolumn{3}{p{13cm}}{
\udl{Note:} 
\begin{itemize}
    \item Where component-level combinations exist (\dots~fine farmer \begin{emp}or\end{emp} revoke \dots), those have to be signaled explicitly by parentheses, or be decomposed into separate logically-combined complete atomic institutional statements. Further details are provided under 
    \hyperlink{tablink:DecompositionOfComponentLevelCombinationsExtendedRegulative}{\begin{emp}General IG Extended Instructions\end{emp}, Item \EntryDecompositionOfComponentLevelCombinations} in \cref{tab:CodingIgExtendedRegulative}.
    \item Ambiguities with respect to the linguistic use of logical operators (exclusive and inclusive or) are to be resolved as part of this process. 
\end{itemize}
}
\\


\newpage

\midrule
\multicolumn{4}{l}{\textbf{General IG Core Instructions for Regulative Statements}} \\
\midrule

\begin{emp}Additional annotations for Attributes, Objects, and Context\end{emp}

\sp

(\udl{Note:} This feature is analogous to the specification of additional annotations for Constituted Entity, Constitutive Function and Constituting Properties components in the context of constitutive statements.)

& 
In addition to the identification of properties embedded in the original statements, components can further be annotated using additional annotation labels. Such labels can follow the categories listed in \cref{sec:Taxonomies}, or be specific to the project objectives. 

\sp

A systematic approach to labelling entities is discussed under \hyperlink{tablink:CrossComponentSemanticAnnotationsLogicoRegulative}{\begin{emp}IG Logico Instructions\end{emp}, Item \EntryCrossComponentSemanticAnnotations} in \cref{tab:CodingIgLogicoRegulative}. This is particularly recommended if annotations are of strong relevance for the coding and of diverse nature. 
& 
\udl{Example statement:} 

\begin{exi}Organic certifier must send farmer notification of compliance.\end{exi} 

\sp

Subject to analytical necessity, additional annotations can for instance relate to the identification of aspects such as the characterisation of encoded objects with respect to their animacy as either animate or inanimate -- signified in brackets in the coded example. Where indicated, the annotation should be separated from the component specification by semicolon and have the structure ``\begin{exi}label=\end{exi}'', followed by the annotation. 

Note that ``\begin{exi}label\end{exi}'' is a characterizing prefix, either based on a specific taxonomy (see \cref{sec:Taxonomies} for an overview), or a custom coder-defined category specification (e.g., defined as part of the project-specific guidelines). In this example, the Animacy Taxonomy has the prefix \begin{emp}\anim\end{emp} as defined in \cref{sec:Taxonomies} and is used correspondingly.

\sp

While exemplified here for regulative statements, this equally applies to constitutive statements. 
& 

\vspace{0.03cm}

\begin{exi}\A{}\customLog{\anim=animate}(Organic certifier) \D{}(must) \I{}(send) \Bind{}\customLog{\anim=animate}(farmer) \Bdir{}\customLog{\anim=inanimate}(notification of compliance).\end{exi} \\

\newpage

\midrule

\hypertarget{tablink:DecompositionOfComponentLevelCombinationsCoreRegulative}{\EntryDecompositionOfComponentLevelCombinations} 

\sp

(Note: This applies to regulative and constitutive statements, and is discussed here with focus on the regulative perspective.)

& {Where combinations of components (component-level combinations) are observed that are not explicitly decomposed as in the case of vertical nesting, or not explicitly identified as component-level combinations (e.g., using parentheses), these can be decomposed into logically-combined statements. Other than for aims, the decomposition is optional for IG Core. 

\sp

Extended details are provided under  \hyperlink{tablink:DecompositionOfComponentLevelCombinationsExtendedRegulative}{\begin{emp}General IG Extended Instructions\end{emp}, Item \EntryDecompositionOfComponentLevelCombinations} in \cref{tab:CodingIgExtendedRegulative}; IG Core Examples can be found in the following.

} & &  \\
 &  &  &  \\
 
\begin{emp}Decomposition of component-level combinations (ctd.)\end{emp} 

& 
{Operationally, combinations of components are evidenced by the presence of multiple attributes, objects, aims or execution constraints within a single institutional statement. Examples of each case are provided in the next column.

\sp

Decomposition essentially entails constructing an individual statement to capture each of the unique components represented in multiples within institutional statements, noting the relation to the original statement in which multiple components are reflected. Information from component fields, other than that containing multiple components, is simply carried over to all related institutional statements.

\sp

Importantly, where decomposition actually changes the meaning of the original institutional statement containing multiple components within a particular syntactic field, the statement should not be decomposed. In such cases, multiples are typically intended to exist in coupled form. An example is provided in the next column.

\sp

\udl{Note:} These guidelines highlight the motivation for the decomposition, and exemplify the process. Depending on the use of annotation means and tool support, the decomposition may be signaled by annotation of component combinations, and thus occur automated without requiring explicit decomposition by users.
} 
& 
\udl{Multiple Attributes Example Statement:} 

\begin{exi}Certifiers and Inspectors must seek accreditation annually.\end{exi}

\sp 

Decomposed as:

\sp

Statement 1: \begin{exi}Certifiers must seek accreditation annually.\end{exi}

\AND

Statement 2: \begin{exi}Inspectors must seek accreditation annually.\end{exi}

\sp

\udl{Multiple Aims Example Statement:} 

\begin{exi}Inspectors must sign and file farm inspection reports following site visits.\end{exi}

\sp

Decomposed as:

\sp

Statement 1: \begin{exi}Inspectors must sign farm inspection reports following site visits.\end{exi}

\AND

Statement 2: \begin{exi}Inspectors must file farm inspections reports following site visits.\end{exi}

\sp

\sp

\udl{Multiple Conditions Example Statement:} 

\begin{exi}Inspectors must conduct site visits in person twice per year.\end{exi}

\sp

Decomposed as:

\sp

Statement 1: \begin{exi}Inspectors must conduct site visits in person.\end{exi}

\AND

Statement 2: \begin{exi}Inspectors must conduct site visits twice per year.\end{exi}


\\

\begin{emp}Decomposition of component-level combinations (ctd.)\end{emp} 

& &

\udl{Coupled Component Example Statement not to be decomposed:} 

\begin{exi}Farmers must pay Certifier \$250 for application and service fees upon entry into contract for certification services.\end{exi}

\sp

The reason for foregoing decomposition in such case lies in the inseparability of application and services fees, since they are reported as a combined fee of \$250.

& \\

\toprule
\caption{Coding Guidance on Syntactic Elements for IG Core as Level of Expressiveness (Regulative Statements)}
\label{tab:CodingIgCoreRegulative}
\end{longtable}%

\vspace{3cm}

\subsubsection{IG Extended Coding of Regulative Statements}
\label{subsubsec:IGExtendedRegulative}

\begin{longtable}{\tabledims}
\toprule
\multicolumn{4}{\tablespan}{\textbf{Level of Expressiveness: IG Extended}

\sp

IG Extended enables more detailed structural analysis of institutional data than IG Core and accommodates computational application to aid in institutional coding and analysis. Encoding at this level is designed to be human readable, moderately computationally tractable, and moderately comprehensive in the detail with which syntactic properties of institutional statements are captured.  

Coding institutional statements on this level enforces many of the features that have been optional in IG Core and affords a fine-grained decomposition of statements. This includes a richer context characterisation based on predefined taxonomies, the expansion of component-level combinations to reconstruct atomic statements and their relationships, the decomposition of components containing embedded institutional statements or statements of fact (component-level nesting). Finally, IG Extended captures the decomposition of hierarchical relationships amongst explicitly highlighted actors, object and their respective properties.

} \\

\newpage

\toprule
{\textbf{Syntactic\newline Component}} & 
{\textbf{Treatment of Syntactic Components by Level of Encoding}} &
{\textbf{Relevant Examples}} &
{\textbf{Complete Syntactic Classification of Examples}} \\

\toprule

\begin{emp}Attributes\end{emp}

&

\udl{Key features:}
\begin{itemize}
\item Property Hierarchy Decomposition
\item Component-level Nesting
\end{itemize}

Building on the IG Core coding that identifies attributes and their respective properties, IG Extended affords a more comprehensive decomposition of the attributes component into Attributes, properties, along with their respective functional descriptors (higher-order properties). In addition, we can identify related attribute objects (so called to maintain the reference to the attribute) that carry their own properties and parameters. This approach is explored in the first example.

Within this property hierarchy, elements may be substituted by an institutional statement or statement of fact, referred to as \begin{emp}component-level nesting\end{emp} (which likewise applies to other components such as objects, conditions and constraints). We explore this approach in the second example.

Note that in all cases, named entities (e.g., United States Department of Agriculture) are not decomposed.

&

\udl{Example Statement:}\newline
\begin{exi}A certified farmer whose certification is suspended by the Secretary under this section may at any time submit a recertification request.\end{exi}

\sp

In this example, the attribute  \begin{exi}``A certified farmer whose certification is suspended by the Secretary under this section \dots''\end{exi} embeds aspects central to the institutional configuration. Beyond the identification of properties of the attribute (\begin{exi}``certified''\end{exi}), it highlights a complex property in the form of a nested institutional statement signaled by the involvement of another entity. Reformulated as \begin{exi}``Secretary suspends certified farmer's certification''\end{exi}, we can construe this second property as a nested statement with the corresponding component characteristics, as shown in the coding of the example. 

\sp

While objects related to attributes are objects, they are coded with a reference to the attribute to ensure unambiguous association with the attributes component in the institutional statement.

&

\vspace{0.05cm}

\begin{exi}A \A{},\propExt{1}(certified) \A{}(farmer) \A{},\propExt{2}\lb whose  \Bdir{}(certification) is suspended \I{}(\lsi suspends\rsi) by the \A{}(Secretary) \Cex{}(under this section)\rb~\D{}(may) \Cac{}(at any time) \I{}(submit) a  \Bdir,\prop{}(recertification) \Bdir{}(request).\end{exi}

\\

\begin{emp}Attributes (ctd.)\end{emp}

& & 
Where properties or parameters apply to multiple objects or entities, the respective references are separated by semicolon (e.g., \A{1},\propExt{1}; \A{2},\propExt{1}). This coding is exemplified in the context of the object component.

Properties can operate on an arbitrary level of depth. For example, properties of properties (second-order properties) are coded as (\A{},\propExt{1},\propExt{1}), where property identifier are unique on each level.

\sp

Using a further example, we can showcase the coding of beliefs or assessments more generally in the form of second-order property hierarchies that rely on component-level nesting as shown before:

\sp

\udl{Example Statement:} 

\begin{exi}Program managers who believe that a certified operation has violated the Act may pursue revocation proceedings.\end{exi}

\sp

In contrast to the previous statement that highlights complex property arrangements on a given level, this statement emphasizes a hierarchical organization amongst properties, where the second-order property contains a complete institutional statement.

& 

\vspace{9.2cm}

\begin{exi}\A{}(Program managers) \A{},\propExt{1}(who believe) \A{},\propExt{1},\propExt{1}\lb that a \A{}(certified operation) \I{}(has violated) the \Bdir{}(Act)\rb~\D{}(may) \I{}(pursue) \Bdir{}(revocation proceedings).\end{exi}

\\

\begin{emp}Attributes (ctd.)\end{emp}

& & 

Nested institutional statement = \begin{exi}certified operation has violated the Act\end{exi}

\sp

Similar to the previous coding, attributes are decomposed into their identifier and properties. Here, the nested statement is captured in the second-order property (\A{},\prop{},\prop{}), since the statement describes the content of the belief (\A,\prop{}). 
& \\

\midrule

\begin{emp}Object\end{emp} &

\udl{Key features:}
\begin{itemize}
\item Object-Property Hierarchy Decomposition
\item Component-level Nesting
\end{itemize}

Building on the differentiation into direct and indirect object along with their respective properties in IG Core, in IG Extended the coding is refined to capture relationships between elements in object specifications. Specific focus lies on the decomposition of hierarchical relationships between objects and properties, respectively. 
In addition, embedded objects without direct functional relationship are identified, i.e., objects that are referred to but are not explicitly acted on in the context of the institutional statement.

As highlighted for the attributes component, elements of the object-property hierarchy may be substituted by an institutional statement, in which case component-level nesting applies for the encoding (which is exemplified following the object-property decomposition).

\sp

\udl{Rules for object-property decomposition:}

\begin{itemize}
    \item The identification of the direct or indirect Object respectively.
\end{itemize}

&

\udl{Example statement:}

\begin{exi}The Program Manager shall send a written notification of proposed suspension or revocation of certification to certified organic farmer.\end{exi}

\sp

Similar to the attributes coding in IG Extended, this statement decomposes complexity embedded in the object component:

\sp

Object = \begin{exi}written notification of proposed suspension or revocation of certification\end{exi}

\sp

The direct object in this statement is the \begin{exi}``notification''\end{exi}, which has the property \begin{exi}``written''\end{exi}, and makes reference to another related functionally independent object \begin{exi}``certification''\end{exi}. Functional independence here refers to the fact that the certification does not depend on the notification. The certification itself is characterised by further dependent objects such as the \begin{exi}``suspension''\end{exi} and \begin{exi}``revocation''\end{exi}. Both objects are dependent, since they depend on the existence of a certification. Both dependent objects share the property of being \begin{exi}``proposed''\end{exi}.

&

\vspace{0.04cm}

\begin{exi}The \A{}(Program Manager) \D{}(shall) \I{}(send) a \Bdir{}, \propExt{}(written) \Bdir{}(notification) of \Bdir{},\propExt{1},\propExt{1},\propExt{}; \Bdir{},\propExt{1},\propExt{2},\propExt{}(proposed) \Bdir{},\propExt{1},\propExt{1}(suspension) or \Bdir{},\propExt{1},\propExt{2}(revocation) of \Bdir{},\propExt{1}(certification) to \Bind{},\propExt{1}(certified) \Bind{},\propExt{2}(organic) \Bind{}(farmer).\end{exi}

\\

\begin{emp}Object (ctd.)\end{emp}

& 

\begin{itemize}
	\item The identification of properties associated with the Object, both including directly functionally dependent (e.g., descriptors) and conceptually independent properties. If an object is a named entity (e.g., United States Department of Agriculture), it is not decomposed.
    \item For each property, functional relationships are identified (i.e., which Object or property they rely on). Where such relationships exist, these properties become child properties of the properties they functionally depend on. This process produces an Object hierarchy that may or may not directly involve the Object component itself.
    \item Where multiple children exist, implicit or explicit logical relationships (conjunction, disjunction, negation) between different branches of the emerging Object hierarchy are retained.
    \item Non-functional relationships are established between the root node of the Object hierarchy and the direct or indirect Object established in the first step.
\end{itemize}

& 

\sp

Establishing these relationships, we can code the object and respective properties as follows:

\sp

Related functionally-independent object = \begin{exi}certification\end{exi}

\sp

Functionally independent objects are identified as individual objects with unique alphabetical index, e.g., (\Bdir{1}), (\Bdir{2}), etc.

\sp

Dependent objects = \begin{exi}suspension, revocation\end{exi}

\sp

Dependent objects are coded with reference to the object they depend on and unique numeric index, e.g., (\Bdir{1}\customExtRaw{,p1}), (\Bdir{1}\customExtRaw{,p2}), etc.

\sp

Dependent object properties = \begin{exi}proposed\end{exi}

\sp

Dependent object properties are identified with reference to objects they relate to, with references to different objects separated by semicolon, e.g., (\Bdir{}\customExtRaw{,p1,}\propExt{};\Bdir{}\customExtRaw{,p2,}\propExt{}), etc. 

\sp

Where multiple properties exist for an object, they are, as in IG Core coding, uniquely identified by numeric index, e.g., (\Bdir{}\customExtRaw{,p1,}\propExt{1}), (\Bdir{}\customExtRaw{,p1,}\propExt{2}), etc.

& \\

\begin{emp}Object (ctd.)\end{emp}

&

\vspace{1.52cm}

Where object hierarchies are specified outside the object component (latent objects, e.g., as part of the context component), the same process applies, even if no reference is made to direct or indirect object. Note, however, that those are coded with lower priority, i.e., following object hierarchies that relate to direct or indirect object.

&

Where higher-order properties exist, these are coded according as specified in the context of the Attributes component.

\sp

\udl{Object hierarchies outside the object component:}

Taking the example \begin{exi}``\A{}(Certifier) \D{}(must) \I{}(send) \Bdir{}(notification) to \Bind{}(inspector) \Cex{}\customExt{ctx=pur}(to produce a written report of the assessment)''\end{exi}, we find an object hierarchy within the context (execution constraint) component.

In this example, first-order object is the report that is written and relates to an assessment, without being functionally dependent -- as reflected in the coding.

& 

\vspace{2.55cm}

\begin{exi}\A{}(Certifier) \D{}(must) \I{}(send) \Bdir{}(notification) to \Bind{}(inspector) \Cex{}\customExt{ctx=pur}\lp to produce a  \Bext{}\customExtRaw{,p1,p1,}\propExt{1}(written) \Bext{}\customExtRaw{,p1,p1}(report) of the \Bext{}\customExtRaw{,p1}(assessment)\rp.\end{exi}

\sp

In this example, we use parentheses to clearly delineate the scope of the execution constraint that contains the external object hierarchy linked to the latent object ``assessment''.

\\

&

\udl{Component-level nesting:}

Another feature of IG Extended is the notion of component-level nesting, i.e., the decomposition of individual components that are expressed in terms of institutional statements or statements of fact. This approach affords the conceptual representation of mental constructs, such as beliefs, opinions or behavioral assessments. Note that component-level applies to various components (e.g., Attributes, Objects, Conditions, Constraints), but is exemplified for objects at this stage. 

&
\udl{Component-level nesting on objects:}

Reviewing the following example ``\begin{exi}Inspectors must ensure that certified organic farmers report annually on their agricultural practices.\end{exi}'', the object of the statement (``\begin{exi}certified organic farmers report annually on their agricultural practices\end{exi}'') reflects nested institutional statement in its own right. 

&

\vspace{0.05cm}

\begin{exi}\A{}(Inspectors) \D{}(must) \I{}(ensure) that \Bdir{}\lb  \A{},\propExt{1}(certified) \A{},\propExt{2}(organic) \A{}(farmers) \I{}(report) \Cex{}(annually) on their \Bdir{}(agricultural practices)\rb.\end{exi}

\\ 

\midrule

\emph{Aim} &

\multicolumn{3}{\tablespanNarrow 
}{
The coding of the Aim is identical to IG Core. See also \hyperlink{tablink:DecompositionOfComponentLevelCombinationsExtendedRegulative}{\begin{emp}General IG Extended Instructions\end{emp}, Item \EntryDecompositionOfComponentLevelCombinations} in \cref{tab:CodingIgExtendedRegulative} for associated decomposition rules.} \\

\midrule

\emph{Deontic} &

\multicolumn{3}{p{13cm}}{
The coding of the Deontic is identical to IG Core.} \\

\midrule

\emph{Context} &

\udl{Key features:}
\begin{itemize}
\item Context Taxonomy
\item Component-level Nesting
\end{itemize}

In IG Extended, Activation conditions (\Cac) and Execution constraints (\Cex) (as specified in IG Core) are further characterised in terms of ontological categories, specifically capturing the nature of circumstances the conditions or constraints describe. Those are grouped into taxonomies comprehensively described in \cref{sec:Taxonomies} (along with subcategories). The central taxonomy in this context is the \begin{emp}Context Taxonomy\end{emp}, an extract of which is provided in the following, along with corresponding annotations:

\sp

\udl{Excerpt from Context Taxonomy:}

\begin{itemize}
    \item Temporal (tmp)
    \item Spatial (loc)
    \item Domain (dom)
    \item Procedural (prc)
    \item Purpose (pur)
    \item \dots
\end{itemize}

A comprehensive overview of the Context Taxonomy including subcategories (e.g., `point in time', `time frame') is provided in \cref{subsec:ContextTaxonomy}, alongside further specific abbreviations (e.g., \begin{emp}tfr\end{emp} for time frame).

&

\udl{Example 1:}

\begin{exi}Upon entrance into agreement with organic farmer to serve as his/her certifying agent, organic certifier must inspect farmer's operation within 60 days.\end{exi}

\sp

This statement is rather complex and includes both component-level nesting, along various further constraint specifications.

\sp

Retracing this coding is best achieved by reformulating this statement as follows:

\sp

\begin{exi}\A{}(Organic certifier) \D{}(must) \I{}(inspect) \Bdir{}(farmer's operation) \Cex{}\customExt{ctx=tfr}(within 60 days)  \Cac{}\customExt{ctx=prc}\lb under the condition that the \A{}(organic farmer) \I{}(enters) \Bdir{},\prop{}(an) \Bdir{}(agreement) \Bind{}(with the organic certifier) \Cex{}\customExt{ctx=pur}(to serve as his/her certifying agent)\rb.\end{exi}

\sp

Here the execution constraint for the initial inspection (\Cex\customExt{ctx=tfr}) is activated by the preceding procedural activation condition (\Cac\customExt{ctx=prc}), expressed as a nested institutional statement, that the farmer enters an agreement, whose purpose is defined as an execution constraint (\Cex\customExt{ctx=pur}).

\sp

&

\vspace{0.05cm}

\begin{exi}\Cac\customExt{ctx=prc}\lb \I{}(Upon entrance) \Bdir{}(into agreement) with \A{}(organic farmer) \Cex\customExt{ctx=pur}(to serve as his/her certifying agent)\rb, \A{}(organic certifier) \D{}(must) \I{}(inspect) \Bdir{}(farmer's operation) \Cex\customExt{ctx=tfr}(within 60 days).\end{exi}

\\

\begin{emp}Context (ctd.)\end{emp}

&

In addition to the contextual characterisation, conditions or constraints can further be expressed as primitive statements (e.g., at 8am), or complex expressions embedding complete institutional statements (e.g., if an organic farmer violates conditions for certification, \dots).

\sp

\udl{Note:} While the use of the annotation prefix label \begin{emp}\ctx\end{emp} is recommended for context annotations (as specified in \cref{subsec:ContextTaxonomy} and further discussed under  \hyperlink{tablink:CrossComponentSemanticAnnotationsLogicoRegulative}{\begin{emp}IG Logico Instructions\end{emp}, Item \EntryCrossComponentSemanticAnnotations} in \cref{tab:CodingIgLogicoRegulative}), where no other labels are applied to activation conditions and execution constraints, the prefix can be omitted as part of the shorthand coding.

\sp

Where logical relationships between conditions or constraints exist, these are made explicit by annotating those as \AND, \OR, and \XOR~ relationships. Where negation exists, this is to be likewise identified (\NOT).

\sp

\udl{Rules:}

\begin{itemize}
    \item Identify involved actions (explicit and tacit) by reformulating statements in active terms, and expand into (potentially multiple) statements.
\end{itemize}
    
& 

\udl{Example 2:}\newline
\begin{exi}When rebuttal is unsuccessful or correction of the noncompliance by certified organic farmer is not completed within the prescribed time period, the Program Manager shall send a written notification of proposed suspension or revocation of certification to certified organic farmer.\end{exi}

\sp

This statement, similar to the previous example, includes the encoding of a nested institutional statement. In contrast, it highlights the combination of multiple conditions statements, where one is of implicit nature and the second one explicit (apart from minor rephrasing).

\sp

Coding this example, we first identify the top-level institutional statement:

\sp

Top-level institutional statement = \begin{exi}the \A{}(Program Manager) \D{}(shall) \I{}(send) a \Bdir{},\prop{}(written) \Bdir{}(notification) of \Bdir{}\customExtRaw{,p1,p1,}\propExt{}; \Bdir{}\customExtRaw{,p1,p2,}\propExt{}( proposed) \Bdir{}\customExtRaw{,p1,p1}(suspension) or  \Bdir{}\customExtRaw{,p1,p2}(revocation) of \Bdir{}\customExtRaw{,p1}(certification) to \Bind{}(certified organic farmer).\end{exi}

\sp

& 

\vspace{0.05cm}

\begin{exi}\lb\Cac\customExt{ctx=evt}\lb When \lsi \A{,}\propExt{1}(certified)  \A{,}\propExt{2}(organic) \A{}(farmer)\rsi~~\I{}(\lsi rebuts\rsi)~   \Cex\customExt{ctx=evt} (unsuccessfully)\rb

\OR 

\Cac\customExt{ctx=tfr}\lb \Bdir{},\prop{}(correction) of the \Bdir{}(noncompliance) by \A{}(certified organic farmer) is not completed \I{}(\lsi does not complete\rsi)~\Cex \customExt{ctx=tfr}(within the prescribed time period)\rb\rb,

the \A{}(Program Manager) \D{}(shall) \I{}(send) a  \Bdir{},\prop{}(written) \Bdir{}(notification) of  \Bdir{}\customExtRaw{,p1,p1,}\propExt{}; \Bdir{}\customExtRaw{,p1,p2,}\propExt{}(proposed) \Bdir{}\customExtRaw{,p1,p1}(suspension) or \Bdir{}\customExtRaw{,p1,p2}(revocation) of \Bdir{}\customExtRaw{,p1}(certification) to \Bind{}(certified organic farmer).\end{exi}

\\

\begin{emp}Context (ctd.)\end{emp}

&

\begin{itemize}
\item Hint: Where statements contain an aim that is linked to an object as noun, it is indicative of a missing or implied actor specification. Such case is signaled by passive tense (e.g., notification is received). The actor specification is either implied from context or potentially signaled by prepositional clauses such as ``\begin{exi}by actor\end{exi}''; Example: ``\begin{exi}Notification by certifier is received by farmer\end{exi}''. In such cases, the statement requires reformulation in active terms along with the injection of an inferred explicit action an explicit action, e.g., ``\begin{exi}certifier sends notification to farmer\end{exi}''. In some cases, this may require the expansion of a single statement into separate statements carrying separate actions (see Example 3).
    \item Identify logical relationships (\AND, \OR, \XOR) and dependencies (sequence) amongst actions.
    \item Reconstruct complete statement using component-level nesting of statements where relevant.
\end{itemize}

& 

In this statement we observe principles of the Attributes/Object-property hierarchy (see \cref{subsec:AttributeObjectHierarchy}), which are encoded based on the instructions provided for objects.

\sp

Context clause = \begin{exi}When rebuttal is unsuccessful or correction of the noncompliance by certified organic farmer is not completed within the prescribed time period\end{exi}

\sp

The context clause (which contains two activation conditions) can be decomposed into two statements that are logically combined (\OR). The passive aim on an object (rebuttal is unsuccessful) signals an implied action, and requires reformulation. 

\sp

First condition statement = \begin{exi}When \lsi \A{}(organic farmer)\rsi~\I{}(\lsi rebuts\rsi)

\Cex\customExt{ctx=met}(unsuccessfully)\end{exi}

\sp

The second condition (``\begin{exi}correction of the noncompliance by certified organic farmer is not completed within the prescribed time period\end{exi}'') can be reformulated in active terms as ``\begin{exi}if certified organic certifier does not complete correction of non-compliance within the prescribed time period\end{exi}''.

& \\

\begin{emp}Context (ctd.)\end{emp}

&

&

Doing so, the structure of the nested statement becomes overt:

\sp

``\begin{exi}if \A{,}\propExt{1}(certified) \A{,}\propExt{2}(organic)  \A{}(certifier) does not \NOT{}~ \I{}(complete)  \Bdir,\propExt{}(correction) of \Bdir{}(non-compliance)  \Cex\customExt{ctx=tfr}(within the prescribed time period)\end{exi}''

\sp

Both conditions are further logically related by an inclusive disjunction (AND/OR), which is annotated explicitly using \AND, \OR, or \XOR, respectively.

The composition of all three statements is shown in the example.

\sp

\udl{Example 3:}\newline
\begin{exi}When an inspection of an accredited certifying agent by the Program Manager reveals any noncompliance with the Act or regulations in this part, a written notification of noncompliance shall be sent to the certifying agent.\end{exi}

\sp

Note: While highlighted here for activation conditions, such logical combination equally applies to statements containing multiple execution constraints.

& 

\vspace{8.1cm}

\begin{exi}
\Cac{}\lb When \lsi \A{}(Program Manager)\rsi~\I{}(reveals) any \Bdir{}(non-compliance) \lsi by the \Bind{},\propExt{1}(accrediting)  \Bind{},\propExt{2}(certifying) \Bind{}(agent)\rsi~ \Cex(with the Act or regulations in this part) \Cac{}\lb\lsi under the condition that\rsi~\A{}(Program Manager) \I{}(\lsi performs\rsi~) \Bdir{}(inspection) of an \Bind{},\propExt{1}(accredited)  \Bind{},\propExt{2}(certifying) \Bind{}(agent)\rb\rb, 

\A{}(\lsi Program Manager\rsi)~\D{}(shall) \I{}(\lsi send\rsi)~a \Bdir{},\propExt{1}(written) \Bdir{}(notification)  \Bdir{},\propExt{2}(of noncompliance) to the  \Bind{},\propExt{1}(certifying) \Bind{}(agent).\end{exi}

\\

\begin{emp}Context (ctd.)\end{emp}

& & 

As highlighted before, for the purpose of coding the reformulation of statements in active terms may be useful to facilitate coding. In the first statement ``\begin{exi}When an inspection of an accredited certifying agent by the Program Manager reveals any noncompliance with the Act or regulations in this part\end{exi}'', the acting party is the Program Manager (actor), but the aim reveals relates to the object inspection. This is indicative of a tacit action (captured as conceptual reification in the object) on the part of the program manager (captured in the prepositional clause). In this case, we can decompose the compound conditional statement into two statements:

\sp

\begin{exi}Program Manager \lsi performs\rsi~inspection\end{exi} (where the performance is tacit) 

\AND

\begin{exi}\lsi Program Manager\rsi~reveals any non-compliance \dots\end{exi} (in which case the attribute is tacit).

\sp

& \\

\begin{emp}Context (ctd.)\end{emp}

& &

In addition, the second statement depends on the activation of the first action. The conditional statement can thus be decomposed into two institutional statements, as follows:

\sp

\begin{exi}``When \A{}(\lsi Program Manager\rsi)~\I{}(reveals) any \Bdir{}(non-compliance) \lsi by the \Bind{},\propExt{1}(accrediting)  \Bind{},\propExt{2}(certifying) \Bind{}(agent)\rsi~ \Cex(with the Act or regulations in this part) \lb\lsi under the condition that\rsi~\A{}(Program Manager) \I{}(\lsi performs\rsi) \Bdir{}(inspection) of an \Bind{},\propExt{1}(accredited)  \Bind{},\propExt{2}(certifying) \Bind{}(agent), \dots''\end{exi}

\sp

The final part of the statement (the main statement), likewise reformulated in active terms, implies that \begin{exi}``[Program Manager] shall [send] a \Bdir{},\propExt{1}(written) \Bdir{}(notification) of  \Bdir{},\propExt{2}(noncompliance) to the \Bind{},\propExt{1}(certifying) \Bind{}(agent)''\end{exi}.

& \\


& &

Structurally, in this statement we observe two levels of component-level nesting in the form \begin{exi}ABDIC\lb ABDIC\lb ABDIC\rb\rb\end{exi}, where the main statement’s (shall send notification; first \begin{exi}ABDIC\end{exi}) activation relies on revealing potential non-compliance (second \begin{exi}ABDIC\end{exi}), which in itself relies on the inspection in the first place (last \begin{exi}ABDIC\end{exi}).

& \\

\midrule

\emph{Or else} & \multicolumn{3}{\tablespanNarrow}{The coding of the Or else is identical to IG Core. The internal structure of nested statements is encoded according to IG Extended instructions (\cref{tab:CodingIgExtendedRegulative}).} \\
\midrule

\newpage

\midrule
\multicolumn{4}{l}{\textbf{General IG Extended Instructions for Regulative Statements}} \\
\midrule

\hypertarget{tablink:DecompositionOfComponentLevelCombinationsExtendedRegulative}{\EntryDecompositionOfComponentLevelCombinations}

&

In the presence of multiple Attributes, actions (aims) and objects in a given statement, such statements are to be decomposed into individual statements that are combined with the corresponding logical operator (following the principles of horizontal nesting). At face value this mirrors the approach taken in the context of or Else statements. However, while in the context of Or else components, combinations designate logical relationships amongst action alternatives, in this context the purpose is to disambiguate the relationship amongst particular actors, actions and objects and to resolve a potential incongruence between the linguistic and logical use of conjunctives. 

When a disaggregation or aggregation fundamentally alters the meaning of the statement (e.g., if a payoff associated with the institutional statement cannot be unambiguously associated with an individual entity), or if expressions are intentionally coupled (e.g., chips and fish) or proper names (e.g., Smith and Sons), the statement is not to be decomposed.

&

\udl{Example Statement:} 

\begin{exi}Certified operations or handlers must comply with organic farming regulations.\end{exi}

\sp

This can be decomposed into

\sp

\begin{exi}Certified operations must comply with organic farming regulations

\AND

Certified handlers must comply with organic farming regulations. \end{exi}

\sp

Note that the interpretation of the logical operator is contextual. In this example, it carries the understanding that the specified obligation applies to both certified operations and certifier handlers.

Naturally, this approach can lead to the decomposition into a large number of additional statements, e.g., attribute and action combinations - as exemplified below. For practical reasons, the explicit coding can be substituted by an additional annotation that reflects the need for decomposition. 

& 

\vspace{0.05cm}

\begin{exi}
\lb \lb \A,\prop{}(Certified) \A{}(operations) \D{}(must) \I{}(comply) with  \Bdir{},\prop{}(organic farming) \Bdir{}(regulations)\rb

\AND

\lb \A{},\prop{}(Certified) \A{}(handlers) \D{}(must) \I{}(comply) with  \Bdir{},\prop{}(organic farming) \Bdir{}(regulations).\rb \rb\end{exi}

\sp

Recall that the minimal institutional statement presumes the existence of context specifications, which -- in absence of specific encoding -- resolves to ``under all circumstances'' (for activation conditions), ``without any constraints'' (for execution constraints).

\\

\begin{emp}Decomposition of component-level combinations (ctd.)\end{emp}

& & 

\udl{Example Statement:} 

\begin{exi}Certified operations or handlers must accept and comply with organic farming regulations.\end{exi}

\sp

Decomposed: 

\sp

\begin{exi}Certified operations must accept organic farming regulations

\AND

Certified handlers must accept organic farming regulations

\AND

Certified operations must comply with organic farming regulations

\AND

Certified handlers must comply with organic farming regulations.\end{exi}

\sp

\underline{Practical considerations:} \newline
If labelling is performed manually, a practical consideration for such decomposition is to keep track of the relationships of such statements, e.g., by introducing sub-identifiers. For example, assuming the coded statement is Statement 10, the decomposed statements could be annotated as 10.1, 10.2, etc.

& 

\vspace{0.05cm}

\begin{exi}\lb\lb\A{},\prop{}(Certified) \A{}(operations) \D{}(must) \I{}(accept) \Bdir{},\prop{}(organic farming) \Bdir{}(regulations)\rb

\AND

\lb\A{},\prop{}(Certified) \A{}(handlers) \D{}(must) \I{}(accept) \Bdir{},\prop{}(organic farming) \Bdir{}(regulations)\rb

\AND

\lb\A{},\prop{}(Certified) \A{}(operations) \D{}(must) \I{}(comply) with  \Bdir{},\prop{}(organic farming) \Bdir{}(regulations)\rb

\AND

\lb\A{},\prop{}(Certified) \A{}(handlers) \D{}(must) \I{}(comply) with  \Bdir{},\prop{}(organic farming) \Bdir{}(regulations).\rb\rb\end{exi}

\\


\begin{emp}Decomposition of component-level combinations (ctd.)\end{emp}

&

& 

Alternatively, annotation can be performed in shorthand form, or rely on tool-specific support offered by the applied text annotation tool that allows the indication of decomposition during the encoding.

In shorthand form, the decomposed statements can be grouped by parentheses to signal component-level combinations as shown below.

\sp

\begin{exa}\udl{Example:} \A{}(\lp Operators \AND~Certifiers\rp)~\D{}(must) \I{}(comply) with \Bdir{}(\lp regulations  \AND~best practices\rp).\end{exa}

\sp

In expanded form (shown on the right), parentheses are then used to signal the association of the decomposed statements.

& 

\udl{Grouping of statements using parentheses (in bold font):}

\sp

\begin{exi}
\lb\lb \A{},\prop{}(Certified) \A{}(operations) \D{}(must) \I{}(accept)  \Bdir{},\prop{}(organic farming) \Bdir{}(regulations)\rb

\AND

\lb\A{},\prop{}(Certified) \A{}(handlers) \D{}(must) \I{}(accept) \Bdir{},\prop{}(organic farming) \Bdir{}(regulations)\rb

\AND

\lb\A{},\prop{}(Certified) \A{}(operations) \D{}(must) \I{}(comply) with  \Bdir{},\prop{}(organic farming) \Bdir{}(regulations)\rb

\AND

\lb\A{},\prop{}(Certified) \A{}(handlers) \D{}(must) \I{}(comply) with  \Bdir,\prop{}(organic farming) \Bdir{}(regulations)\rb\rb\end{exi}

\\

\midrule

\begin{emp}Logical relationships among statement components\end{emp}

&

Where statements make tacit reference to multiple actions, conditions or constraints, these are likewise resolved using the instructions provided as part of the \hyperlink{tablink:LogicalRelationshipsAmongStatementComponents}{\begin{emp}IG Logico Instructions\end{emp} in \cref{tab:CodingIgLogicoRegulative}, Item \EntryLogicalRelationshipsAmongComponents}.

For IG Extended this provision is recommended, but optional.

& & \\

\midrule

\begin{emp}Regulative-constitutive Hybrids\end{emp} 

&

\multicolumn{3}{p{21cm}}{The introduction of constitutive statements as part of IG 2.0 (see \cref{subsec:ConstitutiveStatementCoding}) provides the basis for encoding statements that consist of structural elements both of regulative and constitutive statements. Details are discussed in \cref{subsec:ConstitutiveRegulativeHybrids}.} \\

\toprule

\caption{Coding Guidance on Syntactic Elements for IG Extended as Level of Expressiveness (Regulative Statements)}
\label{tab:CodingIgExtendedRegulative}
\end{longtable}

\vspace{3cm}

\subsubsection{IG Logico Coding of Regulative Statements}
\label{subsubsec:IGLogicoRegulative}

\begin{longtable}{\tabledims}
\toprule
\multicolumn{4}{\tablespan}{\textbf{Level of Expressiveness: IG Logico}

\sp

IG Logico is designed to support semantic analysis of institutional statements wholly relying on computational tools. Encoding at this level is designed to be moderately human readable, computationally tractable and comprehensive in the detail with which syntactic properties of institutional statements are captured. 

In contrast to IG Core and Extended that focus on the encoding of specific grammar components, IG Logico emphasises refinements across individual components and further establishes explicit references to related statements to establish computational tractability, as well as the ability to perform logical transformations on institutional statements. 

} \\

\newpage

\toprule
{\textbf{Syntactic\newline Component}} & 
{\textbf{Treatment of Syntactic Components by Level of Encoding}} &
{\textbf{Relevant Examples}} &
{\textbf{Complete Syntactic Classification of Examples}} \\

\toprule

\begin{emp}Relation-centric Semantic Annotations\end{emp} &

To establish relationships amongst statements and policies more generally, the initial refinement refers to the identification of statement references, e.g., between individual statements, collections thereof, or policies more generally.

To this end, an additional annotation identifies all instances of references to other statements. Note that cross-statement references are not specific to any statement components, but apply across complex component types, including attributes, objects and context. 

\sp

\udl{Rules:}

\begin{itemize}
    \item Identify references in coded statements
    \item Add annotation of structure (\begin{exi}\polref=value\end{exi}), where \begin{exi}\polref\end{exi} signals the reference nature, and value contains an identifier of referenced statement, collection, or policy.
\end{itemize}

&

Exploring this approach, we borrow the complex example statement previously coded in the context of IG Extended:

\sp

\udl{Example Statement:}
\begin{exi} When an inspection of an accredited certifying agent by the Program Manager reveals any noncompliance with the Act or regulations in this part, a written notification of noncompliance shall be sent to the certifying agent.\end{exi}

\sp

\udl{Coded form:}

\begin{exi}\Cac\customExt{ctx=prc}\lb \lp When \lsi \A{}(program manager) \I{}(performs)\rsi~an \Bdir{}(inspection) of an \Bind{},\propExt{1}(accredited)  \Bind{},\propExt{2}(certifying) \Bind{}(agent)\rp~\lsi \AND\rsi~\lp \A{}(Program Manager) \I{}(reveals) any \Bdir{}(noncompliance)  \Cex(with the Act or regulations in this part)\rp \rb, \lsi \A{}(Program Manager)\rsi~a \Bdir{},\propExt{1}(written) \Bdir{}(notification) \Bdir{},\propExt{2}(of noncompliance) \D{}(shall) be sent \I{}(\lsi send\rsi) \Bind{}(to the certifying agent).\end{exi}

&

\vspace{5.55cm}

\begin{exi}\Cac\customExt{ctx=prc}\lb \lb When \lsi \A{}(Program Manager) \I{}(performs)\rsi~an \Bdir{}(inspection) of an \Bind{},\propExt{1}(accredited)  \Bind{},\propExt{2}(certifying) \Bind{}(agent)\rb~\lsi \AND\rsi~\lb  \A{}(Program Manager) \I{}(reveals) any \Bdir{}(non-compliance) \Cex{}\customLog{\polref=\lsi``policy'',``section''\rsi} (with the \lp Act~\lsi \OR{}\rsi~ regulations in this part\rp) \rb\rb, 

\A{}(\lsi Program Manager\rsi) a \Bdir{},\propExt{1}(written) \Bdir{}(notification) \Bdir{},\propExt{2}(of non-compliance) \D{}(shall) be sent \I{}(\lsi send\rsi) \Bind{}(to the certifying agent).\end{exi}

\\

\begin{emp}Relation-centric Semantic Annotations (ctd.)\end{emp}

&

&

This statement makes reference to ``\begin{exi}the Act\end{exi}'' and ``\begin{exi}regulations in this part\end{exi}'', the first of which makes reference to the coded policy in its entirety, whereas the second one focuses on a specific section of the policy.

Both are coded by providing an additional annotation (\begin{exi}ref=value\end{exi}), along with the scope reference (``\begin{exi}value\end{exi}''), i.e., an identifier of relevance in the context of the encoded policy. The identifiers need to be unambiguous within the given document, including statement IDs, section headers, or documents as a whole, etc. The corresponding convention should be decided as part of project-specific coding guidelines.

If occurring in conjunction with existing component classification, the reference specification is appended (e.g., \begin{exi}\Bdir{}\customLog{\polref=``policy''}(Act)\end{exi}).

As stated before, while, in this specific example, cross-statement reference apply to constraints components, such references can likewise occur in other components.

& 

\begin{exi}
\lb\lb \Cac{}\lb When \lsi \A{}(Program Manager)\rsi~\I{}(reveals) \Bdir{},\prop{}(any) \Bdir{}(non-compliance) \lsi by the  \Bind{},\propExt{1}(accrediting) \Bind{},\propExt{2}(certifying)  \Bind{}(agent)\rsi~\Cex{}\customLog{\polref=``policy''} (with the Act)\rb

\sp

\Cac{}\lb \lsi under the condition that\rsi

\sp

\lb \A{}(Program Manager) \lsi \I{}(performs)\rsi~\Bdir{}(inspection) of an \Bind{},\propExt{1}(accredited) \Bind{},\propExt{2}(certifying) \Bind{}(agent)\rb 

\sp

\OR

\sp

\lb \A{}(Program Manager) \lsi \I{}(performs)\rsi~\Bdir{}(inspection) of an \Bind{},\propExt{1}(accredited) \Bind{},\propExt{2}(certifying)  \Bind{}(agent)\rb 

\sp

\OR

\sp

\lb \A{}(Program Manager) \lsi \I{}(performs)\rsi~\Bdir{}(inspection) of an \Bind{},\propExt{1}(accredited) \Bind{},\propExt{2}(certifying) \Bind{}(agent)\rb 

\rb \rb, 

\A{}(\lsi Program Manager\rsi)~\D{}(shall) \I{}(\lsi send\rsi)~a \Bdir{},\propExt{1}(written) \Bdir{}(notification) \Bdir{},\propExt{2}(of non-compliance) to the \Bind{},\propExt{1}(certifying) \Bind{}(agent).\rb

\end{exi}

\sp

\OR~($\leftarrow$ decomposition of ``\begin{emp}Act\end{emp}'' and ``\begin{emp}regulations in this part\end{emp}'')

\dots

\\

\hypertarget{tablink:LogicalRelationshipsAmongStatementComponents}{\EntryLogicalRelationshipsAmongComponents}

&

The objective of decomposition of lists, or other forms of implied conjunctions, such as multiple conditions/constraints is to make logical relationships explicit. In such cases enumerations are decomposed into individual atomic institutional statements and combined using the corresponding logical operator.

Note that this is similar to the expansion of multi-entity components in IG Extended with specific emphasis on attributes and objects. In this case, the review operates across all component types and explicitly focuses on implied logical relationships.

\sp

\udl{Rules:}

\begin{itemize}
    \item Identify action alternatives embedded in lists or enumeration, or in conditions/constraints
    \item Identify associated atomic statement
    \item Establish logical operator and precedence where needed
    \item Expand statement via horizontal nesting to capture individual action alternatives
\end{itemize}

&

We use a modified example previously explored under IG Extended, Context component.

\sp

\udl{Example Statement:}

\begin{exi}When an inspection, review, or investigation of an accredited certifying agent by the Program Manager reveals any noncompliance with the Act or regulations in this part, a written notification of noncompliance shall be sent to the certifying agent.\end{exi}

\sp

Following the decomposition patterns for Context specification established previously (IG Extended, Context), we observe that an inspection of a program manager has to be performed, and may reveal non-compliance, and arrived at the following coding:

\sp

\begin{exi}\Cac{}\lb When \A{}(\lsi Program Manager\rsi)~\I{}(reveals) any \Bdir{}(non-compliance) \lsi by the accrediting)  \Bind{},\propExt{2}(certifying) \Bind{}(agent)\rsi~ \Cex{}(with the \lp Act \lsi \OR{} \rsi regulations in this part\rp ) \Cac{}\lb \lsi under the condition that \rsi~\A{}(Program Manager) \I{}(\lsi performs\rsi)~\Bdir{}(inspection) of an \Bind{},\propExt{1}(accredited) \Bind{},\propExt{2}(certifying) \Bind{}(agent)\rb \rb, \dots \end{exi}

&

\begin{exi}
\dots

\sp

\lb \Cac{}\lb When \A{}(\lsi Program Manager\rsi) \I{}(reveals) \Bdir{},\prop{}(any) \Bdir{}(non-compliance) \lsi by the \Bind{},\propExt{1}(accrediting) \Bind{},\propExt{2}(certifying)  \Bind{}(agent)\rsi~\Cex{} \customLog{\polref=``section''} (with regulations in this part)

\sp

\Cac{}\lb \lsi under the condition that\rsi

\sp

\lb \A{}(Program Manager) \I{}(\lsi performs\rsi) \Bdir{}(inspection) of an \Bind{},\propExt{1}(accredited) \Bind{},\propExt{2}(certifying) \Bind{}(agent)\rb 

\sp

\OR

\sp

\lb \A{}(Program Manager) \I{}(\lsi performs\rsi) \Bdir{}(inspection) of an \Bind{},\propExt{1}(accredited) \Bind{},\prop{2}(certifying) \Bind{}(agent)\rb 

\sp

\OR

\sp

\lb \A{}(Program Manager) \I{}(\lsi performs\rsi) \Bdir{}(inspection) of an \Bind{},\propExt{1}(accredited) \Bind{},\propExt{2}(certifying) \Bind{}(agent)\rb 

\rb \rb, 

\A{}(\lsi Program Manager\rsi)~\D{}(shall) \I{}(\lsi send\rsi) a \Bdir{},\propExt{1}(written) \Bdir{}(notification) \Bdir{},\propExt{2}(of non-compliance) to the \Bind{},\propExt{1}(certifying) \Bind{}(agent).\rb 

\end{exi}

\\

\begin{emp}\EntryLogicalRelationshipsAmongComponents~(ctd.)\end{emp}

&

&

\begin{exi}
\dots~\A{}(\lsi Program Manager\rsi)~\D{}(shall) \I{}(\lsi send\rsi) a  \Bdir{},\propExt{1}(written) \Bdir{}(notification)  \Bdir{},\propExt{2}(of non-compliance) to the \Bind{},\propExt{1}(certifying) \Bind{}(agent).\end{exi}
 
\sp

In this example, the program manager's notification is contingent on an inspection, a review, or investigation, i.e., a variation of instruments for assessment.

Logically, the initial task (inspection) that is prerequisite for further action. The original statement is

\sp

\begin{exi}
\Cac{}\lb \lsi under the condition that\rsi~\A{}(Program Manager) \I{}(\lsi performs\rsi) \Bdir{}(inspection) \Bind{},\propExt{1}(of an accredited) \Bind{},\propExt{2}(certifying) \Bind{}(agent)\rb\end{exi}

\sp

Given that we now have three alternative actions, the statement requires expansion into three separate tasks.

To achieve this, the logical relationship between the tasks needs to be established. Subject to context, the coder can interpret the relationship as either an inclusive disjunction (\OR) or exclusive disjunction (\XOR). 

& \\

\begin{emp}\EntryLogicalRelationshipsAmongComponents~(ctd.)\end{emp}

&

&

Since realistically (based on interpretation of application context) a combination of any of such tasks could equally lead to the detection of non-compliance, suggesting the combination via \OR. This example showcases the importance of coding context and interpretation, which makes an explicit specification necessary for analytical treatment of action alternatives.

In consequence, the statement is expanded into \OR-combined statements as follows (Note: while not necessary in this case, the logical combination of the statements is signaled using surrounding parentheses): 

\sp

\begin{exi}
\dots~\lsi under the condition that\rsi

\sp

\lb 
\lb \A{}(Program Manager) \I{}(\lsi performs\rsi) \Bdir{}(inspection)  \Bind{},\propExt{1}(of an accredited) \Bind{},\propExt{2}(certifying)  \Bind{}(agent)\rb

\sp

\OR

\sp

\lb \A{}(Program Manager) \I{}(\lsi performs\rsi) \Bdir{}(review) of an  \Bind{},\propExt{1}(accredited) \Bind{},\propExt{2}(certifying) \Bind{}(agent)\rb

\sp

\OR

\dots

\end{exi}

& \\

\begin{emp}\EntryLogicalRelationshipsAmongComponents (ctd.)\end{emp}

&

&

\begin{exi}

\dots

\sp

\lb \A{}(Program Manager) \I{}(\lsi performs\rsi) \Bdir{}(investigation) of an \Bind{},\propExt{1}(accredited) \Bind{},\propExt{2}(certifying)  \Bind{}(agent)\rb \rb \end{exi}

\sp

This coding is embedded in the complete statement coding as shown on the right (with formatting adjustments so as to make the discussed coding easily accessible).

Note: While not applicable in this case, where necessary, precedence of specific combinations has to be signaled using parentheses (e.g., if inspection OR review is permitted, or as an exclusive alternative, the investigation, which would be (simplified) represented as \begin{exi}\lp \lp inspection \OR~review\rp~\XOR~investigation\rp \end{exi}.

Another aspect that requires explicit coding in this example is the reference to the scope of violation, i.e., non-compliance with the Act or regulations in this part, the logical relationship of which (here: ``or''), subject to coder interpretation, has to be coded explicitly. 

& \\

\begin{emp}\EntryLogicalRelationshipsAmongComponents~(ctd.)\end{emp}

&

&

In this example the relationship is characterized by an inclusive disjunction (\OR), since regulations in this part are part of the Act. As with all other cases, this annotation is made explicit, and consequently, requires decomposition of the statement by duplication into corresponding statement variants:

\sp

\begin{exi}
\lb When \A{}(\lsi Program Manager\rsi) \I{}(reveals) \Bdir{},\prop{}(any) \Bdir{}(non-compliance) \lsi by the \Bind{},\propExt{1}(accrediting)  \Bind{},\propExt{2}(certifying) \Bind{}(agent)\rsi~\Cex{}\customLog{\polref=``policy''} (with the Act)\rb

\sp

\dots (remainder of statements) \dots

\sp

\OR

\sp

\lb When \A{}(\lsi Program Manager\rsi)~\I{}(reveals) \Bdir,\prop{}(any) \Bdir{}(non-compliance) \lsi by the \Bind{},\propExt{1}(accrediting)  \Bind{},\propExt{2}(certifying) \Bind{}(agent)\rsi~ \Cex{}\customLog{\polref=``section''}(with regulations in this part)\rb

\sp

\dots (remainder of statements) \dots

\end{exi}

& \\

\begin{emp}\EntryLogicalRelationshipsAmongComponents~(ctd.)\end{emp}

&

&

Such decomposition affords a systematic assessment of the individual variants in the context of a specific situation, but likewise offers automation potential. For example, given the assumption that regulations in this part are a subset of the Act, we could ignore the statement variant that assesses the compliance with regulations in this part. However, for the sake of comprehensive illustration of the embedded institutional complexity (and subject to alternative interpretations), we decompose this example comprehensively.

& \\

\midrule

\hypertarget{tablink:CrossComponentSemanticAnnotationsLogicoRegulative}{\EntryCrossComponentSemanticAnnotations}

&

In addition to explicit encoding of properties as specified in the underlying statement (e.g., explicit identification of ``certified'' as a property of an operation), semantic qualities can be enriched by providing additional annotations that capture a differentiation of components with respect to different ontological categories captured in different taxonomies (see \cref{sec:Taxonomies}). In contrast to the property annotations used in IG Core and Extended, the annotations introduced here apply across all components, and explicitly emphasize extensibility both with respect to additional categories within the given taxonomies, as well as specification of further taxonomies, e.g., based on domain-specific or analytical necessities.

Furthermore, multiple annotations of different categories can be applied to a given component at the same time, e.g., animacy descriptors, such as animate, can be combined with role or metatype descriptors. 

For the exemplified taxonomies here, categories include

\begin{itemize}
\itemsep0em
\item Animacy: animate, inanimate
\item Metatype: abstract, concrete
\item Role: Source, Recipient, Possessor, Experiencer, Beneficiary
\end{itemize}

An extended listing of the associated taxonomies along with their description is provided in \cref{sec:Taxonomies}.

&

Complementing the characterisation of context by circumstance, entities can further be annotated by the role they play in a particular setting, as well as further properties, such as physical types. These categorizations can be extended beyond the specified types, apply across all component types, and furthermore allow the introduction of additional taxonomies, beyond the ones highlighted in \cref{sec:Taxonomies}. 

To annotate components with additional categories, the component coding is extended by key-value pairs, with the key specifying the taxonomy, and the value the corresponding categorization(s). In the coding highlighted here, the category specifications are separated from component specification by semicolon. Where multiple categories for a given taxonomy apply, these are separated by comma.

The example on the right highlights this approach with respect to various taxonomies specified in \cref{sec:Taxonomies}.

&

\begin{exi}\Cac{}\lb When \lsi \A{}\customLog{\anim=animate; \role=experiencer; \metatype= concrete}(Program Manager)\rsi~\I{}(reveals) \Bdir{},\prop{}(any) \Bdir{}\customLog{\anim=inanimate; \metatype=abstract}(non-compliance) \lsi by the \Bind{},\propExt{1}(accrediting)  \Bind{},\propExt{2}(certifying) \Bind{}\customLog{\anim=animate; \role=originator; \metatype=concrete}(agent)\rsi  

\Cex{}\customLog{\metatype=abstract}(with the \customLog{\anim=inanimate; \polref=``policy''; \metatype=concrete}~Act \lsi \OR{}\rsi~\customLog{\polref=``section''; \metatype=concrete}regulations in this part), \lsi \A{}\customLog{\anim=animate; \role=originator; \metatype=concrete}(Program Manager)\rsi~\D{}(shall) \I{}(\lsi send\rsi) a \Bdir{},\propExt{1}(written) \Bdir{}\customLog{\anim=inanimate; \metatype=concrete} (notification) \Bdir{},\propExt{2}(of noncompliance) to the  \Bind{},\propExt{1}(certifying) \Bind{}\customLog{\anim=animate; \role=recipient; \metatype=concrete}(agent).\end{exi}

\\

\midrule

\begin{emp}Regulative Function Annotations\end{emp}

&

IG Logico further prescribes annotations of statements with institutional functions in regulative or constitutive form. While other features described above in practice focus on the refined coding of attributes, objects and conditions, institutional actions reflect the regulative function of a statement's action.

\sp

Being fined, for example, reflects the regulative function of ``sanctioning'', adhering to regulation reflects ``compliance''.

\sp

Institutional functions, and here specifically regulative functions, as identified in this specification include the following set, some of which operate complementary and are listed as comma-separated function pairs: 

\begin{itemize}
\item Comply, Violate
\item Reward, Sanction
\item Monitor
\item Detect compliance, Detect non-compliance
\item Delegate
\end{itemize}

The Regulative Function Taxonomy is described in \cref{subsec:regfunc}.

&

Taking the previously used statement as an example (including inferred components, but omitting any annotations), we can identify regulative functions associated with actions.

\sp

\begin{exi}
\lb When \lsi Program Manager\rsi~reveals any non-compliance \lsi by the accrediting certifying agent\rsi~with the Act or regulations in this part\rb , \lsi Program Manager\rsi~shall \lsi send\rsi~a written notification of noncompliance to the certifying agent.\end{exi}

\sp

Actions of institutional relevance include:

\begin{itemize}
    \item Reveal (non-compliance)
    \item Send (notification)
\end{itemize}

In this context revealing non-compliance abstractly corresponds to the ``\begin{exi}detection of a violation\end{exi}'', whereas sending a notification reflects a form of ``\begin{exi}sanctioning\end{exi}''. In consequence, the statement can be annotated with these regulative functions, so as to enable inferences from a purely institutional perspective without concern for the specific operationalization of detecting compliance or sanctioning in a specific scenario.

&

\vspace{2.55cm}

\begin{exi}\Cac{}\lb When \A{}\customLog{\anim=animate; \role=experiencer}(\lsi Program Manager\rsi)~ \I{}\customLog{\regfunc=detect violation}(reveals) \Bdir{},\prop{}(any) \Bdir{}\customLog{\anim=inanimate}(non-compliance) \lsi by the  \Bind{},\propExt{1}(accrediting) \Bind{},\propExt{2}(certifying) \Bind{}\customLog{\anim=animate; \role=originator}(agent)\rsi~\Cex{}(with the (\customLog{\anim=inanimate; \polref=``policy''}Act \OR{}~ (\customLog{\polref=``section''}regulations in this part)\rb, \A{}; \customLog{\anim=animate; \role=originator}(\lsi Program Manager)\rsi) \D{}(shall) \I{}\customLog{\regfunc=sanction}(\lsi send\rsi) a \Bdir{},\propExt{1}(written) \Bdir{}; \customLog{\anim=inanimate}(notification) \Bdir{},\propExt{2}(of noncompliance) to the \Bind{},\propExt{1}(certifying) \Bind{}\customLog{\anim=animate; \role=recipient}(agent).\end{exi}

\\

\begin{emp}Regulative Function Annotations (ctd.)\end{emp}

& & 

The annotation follows the syntactic specification applies for other forms of annotations, i.e., appending a key-value pair to the component coding, where the key is ``\begin{exi}regfunc\end{exi}'' and the value carries the corresponding regulative function specific to the annotated action.

\sp

Providing an additional simplified example, we can explore the use of further annotations:

\sp

\begin{exi}
The Program Manager may initiate revocation proceedings against a certified operation

\sp

\lb When the Program Manager has reason to believe that a \lb certified operation has violated the Act\rb

\sp

\OR

\sp

When a certifying agent fails to take appropriate action to enforce the Act\rb .\end{exi}

\dots

&

\vspace{5.6cm}

\begin{exi}The \A{}(Program Manager) \D{}(may) \I{}\customLog{regfunc=sanction}(initiate) \Bdir{}(revocation proceedings)  \Bind{}(against a certified operation)

\sp

\Cac{}\lb When the \A{}(Program Manager) \I{}\customLog{\regfunc=evaluate}(has reason to believe) that a \Bdir{}\lb \A{}(certified operation) \I{}\customLog{\regfunc=violate}(has violated) the \Bdir{}\customLog{\polref=``policy''}(Act)\rb

\sp

\OR

\sp

When a \A{}(certifying agent) \I{}\customLog{\regfunc=violate}(fails)  \Bdir{}(to take appropriate action) \Cex{},\customLog{ctx=pur; \polref=``policy''}(to enforce the Act)\rb.\end{exi}

\\

\begin{emp}Regulative Function Annotations (ctd.)\end{emp}

& & 

The key actions include:

\begin{itemize}
    \item Initiate (revocation proceedings), corresponding to a sanction
    \item ``Has reason to believe'' reflects an evaluation on the part of the actor
    \item ``Violate'' reflects a violation
    \item ``Fail to take appropriate action'' likewise represents a violation
\end{itemize}

In the coding, these actions can thus be annotated with the corresponding regulative functions.

& \\

\midrule

\multicolumn{4}{\tablespan}{\textbf{General IG Logico Instructions for Regulative Statements}} \\

\midrule

\multicolumn{4}{\tablespan}{
A central objective is to provide a consistent coding that reflects the most fine-granular level of encoding. While the order of encoding is loosely prescribed by the order of specification, in some cases a variation of the order may be indicated. This should be considered as part of the coding preparation. The coding may further require iterative review, specifically with respect to annotations and logical relationships.

While implicit in the multi-pass coding implied for IG Logico, a dedicated review of embedded object hierarchies (encoded as part of IG Extended) and the explication of logical relationships between component elements (e.g., specific execution constraints) is of central concern in IG Logico.}

\\

\toprule

\caption{Coding Guidance on Syntactic Elements for IG Logico as Level of Expressiveness (Regulative Statements)}
\label{tab:CodingIgLogicoRegulative}
\end{longtable}

\end{landscape}

\subsection{Constitutive Statement Coding}
\label{subsec:ConstitutiveStatementCoding}

The principal structure of constitutive statements visualized in \cref{fig:ConstitutiveStatementLevels} (necessary components are held in solid boxes, sufficient components in dashed boxes) highlights the relevant syntactic components (as defined in \cref{tab:SyntacticElementsConstitutive}) using the corresponding symbols specified in \cref{tab:SymbolReference}. Analogous to the introduction of regulative statements, the figure summarizes additional coding refinements introduced across levels of expressiveness, with variations largely relating to the structural decomposition of selected syntax elements and the semantic annotations applicable under IG Logico. 

\begin{figure}[h!]
    \centering
    \framedfigure{\includegraphics[width=0.95\textwidth]{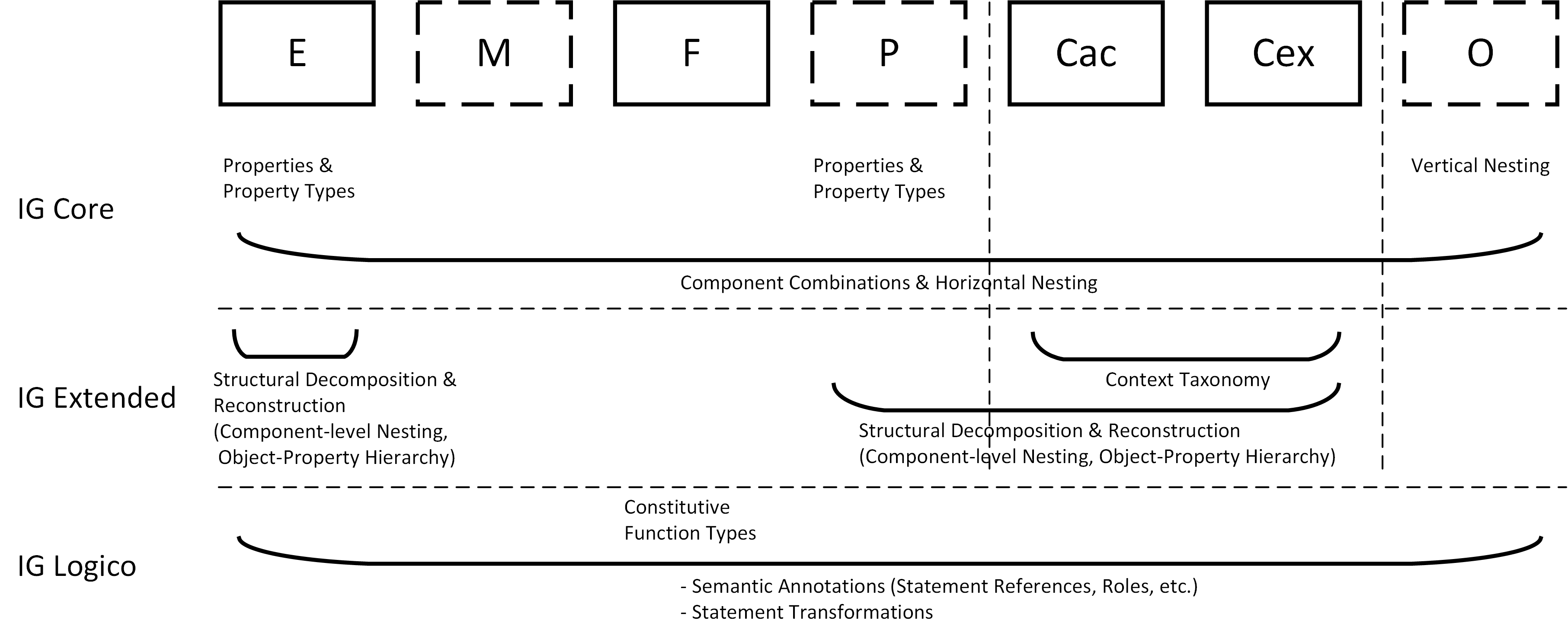}}
    \caption{Syntax and Features of Constitutive Statements by Level of Expressiveness}
    \label{fig:ConstitutiveStatementLevels}
\end{figure}

Motivating the principal coding of constitutive statements (without further elaboration at this stage), a stylized atomic constitutive statement, coded in shorthand form, thus reads: 

\begin{exa}
\Cacconst{}(In the context of organic farming),  \E{},\propconst{}(certified) \E{}(farmers) \F{}(are) \Pp{}(farmers)  \Pp{},\propconst{}(that have undergone a certification process) \Cexconst{}(following relevant procedural guidelines).
\end{exa}

\sp

Mirroring the introduction of coding guidelines for regulative statements, in \cref{tab:CodingIgCoreConstitutive} we provide the corresponding instructions for constitutive statements. Given the feature overlap between regulative and constitutive statements, we make reference to selected feature sets described in the context of regulative statements as part of the coding guidelines. As for regulative statements, symbols are color-coded to signal the association with features specific to IG Core, IG Extended or IG Logico. Symbols associated with IG Core features for constitutive statements are held in \textcolor{\corecolorconst}{\corecolorconstname}. As with regulative statements, symbols associated with IG Extended are held in \textcolor{\extendedcolor}{\extendedcolorname}, and features associated with IG Logico are displayed in \textcolor{\logicocolor}{\logicocolorname} color.

\begin{landscape}

\subsubsection{IG Core Coding of Constitutive Statements}
\label{subsubsec:IGCoreConstitutive}

\begin{longtable}{\tabledims}
\toprule
\multicolumn{4}{p{25cm}}{\textbf{Level of Expressiveness: IG Core}

\sp

IG Core enables basic, structural analysis of institutional statements. Encoding at this level is designed to be human readable and moderately comprehensive in the detail with which syntactic properties of institutional statements are captured.  
} \\

\newpage

\toprule

{\textbf{Syntactic\newline Component}} & 
{\textbf{Treatment of Syntactic Components by Level of Encoding}} &
{\textbf{Relevant Examples}} &
{\textbf{Complete Syntactic Classification of Examples}} \\
\toprule

\begin{emp}Constituted 

Entity\end{emp} 

&

The encoding of the Constituted Entity, which reflects any entity created, modified or otherwise introduced into the institutional setting. Constituted entities can be of physical or virtual nature, reflect concrete or abstract concepts, typically including actors, roles, actions, and objects. Constituted entities can further be differentiated into entity and entity property.

\sp

\udl{Note:} The separation of Constituted Entity and the associated properties operates identical to the decomposition of Attributes and Attributes properties. See corresponding guidelines provided under the Attributes component specification (see \hyperlink{tablink:AttributePropertyHeuristics}{here}).

&

\udl{Example statement:} 

\begin{exi}There is hereby established a public Food Security Advisory Board. \end{exi}

\sp

Entity = \begin{exi}Food Security Advisory Board\end{exi}

\sp

Entity property = \begin{exi}public\end{exi}

\sp

\udl{Example statement:}

\begin{exi}No member of the Council shall be disqualified from holding any public office or employment.\end{exi}

\sp

While reflecting structural patterns of regulative statements, this statement parameterizes members with respect to rights in the context of the Council.

\sp

Beyond the necessary components (\E, \F~and implied Context), the substantive characteristics that do \begin{emp}NOT\end{emp} apply (see negation applied) to the constituted entity are expressed as constituting properties.

\sp

\udl{Additional example:} 

\begin{exi}Established in this Regulation subpart is the right to appeal to a revocation or certification.\end{exi}

&

\vspace{0.05cm}

\begin{exi} There is \Cexconst{}(hereby) \F{}(established) a \E{},\propconst{}(public) \E{}(Food Security Advisory Board). \end{exi}

\vspace{2.45cm}

\begin{exi}\NOTconst{}(No) \E{}(member) of the \E{}, \propconst{}(Council) \M{}(shall) \F{}(be) \Pp{}(disqualified from holding any public office or employment).\end{exi}

\sp

\vspace{6.1cm}

\begin{exi}\F{}(Established) \Cexconst{}(in this Regulation subpart) is the \E{}(right to appeal) to a \E{},\propconst{}(revocation or certification).\end{exi}

\\ 

\begin{emp}Constituted Entity (ctd.)\end{emp}

&

&

Constitutive statements and Implied Attributes: In instances in which a coder is encountering ambiguity in discerning whether she is dealing with a constitutive or regulative statement, one shall consider the wider context of the statement (e.g., implied attribute, type of surrounding statements, etc.). A more detailed discussion can be found in \cref{subsec:SyntacticPolymorphs}.

& \\

\newpage

\midrule

\begin{exi}Constitutive Function\end{exi} &

The constitutive function characterizes the establishment, definition or introduction of a constituted entity into the institutional setting, and where constituting properties exist, functionally link constituted entity and constituting properties. 

&

\udl{Example statement:}

\begin{exi}There is hereby established a public Food Security Advisory Board.\end{exi}

\sp

Constitutive Function: \begin{exi}\lsi is\rsi~\dots~established\end{exi}

\sp

In this context the constitutive function signals the establishment of an entity.

\sp

\udl{Example:} 

\begin{exi}Commissioner of Agriculture and Markets shall be the Chairperson the Council.\end{exi}

\sp

Constitutive Function: \begin{exi}\lsi serve as\rsi\end{exi}

\sp

Here the constitutive function indicates a modified position (\begin{exi}Chairperson\end{exi}) of a specific role (\begin{exi}Commissioner\end{exi}) in a specific organizational context (\begin{exi}Council\end{exi}).

\sp

While diverse in nature, the constituting function can be sensibly organized along a set of patterns discussed in the context of IG Logico.

&

\vspace{0.03cm}

\begin{exi}There is \Cexconst{}(hereby) \F{}(established) a \E{},\propconst{}(public) \E{}(Food Security Advisory Board).\end{exi}

\vspace{3.5cm}

\begin{exi}\Pp{}(Commissioner of Agriculture and Markets) \M{}(shall) \F{}(be) the \E{}(Chairperson the Council).\end{exi}

\\

\newpage

\midrule

\begin{exi}Constituting Properties\end{exi}

&

Constituting properties are optional components in constitutive institutional statements that capture elements functionally linked to the constituted entity by means of the constitutive function. 

Constituting properties may themselves have properties.

\sp

\udl{Note:} The separation of Constituting Properties and the associated nested properties operates identical to the decomposition of Attributes and Attributes properties. See corresponding guidelines provided under the Attributes component specification (see \hyperlink{tablink:AttributePropertyHeuristics}{here}).

&

\udl{Example:} 

\begin{exi}The Committee shall consist of a President, Secretary, and Treasurer.\end{exi}

\sp

Constituting properties: \begin{exi}President, Secretary, and Treasurer\end{exi}

\sp

Here, the council is composed of the members as constituting properties.

\sp

\udl{Example:} 

\begin{exi}A majority of the members of the Council shall constitute a quorum.\end{exi}

\sp

Constituting properties: \begin{exi}majority of the members of the Council\end{exi}

&

\vspace{0.03cm}

\begin{exi}The \E{}(Committee) \M{}(shall) \F{}(consist) of a \Pp{}(President, Secretary, and Treasurer).\end{exi}

\vspace{3.55cm}

\begin{exi}A \Pp{}(majority of the members of the Council) \M{}(shall)
\F{}(constitute) a \E{}(quorum).\end{exi}

\\

\midrule

\begin{emp}Modal\end{emp} &

The Modal signals the extent to which the instruction contained in the constitutive statement is required (necessary), or signals mere possibility, (effectively mirroring the prescriptive and discretionary notion on the regulative side). Linguistically, the expression can be ambiguous between the regulative and constitutive variant (e.g., must, may, must not), but the coder should make an attempt to classify the modal from an epistemic perspective (i.e., focusing on necessity and possibility of the activity expressed in the constitutive function). Note that in constitutive statements the use of the modal can follow legal traditions or be stylistic in kind. The coder should acquaint oneself with such specifics prior to the coding.

&

\udl{Example:} 

\begin{exi}A majority of the members of the Council shall constitute a quorum.\end{exi}

\sp

Modal: \begin{exi}shall\end{exi}

\sp

Here \begin{emp}shall\end{emp} signals the necessity that a quorum means ``majority of the members of the Council'' in the context of the policy. 

\sp

\udl{Example:}

\begin{exi}The Council may have an advisory committee.\end{exi}

\sp

Modal: \begin{exi}may\end{exi}

\sp

Here \begin{emp}may\end{emp} signals the possibility (but not requirement) that the Council has an advisory committee. 

&

\vspace{0.02cm}

\begin{exi}
A \Pp{}(majority of the members of the Council) \M{}(shall) \F{}(constitute) a \E{}(quorum).\end{exi}

\vspace{4.0cm}

\begin{exi}
The \E{}(Council) \M{}(may) \F{}(have) an \Pp{}(advisory committee).\end{exi}

\\

\newpage

\midrule

\begin{emp}Context\end{emp} &

The encoding identifies the Context of the institutional statement. The encoding differentiates between ``Activation Conditions,'' which are contextual clauses that specify preconditions under which the statement applies, and ``Execution Constraints,'' which are contextual descriptors that qualify the constituting function by augmenting the statement with temporal, spatial, procedural, and/or other constraining parameters. &

\udl{Example statement:} 

\begin{exi}From 1st of January onward, food preparation guidelines must adhere to national standards, in addition to communal provisions.\end{exi}

\sp

Context clauses: \begin{exi}From 1st of January onward; in addition to communal provisions\end{exi}

\sp

\udl{Context encoding:} 

\sp

Activation Condition: \begin{exi}From 1st of January onwards\end{exi}

\sp

Execution Constraint: \begin{exi}in addition to communal provisions\end{exi}

\sp

The activation condition signals an event that initiates a discretized setting in which the remaining statement holds.

\sp

The execution constraint characterizes the constitutive function more explicitly. 

&

\begin{exi}\udl{\Cacconst{}(From 1st of January onward)}, \E{}(food preparation guidelines) \M{}(must) \F{}(adhere) to \Pp{}(national standards), \udl{\Cexconst{}(in addition to communal provisions)}.\end{exi}

\\


\midrule

\begin{emp}Or else\end{emp} &

The encoding of Or else statements identifies consequences of non-fulfillment or violation of institutional statements, which, in the context of constitutive statements, can signal consequences addressing specific actors (e.g., consequences in regulative terms), or be of existential nature. Existential nature signals that an establishment or modification of an entity may simply not come about, or that alternative or linked constitutive statements are instantiated. The encoding captures these consequences generally in the form of institutional statements that nest on the leading monitored institutional statement, and can be expressed both in regulative or constitutive form (see \cref{subsec:ConstitutiveRegulativeHybrids} for conceptual details on mixed statement types).

Principles of horizontal and vertical nesting (see \cref{subsec:InstitutionalStatementAssumptions}), as described in the regulative context (see \cref{subsec:RegulativeStatementCoding}), equally apply for constitutive statements, and within or else components.

Vertical nesting is applicable when there is at least one activity that is specified within a distinct institutional statement and activated as a consequence of a non-fulfilled action specified in the leading institutional statement. Horizontal nesting is applicable when there exist two or more activities that can be enacted in some logical combination as shown below. 

&

\udl{Example:} 

\begin{exi}
In student recruitment plans, diversity must mean diversity in race, religion, sexual orientation and gender, or else plan is void.
\end{exi}

\sp

Or else clause comprising statement: \begin{exi}or else plan is void\end{exi}

\sp

The Or else signals an existential consequence for the constituted entity as expressed by the necessity that diversity is defined as prescribed (\begin{emp}``must mean''\end{emp}).

\sp

Naturally, the consequence can also consist of multiple statements that are logically combined (horizontal nesting), as shown below.

\sp

\udl{Example:}

\begin{exi}
In student recruitment plans, diversity must mean diversity in race, religion, sexual orientation and gender, or else plan is void and to be revised within 30 days.\end{exi}

&

\vspace{0.05cm}

\begin{exi}
\Cacconst{}(In student recruitment plans), \E{}(diversity) \M{}(must) \F{}(mean) \Pp{}(diversity in race, religion, sexual orientation and gender), 

or else

\Oeconst{}\lb \E{}(plan) \F{}(is) \Pp{}(void)\rb .\end{exi}

\vspace{6.5cm}

\begin{exi}
\Cacconst{}(In student recruitment plans), \E{}(diversity) \M{}(must)  \F{}(mean) \Pp{}(diversity in race, religion, sexual orientation and gender), 

or else 

\Oeconst{}\lb \lb \E{}(plan) \F{}(is) \Pp{}(void)\rb

\ANDconst

\lb \dots~\E{}(plan) \F{}(is~\dots~to be revised) \Cexconst(within 30 days)\rb \rb .\end{exi}

\sp

In the previous example, horizontal nesting is signaled using braces around statements (as opposed to individual components), and vertical nesting is likewise expressed using braces (\lb~and \rb ) scoped around the compound consequence. 

\\

\begin{emp}Or else (ctd.)\end{emp}

&

These statement combinations can signal 

\begin{itemize}
\item alternative exclusive action options -- \XORconst s -- (e.g., either initiate XOR expand the Council),
\item inclusive action options -- \ORconst s -- (e.g., rights may be assigned AND/OR responsibilities removed when an individual is added to the Council) or 
\item co occurring action options -- \ANDconst s -- (e.g., dissolving a Council AND compensating Council members)
\end{itemize}

&

\udl{Note:}

\begin{enumerate}
\item Where component combinations exist, alternatives are combined (\dots~are void and to be refined fine farmer or revoke \dots), and are subsequently decomposed into separate logically-combined complete atomic institutional statements. 
\item Ambiguities with respect to the linguistic use of logical operators (exclusive and inclusive or) are to be resolved as part of this process. 
\end{enumerate}

&

\\

\midrule

\newpage



\midrule
\multicolumn{4}{l}{\textbf{General IG Core Instructions for Constitutive Statements}} \\
\midrule

\begin{emp}Additional annotations for Constituted Entity, Constitutive Function, Constituting Property and Context\end{emp}

\sp

(\udl{Note:} This feature is analogous to the specification of additional annotations for Attribute, Object and Context components in the context of regulative statements.) 

&

In addition to the identification of properties embedded in the original statements, components can further be annotated using additional annotation labels. Such labels can follow the categories listed in \cref{sec:Taxonomies}, or be specific to the project objectives.

\sp

A systematic approach to labelling entities is discussed under \hyperlink{tablink:CrossComponentSemanticAnnotationsLogicoRegulative}{\begin{emp}IG Logico Instructions\end{emp}, Item \EntryCrossComponentSemanticAnnotations} in \cref{tab:CodingIgLogicoRegulative}. This is particularly recommended if annotations are of strong relevance for the coding and of diverse nature. 

&

\udl{Example:} 

\begin{exi}The Committee shall consist of a President, Secretary, and Treasurer.\end{exi}

\sp

Subject to analytical necessity, additional annotations can for instance relate to the identification of aspects, such as the characterisation of encoded objects with respect to their animacy as either animate or inanimate -- signified in brackets in the coded example. Where indicated, the annotation should be separated from the component specification by semicolon and have the structure ``label='', followed by the annotation. 

Note that ``\begin{exi}label\end{exi}'' is a characterizing prefix, either based on a specific taxonomy (see \cref{sec:Taxonomies} for an overview), or a custom coder-defined category specification (e.g., defined as part of the project-specific guidelines). In this example, the Animacy Taxonomy has the prefix ``\anim'' as defined in \cref{sec:Taxonomies} and is used correspondingly.

\sp

While exemplified here for constitutive statements, this equally applies to regulative statements. 

&

\vspace{0.03cm}

\begin{exi}The \E{}\customLog{\anim=inanimate}(Committee) \M{}(shall)  \F{}(consist of) a \Pp{}\customLog{\anim=animate}(President, Secretary, and Treasurer).\end{exi}

\\

\midrule

\hypertarget{tablink:DecompositionOfComponentLevelCombinationsCoreConstitutive}{\EntryDecompositionOfComponentLevelCombinations}

\sp

(Note: This applies to regulative and constitutive statements, and is discussed here with focus on the constitutive perspective.)

&

Where combinations of components (component-level combinations) are observed that are not explicitly decomposed as in the case of vertical nesting, these can be decomposed into logically-combined statements. 
Other than for constitutive functions, the decomposition is optional for IG Core. 

Operationally, combinations of components are evidenced by the presence of multiple logically-combined tokens or clauses embedded in constituted entities, constitutive functions, constituting properties or context components. 

Decomposition essentially entails constructing an individual statement to capture each of the unique components represented in multiples within institutional statements, noting the relation to the original statement in which multiple components are reflected. Information from component fields, other than that containing multiple components, is simply carried over to all related institutional statements. 

Importantly, where decomposition actually changes the meaning of the original institutional statement containing multiple components within a particular syntactic field, the statement should not be decomposed. In such cases, multiples are typically intended to exist in coupled form. An example is provided in the next column.

&

Details are described in the 
\hyperlink{tablink:DecompositionOfComponentLevelCombinationsExtendedRegulative}{\begin{emp}General IG Extended Instructions\end{emp}, Item \EntryDecompositionOfComponentLevelCombinations} in \cref{tab:CodingIgExtendedRegulative}.

\sp

\udl{Example (Multiple Properties):} 

\begin{exi}The Committee shall consist of a President, Secretary, and Treasurer.\end{exi}

\sp

While expressed in condensed form as ``\begin{exi}The \E{}(Committee)  \M{}(shall) \F{}(consist of) a \Pp{}(\lpconst President \AND{}~Secretary \AND{}~Treasurer\rpconst)\end{exi}'' (note the parentheses),

\sp

it corresponds to the following statement composed of three atomic statements:

\sp

\udl{Statement 1:} \begin{exi}The Committee shall consist of a President\end{exi}

\sp

\ANDconst

\sp

\udl{Statement 2:} \begin{exi}The Committee shall consist of a Secretary.\end{exi}

\sp

\ANDconst 

\sp

\udl{Statement 3:} \begin{exi}The Committee shall consist of a Treasurer.\end{exi}

\sp

\udl{Example (Multiple constitutive functions):} \begin{exi}The form and function of the Council is hereby established.\end{exi}

\sp

&

\udl{Condensed form:}

\begin{exi}The \E{}(Committee) \M{}(shall) \F{}(consist of) a \Pp{}(\lpconst President \ANDconst~Secretary
\ANDconst~Treasurer\rpconst).\end{exi}

\sp

\udl{Expanded form:}

\begin{exi}\lb \lb The \E{}(Committee) \M{}(shall) \F{}(consist of) a \Pp{}(President)\rb


\ANDconst


\lb The \E{}(Committee) \M{}(shall) \F{}(consist of) a \Pp{}(Secretary)\rb


\ANDconst 


\lb The \E{}(Committee) \M{}(shall) \F{}(consist of) a \Pp{}(Treasurer)\rb \rb .\end{exi}

\vspace{8.6cm}

\begin{exi}
\lb \lb \E{}(Council form) is \Cexconst{}(hereby) \F{}(established)\rb


\ANDconst


\lb \E{}(Council function) is \Cexconst{}(hereby) \F{}(established).\rb \rb
\end{exi}

\\

\begin{emp}Decomposition of component-level combinations (ctd.)\end{emp}

&

&

\udl{Note:} These guidelines highlight the motivation for the decomposition, and exemplify it explicitly. Depending on the use of annotation means and tool support, the decomposition may be partially automated, affording a mere annotation for such decomposition without requiring the user to perform statement duplication.

&

\\

\toprule
\caption{Coding Guidance on Syntactic Elements for IG Core as Level of Expressiveness (Constitutive Statements)}
\label{tab:CodingIgCoreConstitutive}
\end{longtable}

\newpage

\subsubsection{IG Extended Coding of Constitutive Statements}
\label{subsubsec:IGExtendedConstitutive}

\begin{longtable}{\tabledims}

\toprule
\multicolumn{4}{\tablespan}{\textbf{Level of Expressiveness: IG Extended}

\sp

Mirroring the progression on the regulative side, IG Extended enables more detailed structural analysis of institutional data than IG Core and accommodates computational application to aid in institutional coding and analysis. Encoding at this level is designed to be human readable, moderately computationally tractable, and moderately comprehensive in the detail with which syntactic properties of institutional statements are captured. 

\sp 

Coding institutional statements on this level enforces many of the features that have been optional in IG Core and affords a fine-grained decomposition of statements. This includes a richer context characterization based on predefined taxonomies, the expansion and combined attributes and aims that reconstruct atomic statements and their relationships, but also decomposes the hierarchical relationships amongst explicitly highlighted Constituted Entities, Constituting Properties and Constitutive Functions, alongside further refinements of contextual descriptors.

\sp

As a central feature IG Extended makes the use of component-level combinations explicit. This specifically facilitates the decomposition of the context component to express institutional content at a more nuanced level. In addition, structural refinements relate to the decomposition of relationships and properties of Constituted Entities and Constituting Properties.

\sp

For constitutive statements, IG Extended features correspond to the regulative side, with the essential difference for the application of refinements on Attributes and Objects, which, in the context of constitutive statements apply to Constituted Entities and Constituting Properties. 

\sp

A specific consideration is the concept of Constitutive-Regulative Hybrids and syntactic polymorphs, both of which are of cross-cutting nature (i.e., affecting both constitutive and regulative statements) and thus discussed in a dedicated section. Their consideration, however, applies to IG Extended.

} \\
\toprule

{\textbf{Syntactic\newline Component}} & 
{\textbf{Treatment of Syntactic Components by Level of Encoding}} & 
{\textbf{Relevant Examples}} &
{\textbf{Complete Syntactic Classification of Examples}} 
\\

\toprule

\begin{emp}
Constituted 

Entity
\end{emp}

&

In IG Extended encoding, Constituted Entities and their properties are decomposed hierarchically following the principles of the Object-Property Hierarchy (\cref{subsec:AttributeObjectHierarchy}) and is applied analogous to ``Attributes'' in IG Extended for regulative statements (\cref{tab:CodingIgExtendedRegulative}). The ensuing overview of constituted properties shows an extended set of examples applied analogously to constituted entities.

&

\udl{Example:}\newline
\begin{exi}There is hereby established a standing public Food Security Advisory Board.\end{exi}

& 

\begin{exi}There is \Cexconst \customExt{ctx=met}(hereby) \F{}(established) a \E{},\propExt{1}(standing) \E{},\propExt{2}(public) \E{}(Food Security Advisory Board).\end{exi} 

\\

\midrule

\begin{emp}
Constituting 

Property
\end{emp}

&


Analogous to the decomposition of object properties in the context of regulative statements, constituting properties are likewise decomposed following the principles of the Object-Property Hierarchy (as introduced in \cref{subsec:AttributeObjectHierarchy} and applied in the context of ``Objects'' in IG Extended for regulative statements in \cref{tab:CodingIgExtendedRegulative}). 


&

\udl{Example:}\newline
\begin{exi}The Council consists of  elected officials resident in the electorate.\end{exi}

\sp

In this example, the individual properties of the constituting property \begin{emp}officials\end{emp}, namely \begin{emp}elected\end{emp} and \begin{emp}resident in the electorate\end{emp}, are uniquely identified as properties.

\sp

Another feature is the richer hierarchical structure embedded in phrase expressing compound property characterizations. 

\sp

\udl{Example:}\newline
\begin{exi}A majority of the members of the Council shall 
constitute a quorum.\end{exi}

\sp

In this example, the constituting property is captured in the phrase  \begin{emp}A majority of the members of the Council\end{emp}. While the entire phrase represents the constituting property (and is coded as such on IG Core), the embedded hierarchy, i.e., members are a property of the Council, and the majority is a property of the members, can be explicitly captured using hierarchical property annotations as shown on the right. 

Where properties are not functionally dependent on another property, they are signaled using unique identifiers (e.g., \PpExt{a}, \PpExt{b}) equivalent to ``Object'' decomposition highlighted in \cref{tab:CodingIgExtendedRegulative} and exemplified in the following.

&

\vspace{0.02cm}

\begin{exi}The \E{}(Council) \F{}(consists of) \Pp{},\propExt{1}(elected) \Pp{}(officials) \Pp{},\propExt{2}(resident in the electorate).\end{exi}

\vspace{5.6cm}

\begin{exi}A \Pp{},\propExt{1},\propExt{1}(majority) of the \Pp{},\propExt{1}(members) of the \Pp{}(Council) \M{}(shall) \F{}(constitute) a \E{}(quorum).\end{exi}

\\

\begin{emp}
Constituting 

Property (ctd.)
\end{emp}

&

&

Collections of functionally independent entities are represented as a compound constituting property signaled by parentheses. Individual compound properties can be uniquely identified, alongside potential further properties shared across all embedded entities.\footnote{Specific data structure patterns commonly found in institutional statements are revisited in \cref{subsec:DataStructurePatterns}.}

\sp

\udl{Example:}\newline
\begin{exi}The Committee shall consist of a President, Secretary, and qualified Treasurer appointed by the public.\end{exi}

\sp

In this example, properties specific to an entity are called out with reference to the entity (\begin{emp}qualified\end{emp}), whereas shared properties are associated with all entities (\begin{emp}appointed by the public\end{emp}).

& 

\vspace{4.1cm}

\begin{exi}The \E{}(Committee) \M{}(shall) \F{}(consist of) a \Pp{}(\lpconst  \PpExt{1}(President) \ANDconst~\PpExt{2}(Secretary)
\ANDconst~\PpExt{3},\propExt{1}(qualified) \PpExt{3}(Treasurer)\rpconst )~\Pp,\propconst{}(appointed by the public).\end{exi}

\\

\midrule

\begin{emp}
Context
\end{emp}

&

\multicolumn{3}{\tablespanNarrow}{
See ``Context'' in IG Extended for regulative statements (\cref{tab:CodingIgExtendedRegulative})

} \\

\midrule
\begin{emp}
General IG 

Extended Instructions
\end{emp}
&

\multicolumn{3}{\tablespanNarrow}{
See ``General IG Extended Instructions'' in IG Extended for regulative statements (\cref{tab:CodingIgExtendedRegulative})
} \\

\midrule

\begin{emp}
Constitutive-

regulative Hybrids
\end{emp}

&

\multicolumn{3}{\tablespanNarrow}{
The introduction of constitutive statements as part of IG 2.0 (see \cref{subsec:ConstitutiveStatementCoding}) provides the basis for encoding statements that consist of structural elements both of regulative and constitutive statements. Details are discussed in \cref{subsec:ConstitutiveRegulativeHybrids}.

} \\

\toprule

\caption{Coding Guidance on Syntactic Elements for IG Extended as Level of Expressiveness (Constitutive Statements)}
\label{tab:CodingIgExtendedConstitutive}
\end{longtable}

\newpage

\subsubsection{IG Logico Coding of Constitutive Statements}
\label{subsubsec:IGLogicoConstitutive}

\begin{longtable}{\tabledims}
\toprule
\multicolumn{4}{\tablespan}{\textbf{Level of Expressiveness: IG Logico}

\sp

IG Logico is designed to support semantic analysis of institutional statements wholly relying on computational tools. Encoding at this level is designed to be moderately human readable, computationally tractable and comprehensive in the detail with which syntactic properties of institutional statements are captured. 

\sp

In contrast to IG Core and Extended that focus on the encoding of specific grammar components, IG Logico emphasises refinements across individual components and further establishes explicit references to related statements to establish computational tractability, as well as the ability to perform logical transformations on institutional statements. 

\sp

While largely equivalent for regulative and constitutive statements, the only variant to the instructions provided in the context of regulative statements is the discussion of Constitutive Function taxonomies (as opposed to Institutional Functions in the context of regulative statements) as outlined below.

}\\

\newpage

\toprule

{\textbf{Syntactic\newline Component}} & 
{\textbf{Treatment of Syntactic Components by Level of Encoding}} &
{\textbf{Relevant Examples}} &
{\textbf{Complete Syntactic Classification of Examples}} \\

\toprule

\begin{emp}Constitutive Function Annotations\end{emp} 

&

Complementing the content characterization for other components, the constitutive function maintains the central role as a descriptor of constituted entities, and where constituting properties exist, links those to constituted entities. 

\sp

In an attempt to characterize the function of the constitutive statement as expressed in the constitutive function more generally, we propose a taxonomy capturing common relationships more generally. Doing so, we differentiate between statements that characterize the constituted entity as newly introduced into the institutional setting, and a commonly found alternative, that is, the characterization of the policy that contains the statements itself.

\sp

Entities, such as novel actors, objects, roles or action, can be 
\begin{itemize}
\item defined explicitly (``is'', ``does''), 
\item defined based on relationships, such as composition (``consists of''), organizational embedding (``is embedded in'', ``relates to''), 
\end{itemize}

&

\udl{Example:} 

\begin{exi}
Starting January 1st, the Connecticut Food Policy Council shall be within the Department of Agriculture.
\end{exi}

\sp

In this example the constitutive function signals the constituted entity (Connecticut Food Policy Council) as an organizational unit.

\sp

\udl{Example:} 

\begin{exi}
The Committee shall consist of a President, Secretary, and Treasurer.
\end{exi}

\sp

The constitutive function signals a composition of the constituted entity (\begin{exi}Committee\end{exi}) based on constituting properties. 

\sp

\udl{Example:} 

\begin{exi}The purpose of this Part is to establish standards for net metering in accordance with the requirements of Section 16-107.5 of the Act.\end{exi}

\sp

In this example, the constitutive function identifies the entity as a policy and signals the intent underlying the policy. 

&

\vspace{0.03cm}

\begin{exi}
\Cacconst{}(Starting January 1st), the \E{}(Connecticut Food Policy Council) \M{}(shall) \F{}\customLog{\confunc=organization}(be within) the \Cexconst{}(Department of Agriculture).\end{exi}

\vspace{2.5cm}

\begin{exi}
The \E{}(Committee) \M{}(shall) \F{}\customLog{\confunc=composition}(consist of) a \Pp{}(\lpconst President \ANDconst~Secretary
\ANDconst~Treasurer\rpconst).\end{exi}

\vspace{1.55cm}

\begin{exi}
The \E{}(purpose of this Part) \F{}\customLog{\confunc=intent}(is) \Pp{}(to establish standards for net metering in accordance with the requirements of \customLog{\polref=Section/16-107.5}(Section 16-107.5) of the Act).\end{exi}

\\

\begin{emp}Constitutive Function Annotations (ctd.)\end{emp}

&

\begin{itemize}
\item defined based on lifecycle stages (``established'', ``terminated''), and finally 
\item characterized by the conferral of status in the form of rights, authority, or exertion of institutional power more generally.
\end{itemize}

Policies as constituted entities in institutional statements, in contrast, are generally referred to with respect to the 
\begin{itemize}
\item lifecycle stage they are involved in (``come into force''), 
\item relationship between and to other statements or policies (``amends'', ``substitutes''), 
\item intent in the form of purpose of a specific policy, and appear as 
\item information statements that offer information about the policy itself.
\end{itemize}


Naturally, these characterizations are not exhaustive and can carry more specific forms. An overview of the different characterizations, alongside the labels used in this context is provided in \cref{sec:Taxonomies}.

& 

\udl{Example:}

\begin{exi}In department's university plan, diverse population means diversity in religion, sexual orientation and race.\end{exi}

\sp

In this example, the constituted entity is defined intensionally, that is in terms of its underlying interpretations.

& 

\vspace{0.03cm}

\begin{exi}
\Cacconst{}\customExt{ctx=dom}(In department's university plan), \E{}(diverse population) \F{}\customLog{\confunc=definition}(means) \Pp{}(\lp diversity in religion \AND{}~sexual orientation \AND{}~race\rp).\end{exi}

\\

\toprule

\caption{Coding Guidance on Syntactic Elements for IG Logico as Level of Expressiveness (Constitutive Statements)}
\label{tab:CodingIgLogicoConstitutive}
\end{longtable}

\end{landscape}

\subsection{Hybrid \& Polymorphic Institutional Statements}
\label{subsec:ConstitutiveRegulativeHybrids}

In addition to the specific treatment of regulative and constitutive statements as part of the coding guidelines, an aspect that demands specific attention is the combined use of both statement types. While distinctively different in their function, regulative and constitutive statements, of course, share structural patterns as outlined in the context of the operational coding across varying levels of expressiveness.

\sp

However, an operational concern that links both statement types is the interleaved use in practice. In addition to the commonly found organization of primarily constitutive and regulative statements into distinctive sections of documents (e.g., constitutive statements as part of the preamble), in regulative statements we may encounter inline specifications of entities that are positioned in the institutional setting and are thus of relevance for subsequent statements. Conversely, in constitutive statements, we can potentially encounter embedded regulative elements that regulate behavior of the constituted entities. Linking the nested relationships of institutional statements across both types, we recognize the concept of \begin{emp}hybrid institutional statements\end{emp}, in which the combined use of constitutive and regulative statements can either take constitutive-regulative form (where the overall statement is of constitutive nature) or be a regulative-constitutive hybrid (where the leading statement is of regulative nature). 
Where existing, their resolution is a central feature of IG Extended onwards (and optional for IG Core). 

In contrast to statements that embed a combination of parameterizing and regulating features, in specific instances statements may be encodeable in both regulative and constitutive form. While such variable coding often reflects a concession to analytical objectives, potential coding inconsistencies can be resolved by a priori specification of the interpretational scope applied in the coding process (see \cref{subsec:StatementTypesHeuristics}). Where a generic coding of statements, agnostic of analytical uses, is desired, statements can be coded as \begin{emp}polymorphic institutional statements\end{emp} that allow the concurrent encoding in different statement forms. This special form of institutional statements is discussed in \cref{subsec:SyntacticPolymorphs}.

Initially, however, we will exemplify both variants of statement hybrids.

\subsubsection{Regulative-Constitutive Statements}

A typical reflection of hybrids stems from the introduction of novel entities as part of a regulative statement, as shown in the following example (\cref{fig:RegulativeConstitutiveHybridRaw}):

\begin{figure}[h!]
    \centering
    \framedfigure{\includegraphics[width=0.8\textwidth]{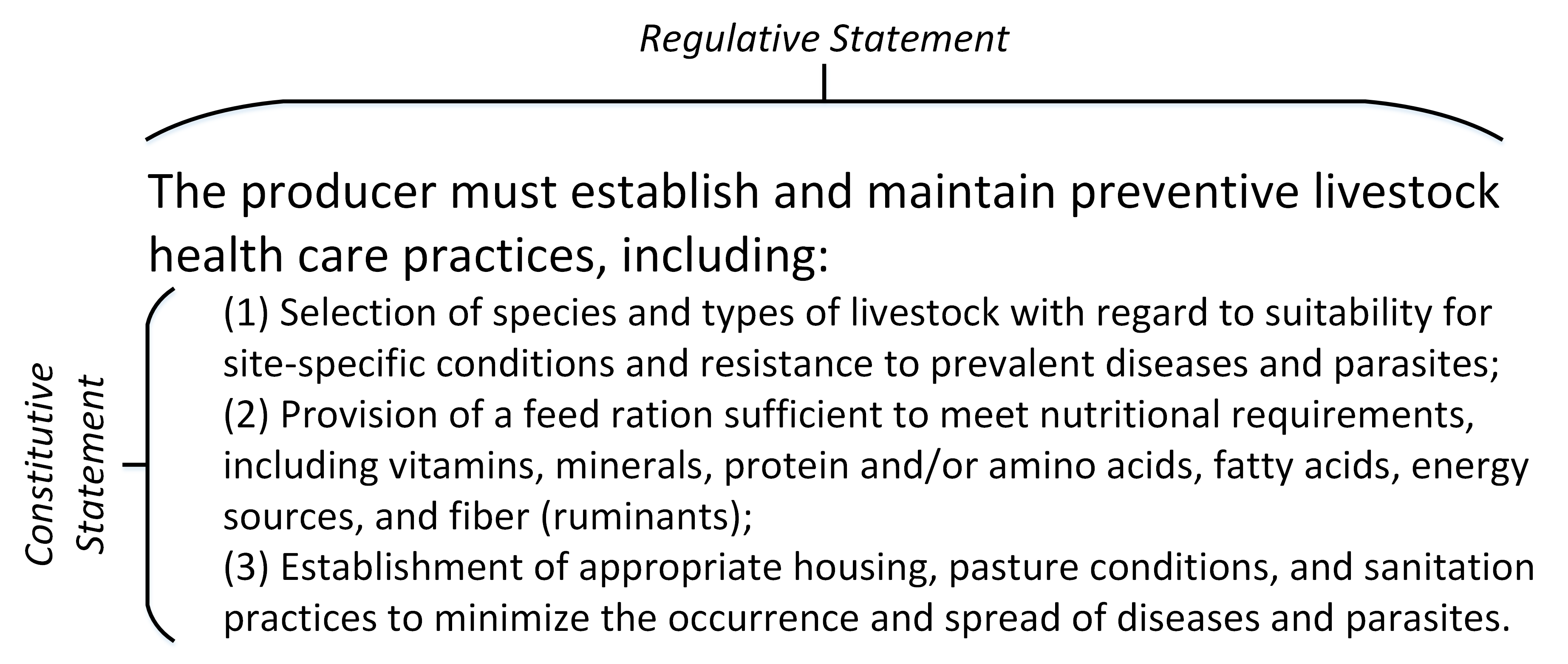}}
    \caption{Regulative-Constitutive Hybrid Example}
    \label{fig:RegulativeConstitutiveHybridRaw}
\end{figure}

As signaled visually, this example highlights a regulative statement capturing an actor's obligations, with the latter defined in an embedded constitutive statement, reflecting a regulative-constitutive hybrid. In this example, the constitutive statement is nested in a specific component of the regulative statement, such as the object as shown in the example in \cref{fig:RegulativeConstitutiveHybridCoded}. Note that the following figures use the same color-coding used in the preceding sections: Symbols associated with IG Core features for regulative statements are displayed in \textcolor{\corecolor}{\corecolorname}, whereas symbols signaling constitutive statements are held in \textcolor{\corecolorconst}{\corecolorconstname}.\footnote{In the context of this section, parentheses and logical operators are color-coded to emphasize the association with the corresponding institutional statement type.} 

\begin{figure}[h!]
    \centering
    \framedfigure{\includegraphics[width=0.8\textwidth]{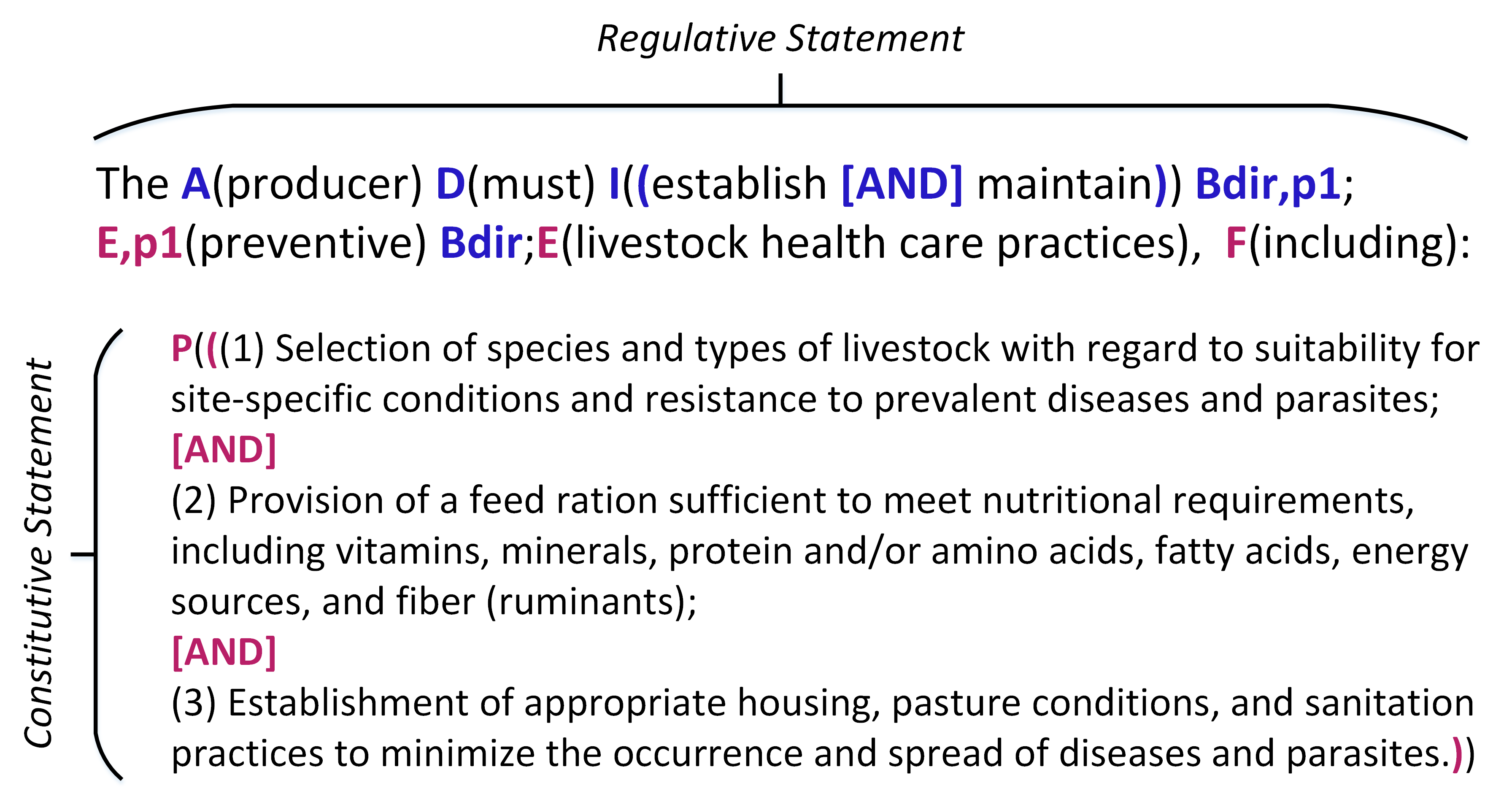}}
    \caption{Coded Regulative-Constitutive Hybrid Example}
    \label{fig:RegulativeConstitutiveHybridCoded}
\end{figure}

In many instances (such as the one shown here), this interleaved representation  affords a decomposition or transformation of hybrids into individual statements by separating the statements by syntactic components, and replication of overlapping components in the individual statements. This decomposition is exemplified in \cref{fig:RegulativeConstitutiveHybridDecomposed}.

\begin{figure}[h!]
    \centering
    \framedfigure{\includegraphics[width=0.8\textwidth]{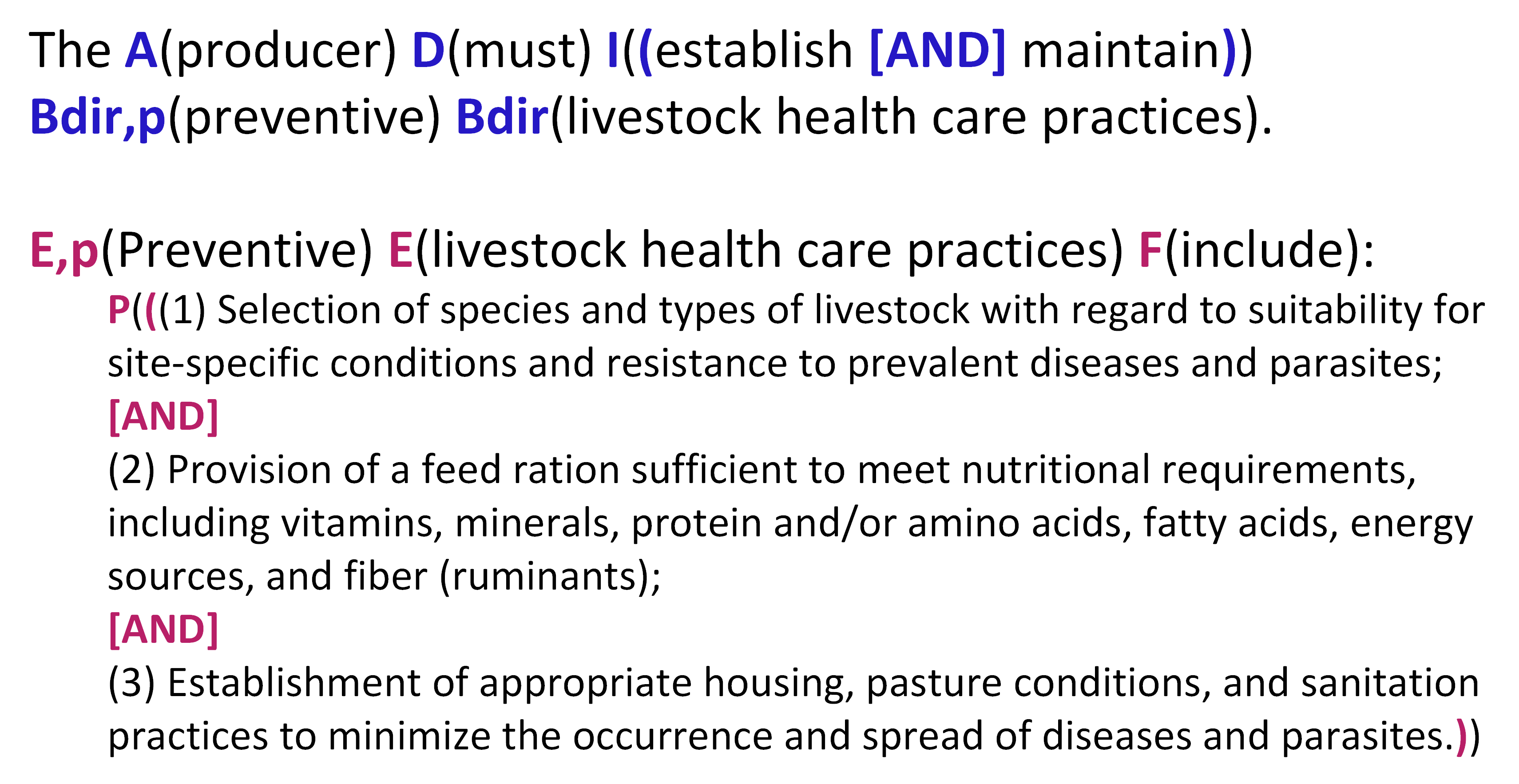}}
    \caption{Decomposed Regulative-Constitutive Hybrid Example}
    \label{fig:RegulativeConstitutiveHybridDecomposed}
\end{figure}

\newpage

\subsubsection{Constitutive-Regulative Statements}

Contrasting the embedding of constitutive statements in regulative settings, we can likewise observe the embedding of regulative statements in constitutive ones. In the following example (\cref{fig:ConstitutiveRegulativeHybridExample}), the consequence of breaching a constitutive statement\footnote{This statement signals the requirement that diversity is understood as specified.} is expressed in regulative terms, following the principles of statement-level nesting. While the leading statement is coded as a constitutive statement, the consequences are a combination of regulative statements.

\begin{figure}[h!]
    \centering
    \framedfigure{\includegraphics[width=0.9\textwidth]{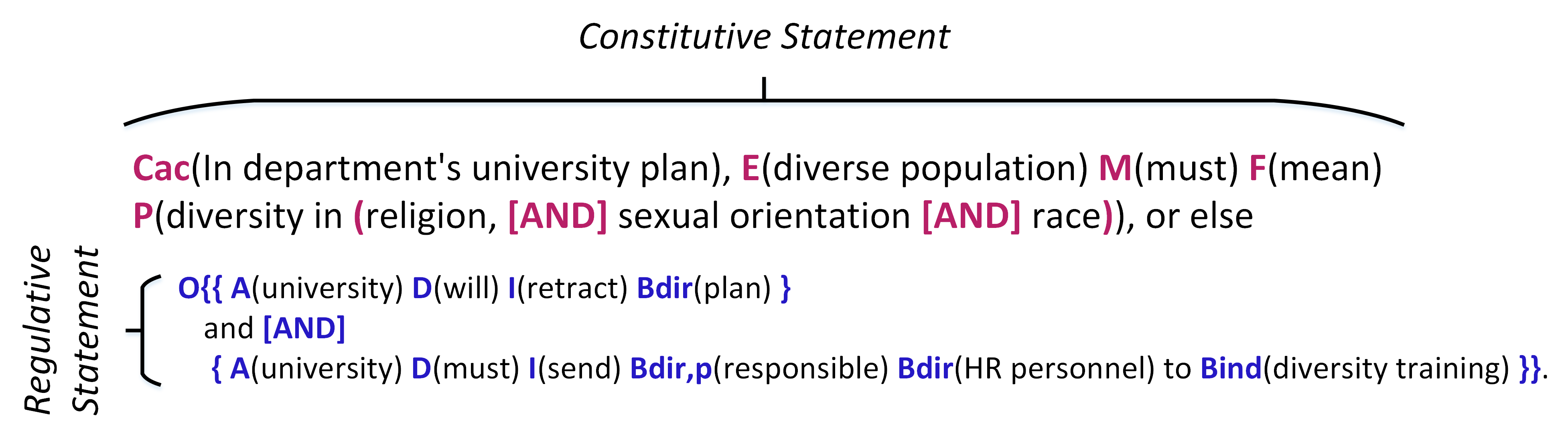}}
    \caption{Constitutive-Regulative Hybrid Example}
    \label{fig:ConstitutiveRegulativeHybridExample}
\end{figure}

\subsubsection{Second-Order Constitutive Statements}

In addition to the combined characterisation of constitutive and regulative hybrids, we can further observe component-level nesting of constitutive statements as shown below. While, in principle, equally admissible for regulative statements, specifically the higher-order decomposition of constitutive statements is commonly found. Higher-order decomposition thereby implies the nesting of constitutive statements within individual components, such as property items. Naturally, as motivated in the earlier example, this can occur in conjunction with hybrid statements and independent of the regulative or constitutive nature of the leading institutional statement. 

\begin{figure}[h!]
    \centering
    \framedfigure{\includegraphics[width=0.8\textwidth]{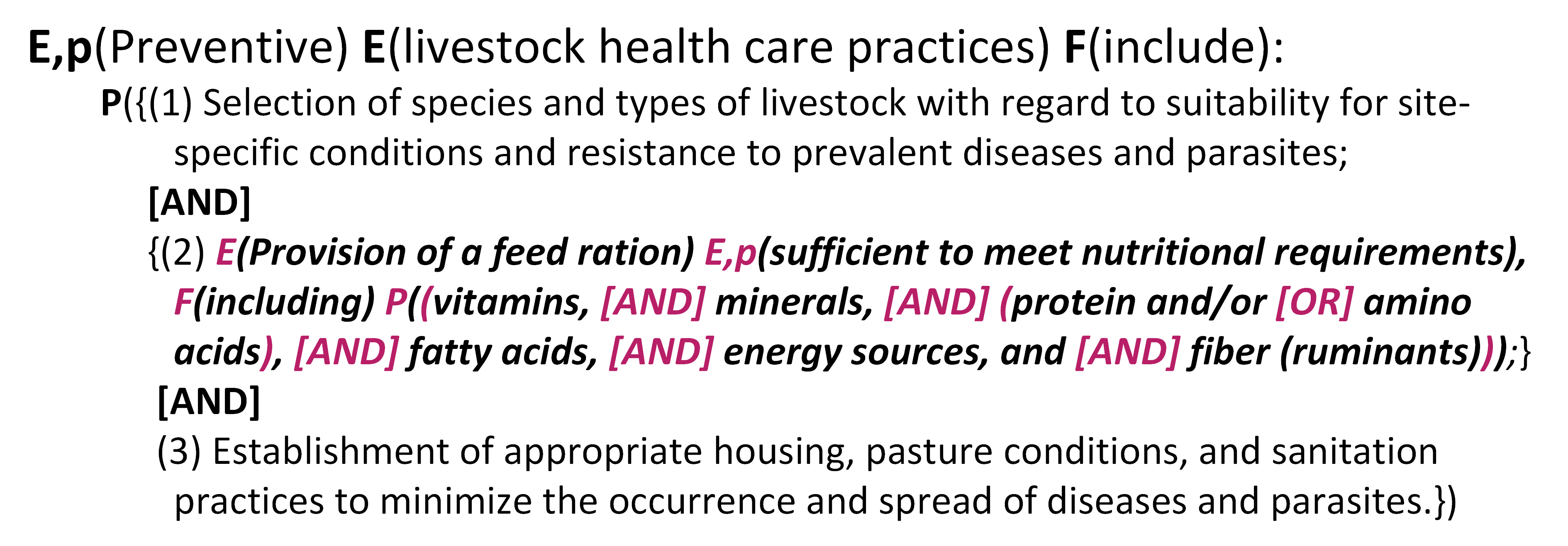}}
    \caption{Second-order Constitutive Statement Example}
    \label{fig:SecondOrderHybridExample}
\end{figure}

Reviewing the example above (\cref{fig:SecondOrderHybridExample}, second-order statement in italicized bold font), we note the reference to feed rations as a property element of livestock health care practices that is defined in terms of an embedded constitutive statement that in itself captures a complex set of properties that constitute a feed ration. 

\clearpairofpagestyles
\chead{\headmark}
\automark[subsubsection]{subsubsection}
\ifoot{\ifootertext}
\ofoot{\ofootertext}
\cfoot{\thepage}

\subsubsection{Polymorphic Institutional Statements}
\label{subsec:SyntacticPolymorphs}

Another aspect of discussion of the combined nature of statements relates to the identification of constitutive and regulative statements. The function of either statement type is generally well defined based on the parameterizing role (e.g., introduction or modification of entities or endowment of rights/authority) of constitutive statements, and the behavioral regulation (specification of operational duties and constraints for given actors or actor interaction) captured by regulative statements. While the heuristics provided in \cref{subsec:StatementTypesHeuristics} lend support for an unambiguous characterization of institutional statements as either constitutive or regulative, the characterization may not for all instances be non-contentious, or may make tacit analytical (or coding) preferences overt. While such preferences should inform the coding process by specifying the interpretational scope applied to individual institutional statements (see \cref{subsec:StatementTypesHeuristics}), where the data set is to be encoded in generic form (i.e., without concessions to specific analytical objectives),  
statements can be considered \begin{emp}polymorphic institutional statements\end{emp}, or \begin{emp}syntactic polymorphs\end{emp}. This means, they are encoded in both forms, and the analyst can draw on either shape depending on their desired analytical use. 

To substantiate this approach with an example, let us use the statement \begin{exa}``The functions of the Board shall be: (a) give effect to the decisions and policies of the Health Assembly; (b) perform any other functions entrusted to it by the Health Assembly.''\end{exa}

This statement can be read in constitutive terms, since it parameterizes the institutional system with respect to a specific entity, namely the characterization of the board in terms of its assigned functions and endowed responsibilities. The statement, however, can also be read in regulative terms, in which the functions of the board are expressed as obligations (and thus as regulated behavior) in operational terms. The same statement can thus be encoded as a polymorph by drawing on syntactic components of both statement types, as visualized in the following (including color coding to signal regulative (\textcolor{\corecolor}{\corecolorname}) and constitutive components (\textcolor{\corecolorconst}{\corecolorconstname}), respectively).

\sp

\fcolorbox{black}{\boxbgcolor}{\begin{minipage}{16.0cm}

\begin{exa}
The \E{}(functions) of the \A{}(Board) \D{}/\M{}(shall) \F{}(be): 

\Pp{}\lp \{

\hspace{0.2cm} (a) \I{}(give) \Bdir{}(effect) to the \Bind{}(decisions) of the \Bind{},\prop{}(Health Assembly); 

\hspace{0.2cm} \AND

\hspace{0.2cm} (b) \I{}(perform) \Bdir{}(any other functions)  \Bdir{},\prop{}(entrusted to it by the Health Assembly)

\hspace{0.1cm} \}\rp.
\end{exa}

\end{minipage}}

\sp

The coding shows overlap, but the central components for regulative and constitutive statements (namely constituted entity, attributes, constitutive function, aim, as well as constituting properties) showcase the varying focal aspects of different statement types. Subject to emphasis on either the actor (i.e., `the Board') or its functions (i.e., `the functions') can determine the coding, or even admit both approaches for analytical purposes.

A specific value of the polymorphic representation is to make the linkage between configurational aspects that affect the wider institutional setting, and behavioral regulation that specifically embeds actors in this institutional setting, both configurationally, as well as operationally, explicit.

While the notion of polymorphic structure is explicit in the previous example, the coding of statements in both forms can be more complex and may even afford reconstruction, as shown in the following example, with the same statement coded separately in constitutive and regulative forms: \begin{exi}``Members of the Council shall have the right to hold any public office or employment.''\end{exi}

In constitutive terms the statement reflects the Council members endowment with a right, namely the right to hold other public positions in addition to a Council membership, and is encoded as follows: 

\sp

\fcolorbox{black}{\boxbgcolor}{\begin{minipage}{15.5cm}
\begin{exi}\E{}(Members) of the \E{},\propconst{}(Council) \M{}(shall)  \F{}(have the right) \Pp{}(to hold any other public office or employment).\end{exi}
\end{minipage}}

\sp

Here, the Council member is at the center of the encoding. Depending on the analytical use (e.g., assumed actor perspective), constitutive statements reflect an abstract conception of a right or a property. Expressing this assurance in regulative terms by invoking the principle of \begin{emp}jural correlatives\end{emp}~\citep{Hohfeld1913SomeReasoning} requires the representation of behavioral constraints by introducing a conception of actorship in the form of a tacit (and in this case unidentified) actor whose behavior is regulated as corresponding duty, alongside further structural adaptations. Reconstructing the statement in regulative terms\footnote{Explicit transformation rules associated with the reconstruction of statements of various types are subject to forthcoming work.} thus produces the following coding:\footnote{Note that the following coded samples call out negations of deontics and modals explicitly for clarity.}

\sp

\fcolorbox{black}{\boxbgcolor}{\begin{minipage}{15.5cm}
\begin{exi}\A{}([Attribute])~\D{}([must]) \NOT{}~\I{}(disqualify)  \Bdir{}(member of the council) \Cex{}(from holding any other public office or employment).\end{exi}
\end{minipage}}

\sp

Exemplifying the use of polymorphic structures and their practical implications further, the following statement constellation highlights interpretational challenges associated with statement references. 

Assessing the set of statements shown below, we specifically turn our attention to the final statement. This statement is embedded within a paragraph of statements that regulate operational activity, but this statement specifically introduces further constraints that apply to all other statements. 

Encoding this statement using a \begin{emp}narrow interpretational scope\end{emp} (as described in \cref{subsec:StatementTypesHeuristics}), the statement can be interpreted as parameterizing the wider institutional setting by affecting a series of provisions (``Paragraphs in this section'') and furthermore clearly signaling the requirement that the paragraphs do not apply (see \begin{emp}Deontic vs.~Epistemic Modal\end{emp} heuristic in \cref{tab:SemanticStatementTypeHeuristics} in \cref{subsec:StatementTypesHeuristics}). Consequently, the statement is coded as constitutive (as exemplified below). 

\sp
\fcolorbox{black}{\boxbgcolor}{\begin{minipage}{15.5cm}
\begin{exi}
(1) Organic farming operations must not utilize genetically-modified seeds.

(2) Organic farming operations may not process crops other than the ones specified in Appendix A.

\dots

(10) \E{}(Paragraphs in this section) \M{}(do) \NOTconst{}~\F{}(apply) \Pp{}(to traces) \Pp{},\propconst{}(of genetically modified material).
\end{exi}
\end{minipage}}
\sp

However, given the evident regulative nature of the referenced statements, coding such statement with \begin{emp}wide interpretational scope\end{emp} in mind leads to a resolution of the semantic links to the referenced statements, and doing so, affords a reconstruction of the statement in regulative form -- rendering the following encoding (simplified for illustrative purposes):

\sp
\fcolorbox{black}{\boxbgcolor}{\begin{minipage}{15.5cm}
\begin{exi}
\A{}(Organic handling operations) \D{}(may) \NOTconst{}~\I{}(apply)  \Bdir{}(paragraphs in this section) to \Bind{}(traces) \Bind{},\prop{}(of genetically modified material).
\end{exi}
\end{minipage}}
\sp

Similar to the previous case, the coding requires reformulation of the statement in regulative terms. However, unlike the previous case, the specification of the responsible actor is explicit in the referenced institutional statements, thus affording a reliable encoding in a form that is aligned with the syntactic structure of the related institutional statements. As indicated for all cases above, the coding of such statement can selectively take regulative and constitutive form where interpretational scope favors narrow and wide resolution of semantic linkages. As showcased for the first example, where coding is independent from analytical use, the statement can be concurrently encoded in both forms. 

\udl{Note:} As discussed in the context of the interpretational scope applied by the coder (see \cref{subsec:StatementTypesHeuristics}), encoding statements in constitutive and/or regulative forms may invoke varying levels of complexity with respect to necessary adaptations of the statements, and, depending on analytical objectives, potential modifications of the statement semantics (in the given example, a concrete actorship for regulative statements is presumed, whereas the right more generally is expressed in constitutive terms). This can require a reformulation that could be challenged on methodological grounds and thus considered during design of the study. In addition to considering a dual annotation in the first place, the potential mischaracterization of a statement as either regulative or constitutive may be sensibly assessed in small-scale inter-coder reliability tests to resolve disagreements and misconceptions early in the encoding process.

\newpage

\subsection{Data Structure Patterns}
\label{subsec:DataStructurePatterns}

\begin{emp}Note that this section is under ongoing development and will likely experience significant extension in upcoming releases.\end{emp}

\vspace{0.5cm}

Another aspect related to the features introduced for both regulative and constitutive statements is the recognition of data structure patterns. A pattern commonly observed is the notion of collections, such as composition of committees, definition of practices in terms of underlying activities, delineation of goals or outcomes to be pursued, etc.

\sp 

\udl{Collections:} While dominant in objects and constituting properties, respectively, those can occur across other components (e.g., context). Central here is the identification of a collection descriptor and the corresponding elements, along with the explicit specification of logical operators that link the individual elements.

\sp

Example: \begin{exi}\E{}(Health care practices) \F{}(consist of) \Pp{}(\lpconst preventative measures [\ANDconst] appropriate nutrition [\ANDconst] rest\rpconst)\end{exi}. 

\sp

\udl{Complex elements:} Where quantitative information is expressed, or listed (and thus potentially embedded in collections or referred to as a single item), the elements commonly follow a schematic structure specific to individual documents (e.g., based on style or disciplinary background), but follow general patterns, such as variations of the following: \begin{exi}\textbf{[qualifier] [comparator] [quantity] [unit] [object property] [object]}\end{exi}. 

\sp

Example: \begin{exi}significantly (qualifier) more than (comparator) 10 (quantity) tons (unit) high-quality (object property) building material (object)\end{exi}

\sp

While those patterns can vary in extent and detail, their consideration in project-specific coding guidelines can be useful in as far as they are relevant for analytical purposes.

\sp

\udl{Comparisons/Equations:} In many instances, statements, and more specifically, activation conditions, are expressed in terms of comparisons, such as ``if gender distribution within department staff substantially differs from distribution within the wider organization, \dots''. Comparisons of such kind can be expressed in equation form, following the general pattern \textbf{[left-hand side] [comparison operator] [right-hand side]}, where left- and right-hand sides can be of simple form (e.g., numeric) or complex (e.g., consisting of complex elements as introduced above). Comparison operators include the following: $<$, $\leq$, $\lesssim$, $\approx$, $=$, $\gtrsim$, $\geq$, $>$, amongst others such as $\neq$. Oftentimes, comparators can further be complemented by additional qualification of the relationships. 

\sp

Example: \begin{exi}if gender distribution (object) within department staff (object property) substantially (qualifier) differs ($\neq$) from distribution (object) within the wider organization (object property)\end{exi}

\sp

Naturally, generic mathematical expressions are likewise conceivable, but not subject to further elaboration at this stage. Where coding on finer levels of granularity (e.g., resolving comparative relationships) is relevant, level and specific form of decomposition should be considered as part of the study design.
\newpage

\clearpairofpagestyles
\chead{\headmark}
\automark[subsection]{subsection}
\ifoot{\ifootertext}
\ofoot{\ofootertext}
\cfoot{\thepage}

\subsection{IG Configurations (Institutional Grammar Profiles)}
\label{subsec:IgProfiles}

As discussed before, the various coding levels of IG 2.0 accommodate different analytical needs as well as complexity of the encoded documents. Commitment to a level, including all the associated features, may in some cases be too coarse-grained to accommodate analytical needs. This can include, for example, the omission of components entirely, as well as the selective considerations of features of higher coding levels. To capture the considered feature set, a coded dataset should be accompanied with the applied coding configuration. 

For this purpose, we specify a fine-grained configuration syntax that allows the choice of features across levels of IG 2.0. The features for individual levels as specified in this document are captured in \cref{tab:IGProfilesRegulative} for regulative statements, and \cref{tab:IGProfilesConstitutive} for constitutive statements, along with a symbol specification used for the definition of specific IG configurations, or profiles.

\begin{table}[h!]
\centering
\begin{threeparttable}
\begin{tabular}{p{3cm} p{9cm} p{2cm}}
\toprule
\textbf{Coding Level} & \textbf{Feature} & \textbf{Symbol} \\
\toprule
IG Core & \multicolumn{1}{p{10em}}{Attributes} & A \\
IG Core & \multicolumn{1}{p{10em}}{Object\tnote{1}} & B \\
IG Core & \multicolumn{1}{p{10em}}{Deontic} & D \\
IG Core & \multicolumn{1}{p{10em}}{Aim} & I \\
IG Core & \multicolumn{1}{p{10em}}{Context\tnote{2}} & C \\
IG Core & \multicolumn{1}{p{10em}}{Or else} & O \\
\midrule
IG Extended & \multicolumn{1}{p{10em}}{Attributes refinements} & A$\subtxtNormal{Ext}$ \\
IG Extended & \multicolumn{1}{p{10em}}{Object refinements\tnote{1}} & B$\subtxtNormal{Ext}$ \\
IG Extended & \multicolumn{1}{p{10em}}{Context refinements\tnote{2}} & C$\subtxtNormal{Ext}$ \\
\midrule
IG Logico & Statement references & R \\
IG Logico & Logical relationship annotations & L \\
IG Logico & Semantic annotations & S \\
IG Logico & Regulative function annotations\tnote{3} & \Ureg \\
\toprule
\end{tabular}

\begin{tablenotes}
\footnotesize
\item[1] Where the configuration is specific to direct or indirect object component, those can be indicated using \Bdirraw~and \Bindraw, respectively; where applied in combination with the specification of IG Extended features, `$\subtxtNormal{,Ext}$' is appended accordingly (e.g., \Bdirraw$\subtxtNormal{,Ext}$).
\item[2] Where the configuration is specific to activation condition or execution constraint component, this can be indicated using \Cacraw~and \Cexraw, respectively; where applied in combination with the specification of IG Extended features, `$\subtxtNormal{,Ext}$' is appended accordingly (e.g., \Cacraw$\subtxtNormal{,Ext}$).
\item[3] Where institutional functions are annotated for both regulative and constitutive statements, the symbol U (without subscript) is used.
\end{tablenotes}
\end{threeparttable}

\caption{IG Feature Specifications for Regulative Statements}
\label{tab:IGProfilesRegulative}
\end{table}

\begin{table}[h!]
\centering

\begin{threeparttable}
\begin{tabular}{p{3cm} p{9cm} p{2cm}}
\toprule
\textbf{Coding Level} & \textbf{Feature} & \textbf{Symbol} \\
\toprule
IG Core & Constituting Properties & P \\
IG Core & Modal & M \\
IG Core & Constituted Entity & E \\
IG Core & Constitutive Function & F \\
IG Core & Context\tnote{1} & C \\
IG Core & Or else & O \\
\midrule
IG Extended & Constituting Properties refinements & P$\subtxtNormal{Ext}$ \\
IG Extended & Constituted Entity refinements & E$\subtxtNormal{Ext}$ \\
IG Extended & Context refinements\tnote{1} & C$\subtxtNormal{Ext}$ \\
\midrule
IG Logico & Statement references & R \\
IG Logico & Logical relationship annotations & L \\
IG Logico & Semantic annotations & S \\
IG Logico & Constitutive function annotations\tnote{2} & \Ucon \\
\toprule
\end{tabular}

\begin{tablenotes}
\footnotesize
\item[1] Where the configuration is specific to activation condition or execution constraint component, this can be indicated using \Cacraw~and \Cexraw, respectively; where applied in combination with the specification of IG Extended features, `$\subtxtNormal{,Ext}$' is appended accordingly (e.g., \Cacraw$\subtxtNormal{,Ext}$).
\item[2] Where institutional functions are annotated for both regulative and constitutive statements, the symbol U (without subscript) is used.
\end{tablenotes}
\end{threeparttable}

\caption{IG Feature Specifications for Constitutive Statements}
\label{tab:IGProfilesConstitutive}
\end{table}

Using coding levels, along with \textbf{--} and \textbf{+} symbols in combination with specific features references as listed in the table, we can express specific coding configurations, or coding profiles, that allow the omission or inclusion of features across all levels, or the selective coding of specific components based on lower coding levels.

Abstractly specified, a configuration has the following structure (where $<$ and $>$ embeds symbol characterizing features omitted from or added to the baseline coding level): 

\sp

{\small
\begin{emp}\textbf{$<$Baseline level$>$--$<$omitted features from baseline level$>$+$<$additional features from higher level$>$}\end{emp}}

\sp

\noindent\udl{Examples:} To capture the commitment to IG Core, along with the Context coding from IG Extended (e.g., component-level nesting, use of taxonomies, is specified as the configuration \begin{emp}\textbf{IG Core+C$\subtxtNormal{Ext}$}\end{emp}, where the \begin{emp}\textbf{+C$\subtxtNormal{Ext}$}\end{emp} signals features from the next higher level (IG Extended). 

Conversely, we can specify coding on IG Core level as the baseline, but without the consideration of Or else components as \begin{emp}\textbf{IG Core--O}\end{emp}. Where multiple components are omitted, we can specify \begin{emp}\textbf{IG Core--IO}\end{emp}, where features should be referred to in the order specified in the table (here: \begin{emp}Aim\end{emp} before \begin{emp}Or else\end{emp}). Selectively capturing features from IG Logico in IG Core-based coding, \begin{emp}\textbf{IG Core+R}\end{emp} indicates the coding of statement relationships in addition to the base IG Core coding.

Finally, omission and extensions can be combined, with omissions specified first, followed by feature additions, such as \begin{emp}\textbf{IG Extended-B\Cexraw+S\Ureg}\end{emp}, to signal the coding of Object and execution constraints on IG Core level, while considering semantic annotations and institutional functions (here for regulative statements only) in addition to this (reduced) IG Extended baseline. Complementing this discussion for the highest level, \begin{emp}\textbf{IG Logico-S}\end{emp} would imply complete coding on IG Logico level under omission of semantic annotations only.

Combining both omission and extension leverages complete flexibility with respect to the composition of features, and, where analytically useful, in principle even foregoing the inclusion of components defined necessary for institutional statements (i.e., A, I, and C component for regulative statements; F, E, and C for constitutive statements). For example, modeling the selective omission of components entirely, along with the inclusion of advanced features, \begin{emp}\textbf{IG Core-AI+C$\subtxtNormal{Ext}$SU}\end{emp} signals IG Core baseline encoding under omission of Attributes and Aim, while adding refined coding of Context (based on IG Extended), along with features from IG Logico, namely semantic annotations and institutional functions (here both for regulative and constitutive statements).

A common encoding level that offers the smallest possible extension to previous coding practice of institutional statements based on Crawford and Ostrom's original grammar is \begin{emp}\textbf{IG Core+C$\subtxtNormal{Ext}$}\end{emp}.

\noindent\udl{Custom refinements:} Where coders seek more fine-granular refinements (e.g., applying a subset of the features of a given configuration (e.g., coding objects without properties), such modifications should be indicated alongside the specified configuration. Similarly, extensions (e.g., additional taxonomies, or extensions or substitution of existing ones) should be documented alongside the configuration.

\clearpairofpagestyles
\chead{\headmark}
\automark[section]{section}
\ifoot{\ifootertext}
\ofoot{\ofootertext}
\cfoot{\thepage}

\section{Taxonomies}
\label{sec:Taxonomies}

This section provides an overview of the taxonomies for the categorization of components, parts thereof, or annotation schemes including (but not limited to) the ones referred to from the coding guidelines in \cref{sec:CodingGuidelines}. The overview is largely summarizing, with essential specification of labels, but limited conceptual elaboration, which is provided in the corresponding guidelines (\cref{sec:CodingGuidelines}). The taxonomies further specify the \begin{emp}label prefixes\end{emp} used to ensure unambiguous reference to the respective taxonomy/ies. Where only the circumstances taxonomy is used, the use of labels is optional.

The extension of existing taxonomies and introduction of additional taxonomies to draw seek epistemological linkage to the field of concern (e.g., to accommodate domain-specific characteristics or analytical necessities) is explicitly permitted. This includes the use of ontologies to support conceptually richer classification of entities. Any adjustment should be clearly indicated as part of the dataset specification (see \cref{sec:Checklist}). 

\begin{emp}Note that the taxonomies in this section are subject to ongoing empirical evaluation and may experience refinement in the future.\end{emp}

\subsection{Context Taxonomy}
\label{subsec:ContextTaxonomy}

The Context Taxonomy captures contextual characterizations with respect to temporal, spatial and various further descriptors that capture institutional context more accurately. It is a systematic extension of the descriptors of the \begin{emp}Conditions\end{emp} component as highlighted in the original grammar specification~\citep{Brady2018InstitutionalGuidelines}. Note that the listed categories include an embedded hierarchy, with more specific labels indented. 

\sp 

\noindent \udl{Specific considerations:}

\begin{itemize}
\item Where possible (and analytically useful/specified in project-specific guidelines), the most specific annotation of a given category should be used (e.g., `point in time', as opposed the more general `temporal' category).
\item Note that selected categories are more general in nature (e.g., the `state' category, since it potentially captures temporal, spatial and domain characterizations). As before, the coder should attempt to classify context in more specific terms.
\item Multiple annotations are explicitly suggested in cases where the characterizations overlap for an annotated element (e.g., `temporal' and `event'). Where possible, the coder may further consider the separation of distinctive activation conditions/execution constraints by category and annotate those correspondingly.
\end{itemize}

\noindent The suggested annotation label prefix -- if applied -- is \begin{emp}\ctx\end{emp}.

\sp

\noindent \udl{Categories:}

\begin{itemize}
\item[] Substantive Context

\begin{itemize}
\item Temporal (tmp) -- Conditions/Constraints associated with time - the when
	\begin{itemize}
	\item Point in time (tim) -- References to specific points in time
		\begin{itemize}
		\item Beginning (e.g., ``from 1st January'')
		\item End (e.g., ``until 31st January'')
		\end{itemize}	
	\item Time frame (tfr) -- References to time frames
	\item Frequency (frq) -- References to frequencies (e.g., ``annually'') 
	\end{itemize}
\item Spatial (spt) -- Conditions/Constraints associated with spatial representations -- the where
	\begin{itemize}
	\item Location (loc) -- References to specific locations
		\begin{itemize}
		\item Beginning
		\item End
		\end{itemize}
	\item Direction (dir) -- References to directions, inclusion of intermediary locations (e.g., ``via'')
	\item Path (pth) -- References to pathways (e.g., ``through the valley'')
	\end{itemize}
\item Domanial (dom) -- Conditions/Constraints applying to a specified activity, topical or existential realm. 
\begin{itemize}
\item Activity realm  (e.g., ``during decision-making'')
\item Topical realm (e.g., ``for drinking water'')
\item Existential realm (e.g., ``during childhood'')
\end{itemize}
\end{itemize}
\end{itemize}

\begin{itemize}
\item[] Procedural Context

\begin{itemize}
\item Order (prc) -- Conditions/Constraints associated with explicit or implied execution (procedural) order (e.g., ``\dots according to the following stages: \dots''). Operationally, this can include expressions of input into the activity identified in the institutional statement (e.g., ``with input from \dots''). Procedural order can further include the explicit specification of required actions at any step in the procedural execution. 

\item Method (met) -- Conditions/Constraints associated with means or method by which an action is performed, recognizing the following specializations:
	\begin{itemize}
	\item Means -- Action as method (e.g., ``by handshake'')
	\item Instrument -- Artifact as method (e.g., ``by car'')
	\end{itemize}
Note that the method specification is more general than the characterization of distinctive procedural steps under the Procedural `Order' category.
\end{itemize}
\end{itemize}

\begin{itemize}
\item[] Aspirational Context

\begin{itemize}
\item Purpose/Function (pur) -- Conditions/Constraints describing the purpose or intent of an aim or constitutive function; generally output of action (e.g., ``for the purpose of reducing pollution levels'')
\end{itemize}
\end{itemize}

\begin{itemize}
\item[] Situational Context

\begin{itemize}
\item State (ste) -- References to a specific state environmental state (e.g., ``when traffic light is red''); note that state characterization is general in kind and commonly associated with an entity that holds the referenced state (e.g., traffic light).  Where possible, a more specific annotation should be chosen.


\item Event (evt) -- Conditions/Constraints referencing specific events (e.g., ``on arrival at the airport \dots''). In contrast to the state characterization, an events is instantaneous in nature, whereas a state can persist over longer time frames and is associated with an entity whose state is described.

\end{itemize}
\end{itemize}



\newpage

\subsection{Animacy Taxonomy}

The Animacy Taxonomy captures differentiates between animate and inanimate entities, maintaining compatibility with annotation conventions commonly adopted in datasets coded according to the previous IG coding guidelines. The suggested annotation label prefix is \begin{emp}\anim\end{emp}.

\begin{itemize}
\item Animate -- Living entities
\item Inanimate -- Non-living entities, both real and mental constructs
\end{itemize}

\subsection{Metatype Taxonomy}

The Metatype Taxonomy differentiates between concrete physical and abstract entities, e.g., mental constructs. This also applies to component-level nesting (see \cref{subsec:ComponentLevelNesting}) for cases in which an institutional statement is nested in a statement component, such as an object, in which case the object is characterized as abstract in nature. The suggested annotation label prefix is \begin{emp}\metatype\end{emp}.

\begin{itemize}
\item Abstract -- Abstract entities 
\item Concrete -- Concrete entities
\end{itemize}

\subsection{Role Taxonomy}

The role taxonomy serves the annotation of attributes and objects with additional labels to capture their role within a statement structure with respect to the action, organized by relevant role of actor and effect in terms of affected subjects. The suggested annotation label prefix is \begin{emp}\role\end{emp}.

\begin{itemize}
\item Role Characterizations
\begin{itemize}
    \item Originator/Causer/Agent -- Entity from which action originates
    \item Recipient -- Recipient of an artifact/sanction
    \item Possessor -- Owner of an object/entity (e.g., ``house owner'')
\end{itemize}
\item Effect Characterizations
\begin{itemize}
    \item Experiencer -- Indirectly affected actor (e.g., ``observer of non-compliance'')
    \item Advantaged -- Beneficiary distinctively advantaged by referenced activity/function; may not necessarily be recipient (e.g., ``welfare recipient'')
    \item Disadvantaged -- Maleficiary distinctively burdened by referenced activity/function; may not necessarily be recipient
\end{itemize}
\end{itemize}

\subsection{Regulative Functions Taxonomy}
\label{subsec:regfunc}

Some types of syntactic annotations can aid the coder in discerning and capturing information that signals the broader function of institutional statements, as indicated by components of which they are comprised. Those are referred to as ``institutional function'', and more specifically ``regulative functions'' in as far as regulative statements are concerned (Constitutive functions are discussed in \cref{subsec:confunc}). Regulative functions facilitate the annotation of aims in order to capture the correspondence of aims to analytical functions of relevance through specific theoretical lenses. Exemplifying the use of the institutional grammar for the analysis from a regulatory compliance perspective, compliance and violation behaviour is of specific concern, whereas institutional life cycles may require the annotation of action verbs signalling the initiation of termination of institutional arrangements. Note that the offered taxonomy provides examples for regulative functions organized by categories (alongside potential specializations), but does not claim exhaustiveness. The suggested annotation label prefix is \begin{emp}\regfunc\end{emp}.

\begin{itemize}
\item Compliance action -- action reflecting compliance behavior
	\begin{itemize}
	\item Comply -- action reflecting compliance
	\item Violate -- action reflecting violation
	\end{itemize}

\item Monitor -- action reflecting the institutional function of monitoring
	\begin{itemize}
	\item Detect compliance -- action reflecting the detection of compliance
	\item Detect violation -- action reflecting the detection of violation
	\end{itemize}

\item Enforce -- action reflecting enforcement acts
	\begin{itemize}
	\item Reward -- action reflecting rewarding behaviour (regulative-incentivizing)
	\item Sanction -- action reflecting sanctioning behaviour (regulative-punitive)
	\end{itemize}

\item Enforcement response -- action reflecting responses to enforcement outcomes
	\begin{itemize}
	\item Accept -- action reflecting acceptance of enforcement outcome
	\item Reject -- action reflecting rejection of enforcement outcome
		\begin{itemize}
		\item Appeal (specialization of reject) -- action reflecting appeal against enforcement outcome
		\end{itemize}
	\end{itemize}

%
%
%
%
%
\end{itemize}

As noted above, the regulative functions concept is intended to draw the epistemological linkage to analytical frameworks and/or concepts associated with a given domain. 

\subsection{Constitutive Functions Taxonomy}
\label{subsec:confunc}

The constitutive function taxonomy is complementary to the regulative functions taxonomy under the shared  institutional functions umbrella. Constitutive function annotations offer a semantic characterization of specific constitutive functions in relation to the constituted entity and/or the linkage of constituted entity and constituting properties, respectively. The top-level distinction of constitutive functions is the identification of the constituted entity as either an \begin{emp}entity\end{emp} established, modified or otherwise referenced in the context of the policy document, or the \begin{emp}policy\end{emp} itself. On a more fine-grained level, the categories capture the role of the constituting function with respect to the constituted entity (i.e., a specific entity, or the policy). The suggested annotation label prefix is \begin{emp}\confunc\end{emp}.

The structure, alongside specific annotations, is visualized in \cref{fig:ConstitutiveFunctions} and described in the following. 
\begin{figure}[h!]
    \centering
    \framedfigure{\includegraphics[width=0.75\textwidth]{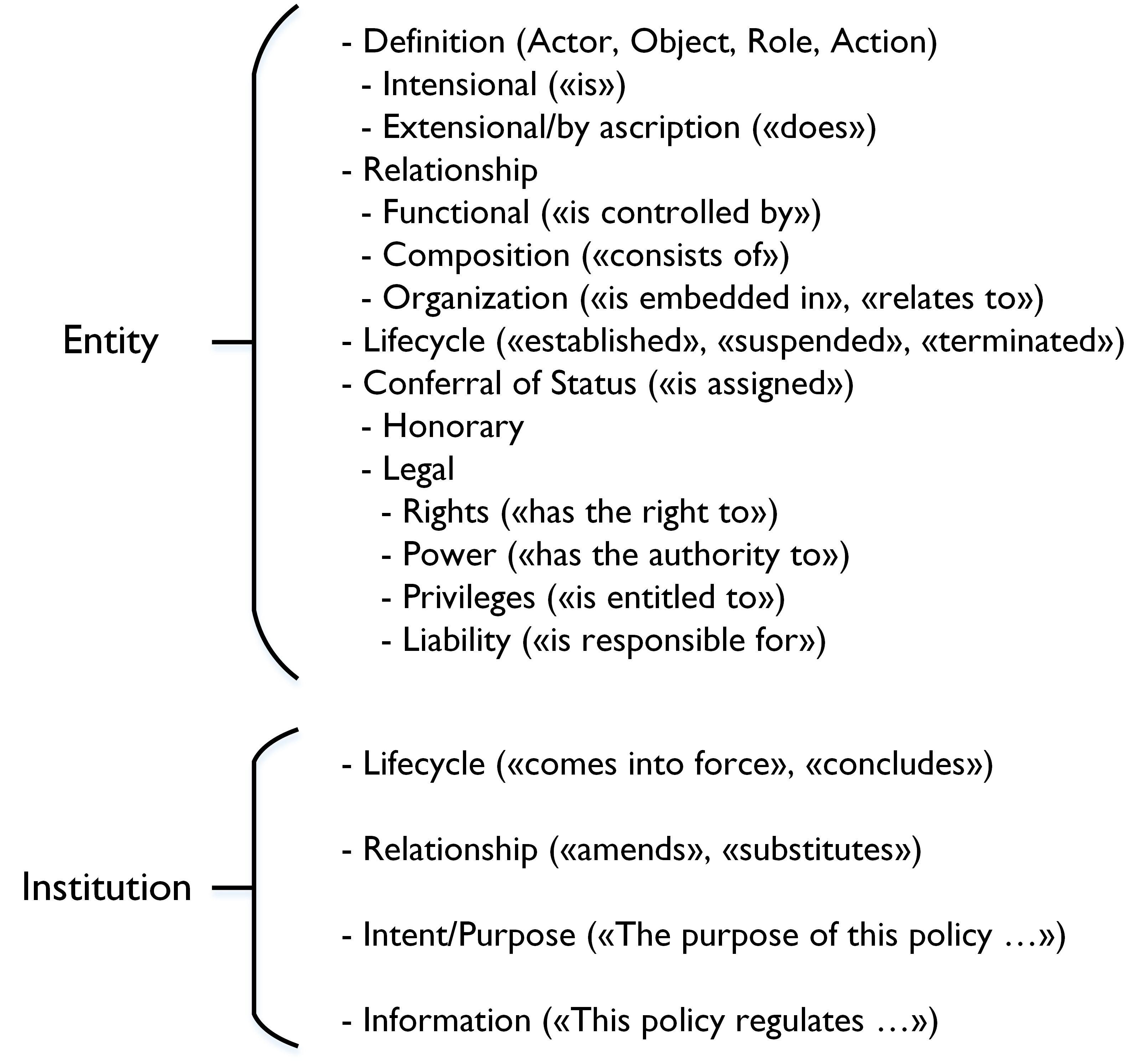}}
    \caption{Constitutive Functions}
    \label{fig:ConstitutiveFunctions}
\end{figure}
Where entities are subject to the statement, constitutive functions can capture the definition of constituted entities as relevant for the parameterization of the institutional setting, including actors, object/artifacts, role specifications or actions, alongside status characterizations. This definition is often signaled intensionally (i.e., by explicit definition), or implied by ascription (e.g., implicit characterization of entity based on exhibited behavior, etc.).

In addition to the definition of entities, constitutive functions can capture relationships of different kinds, including organizational relationships in the form of compositions (e.g,. specification of committee membership) and further reflect hierarchical relationships (e.g., leadership structures, embedding positions within organizations).

Possible constitutive functions further include the initiation and termination of entity lifecycles (e.g., dates of appointment termination, etc.). 

Finally, the explicit conferral of status (e.g., in the form of institutional power, such as authority or  competence; rights; privileges, or liability) is a central application of constitutive statements. 

Complementing the perspective on entity specification as part of constitutive statements, constitutive functions that characterize the policy or document itself, have a varying set of functions. Typical characterizations include the policy lifecycle (e.g., date of enactment), as well as its relationship to other policy (e.g., amending or superseding it). Further statements refer to the purpose or intent underlying a given policy. Informational statements offer supplementary information about the document, or state institutional facts contextualizing the policy or domain of concern. Note that the latter is often expressed in terms of statement of fact, rather than institutional statements. 

As with the preceding taxonomies, the constitutive functions taxonomy is subject to further refinement based on ongoing empirical validation efforts.

\section{Statement-Level Annotations}

In addition to annotations pertaining to individual components or component types, annotations may also apply to statements as whole. Most notably, this includes the characterization of statements based on their function in the context of an interlinked institutional statement.

\subsection{Governance Types (Vertical Nesting Annotations)}

The governance annotation relates to the nature of the statement as either \begin{emp}monitored\end{emp} or \begin{emp}consequential\end{emp} to signal regulated, or monitored part of a statement (e.g., \begin{emp}
``citizen must comply with the law, \dots''\end{emp}), in contrast to the consequential statement that signals the associated consequence (e.g., \begin{emp}
``\dots~or else official must sanction citizen.''\end{emp}). Augmenting this, a \begin{emp}monitoring\end{emp} statement highlights the potentially dissociated function of a third-party observer (e.g., bystander, appointed monitor), who may or may not be the enforcer. As a consequence, statements can have multiple statement-level annotations. 

\subsection{Consequence Types}

A second annotation relates to the nature of potential consequences. The main differentiation includes consequences of \begin{emp}existential\end{emp}, and \begin{emp}non-existential\end{emp} kind. 

Consequences of the first kind are understood as ontologically existential, i.e., reflecting whether an entity does not come about, is invalidated, etc. This can further include the institution itself (e.g., lacking preconditions for its continuation), hence leveraging \begin{emp}configurational consequences\end{emp}. 

\begin{emp}Social consequences\end{emp}, as common institutional consequences, are related to economic situation, status, or penalties. As far as the institution is concerned, these are considered non-existential. 

As with other taxonomies, the analyst may introduce subdifferentiations that respond to specific analytical objectives, or establish epistemological linkages to the domain of concern.

The listing of taxonomies complements and concludes the supplementary coding guidelines for the Institutional Grammar 2.0.


\clearpairofpagestyles
\chead{\headmark}
\automark[section]{section}
\ifoot{\ifootertext}
\ofoot{\ofootertext}
\cfoot{\thepage}

\section{Conclusion}
\label{sec:Conclusion}

This codebook provides operational guidance for the planning and execution of encoding institutions using the Institutional Grammar 2.0. This information is supplementary to the conceptual introduction provided in \citet{Frantz2021,Frantz2022InstitutionalGrammar}. 

Following a general introduction of the principles and motivations of the Institutional Grammar, the codebook initially outlines the theoretical concepts underlying IG 2.0, including the coding on multiple levels of expressiveness (IG Core, IG Extended, IG Logico) in \cref{sec:Definitions}, before supporting the navigation through the codebook based on the reader's objectives and background in the form of a Reader's Guide in \cref{subsec:ReadersGuide}. This is followed by the discussion of essential document preparation steps (pre-coding steps) in \cref{sec:PrecodingSteps} that complement the broader design considerations outlined in \citet{Frantz2022InstitutionalGrammar}, central aspects of which are captured in the Study Design Checklist appended to this codebook (see Appendix \ref{sec:Checklist}).

Following the general methodological orientation, the codebook provides detailed coding guidelines for regulative and constitutive statements across all IG 2.0 levels of expressiveness (\cref{sec:CodingGuidelines}). This is followed by advanced concepts, such as the discussion of encoding hybrid institutional statements (\cref{subsec:ConstitutiveRegulativeHybrids}), i.e., institutional statements consisting of both regulative and constitutive components, polymorphic institutional statements (\cref{subsec:SyntacticPolymorphs}), and the discussion of structural patterns (\cref{subsec:DataStructurePatterns}). 

A mechanism to document the parameterization of IG 2.0 studies based on the selective application of IG 2.0 components and features -- in the form of IG 2.0 Profiles -- is provided in \cref{subsec:IgProfiles}. The listing of taxonomies to support the semantic annotation of institutional information in the previous section (\cref{sec:Taxonomies}) concludes the substantive part of the codebook.

It is important to note that the guidelines provided in this codebook are of general nature and emphasize the \begin{emp}operational use\end{emp} of IG 2.0, with specific focus on the encoding of institutional information. Doing so, they may not capture specifics potentially relevant for a given project, let alone the detailed introduction of the underlying theoretical concepts. For an in-depth introduction into the Institutional Grammar, existing applications, challenges that motivate the IG 2.0 and detailed conceptual foundations, the reader is referred to companion literature~\citep{Frantz2021,Frantz2022InstitutionalGrammar}. 

However, the codebook aims to support the development of supplementary project-specific guidelines that consider application context (e.g., domain, language, types of documents, legal traditions, etc.) and analytical objectives (i.e., evaluation of encoded statements) more explicitly, aspects that a general codebook cannot consider. As indicated above, the reader's guide in \cref{subsec:ReadersGuide} intends to highlight relevant sections. It is furthermore important to note that the coding instructions provided here are intentionally tool-agnostic, and open to adaptation to arbitrary encoding means (e.g., spreadsheets, text annotation tools, etc.), which themselves may be augmented with specific guidance. 

For a complete overview of relevant resources, including a theoretical treatment of the underlying concepts and principles, as well as resources that support operational coding (e.g., cheat sheet, tool-specific guidance, software), please refer to \url{https://newinstitutionalgrammar.org/resources}. Given the ongoing development and application efforts, the website is updated whenever relevant new material becomes available.

Please further note that these guidelines will be continuously refined based on theoretical developments, feedback from users as well as ongoing empirical validation efforts. To retrace subsequent changes, please note the specific version and version history of these guidelines outlined at the beginning of this document. Irrespective of refinements, all revisions of the codebook will be retained for future reference.


\printbibliography[heading=bibintoc]

\newpage

\begin{appendices}
\section{Study Design Checklist}
\label{sec:Checklist}

This checklist summarizes considerations relevant for any study design using IG 2.0. The checklist is not prescriptive, but rather a collation of the methodological aspects introduced throughout the codebook, serving the development of a project-specific coding protocol. Reflecting on these aspects in the study design phase aims at ensuring high levels of reliability and validity of the coded institutional information. 

Overarching the development of a project-specific codebook adaptation is the question as to whether you intend to develop a generically useful dataset (useful for later adoption using specific techniques or for comparative studies), or whether your coding is purpose-specific (i.e., aimed at addressing specific needs as well as relying on specific analytical techniques). This will inform choices with respect to statement types, encoding heuristics, as well as help establish the relevant IG feature set as well as the scope of encoding, both in terms of number of statements and interpretational scope (see below).

\begin{itemize}
\item Identify applicable statement types (regulative, constitutive, hybrids)
\item Identify level of expressiveness (IG Core, IG Extended, IG Logico), or customized characterizations
\begin{itemize}
\item See \cref{subsec:IgProfiles}
\end{itemize}
\item Decide on the interpretational scope of statement encoding (narrow vs.~wide)
\begin{itemize}
\item Note: This is of specific relevance when coding both constitutive and regulative statements, since it influences the decision heuristics.
\item See \hyperlink{tablink:InterpretationalScopeOfInstitutionalStatements}{here} (\cref{subsec:StatementTypesHeuristics})
\end{itemize}
\item Decide on applicable heuristics for Attributes/Attributes Properties, Objects/Object properties, Constituted Entity/Constituted Entity Properties, and Constituting Properties/Constituting Properties Properties respectively
\begin{itemize}
\item See \hyperlink{tablink:AttributePropertyHeuristics}{here} (\cref{subsubsec:IGCoreRegulative})
\end{itemize}
\item Decide on specific pre-coding/processing steps based on the suggestions listed in \cref{sec:PrecodingSteps}
\begin{itemize}
\item This is particularly relevant where potential reconstruction of institutional statements (conceptual reification) is considered prior to coding.\footnote{For a detailed discussion of conceptual reification, see also Chapters 3 and 5 in \citet{Frantz2022InstitutionalGrammar}.}
\item Decide on aspects such has component scoping during encoding. This includes the decision whether prepositions (`by', `above', `beyond') and articles (`a', `the') are included in the encoding. For instance, if employing modelling techniques, considering such information is counter-productive in the encoding process. An alternative vehicle is the use of semantic annotations to ensure unambiguous labelling of specific component values, especially where you experience divergence in expression (e.g., `EU member states', `member states', `members').
\item Further consider project-specific preparation and coding conventions under consideration of domain-specific and epistemic considerations.
\end{itemize}
\item Decide on applicable taxonomies (see \cref{sec:Taxonomies}), as well as possible extensions and/or additional taxonomies (derived from domain-specific theories or frameworks, for instance).
\end{itemize}

\end{appendices}

\end{document}